\documentclass[11pt]{article} \usepackage{mystyle-new}
\usepackage{epsfig,amsmath} \usepackage{graphics}
\usepackage{hyperref}
\usepackage{cite,./mcite}
\usepackage{epsfig,amsmath} 
\usepackage{hepnames,hepunits}
\usepackage{comment}
\usepackage{authblk}
\usepackage{xspace}
\usepackage[export]{adjustbox}

                \def\lsim{\mathrel{\rlap{\lower4pt\hbox{\hskip1pt$\sim$}}
    \raise1pt\hbox{$<$}}}                \def\gsim{\mathrel{\rlap{\lower4pt\hbox{\hskip1pt$\sim$}}
    \raise1pt\hbox{$>$}}}

\def\prp{t}

\def\kt{\ensuremath{k_{\prp}}\xspace}

\newcommand{\pt}{p_{t}}

\def\RAPGAP{{\sc Rapgap\xspace}}
\def\ARIADNE{{\sc Ariadne\xspace}}

\def\HERWIG{{\sc Herwig\xspace}}

\def\herwig{{\sc Herwig\xspace}}

\def\ktclus{{\sc ktclus}\xspace}
\def\fastjet{{\sc FastJet}\xspace}

\newenvironment{tolerant}[1]{\par\tolerance=#1\relax}{ \par }

\newcommand{\thischapter}{ }
\newcommand{\thissection}{ }

\def\hztool {HZTool\xspace}
\def\rivet {Rivet\xspace}
\def\rivethztool {RivetHZTool\xspace}
\def\hera{HERA\xspace}

\usepackage{lineno}
\makeindex
\begin{document}

\title{\rivet, \rivethztool and \hera  \\ \vspace*{0.2cm}
A validation effort for coding \hera\ measurements for \rivet }

\author[1]{M.~I.~Abdulhamid}
 \affil[1]{\small Department of Physics, Tanta University, Egypt}
\author[2]{A.~Achilleos}
  \affil[2]{Department of Physics and Astronomy, University of Manchester, UK}
\author[3]{A.~Bermudez~Martinez} 
  \affil[3]{Deutsches Elektronen-Synchrotron DESY, Germany}
  \author[4]{C.~Bierlich}
  \affil[4]{ Department of Astronomy and Theoretical Physics, Lund University, Sweden} 
\author[5]{Giorgia~Bonomelli}
\affil[5]{Università degli Studi Milano-Bicocca, Italy}
\author[6]{A.~Borkar}
   \affil[6]{Department of Physics, Sardar Vallabhbhai National Institute of Technology, Surat, India}
\author[7]{A.~Buckley}
  \affil[7]{School of Physics \& Astronomy, University of Glasgow, Scotland}   
\author[8]{J.~M.~Butterworth}
  \affil[8]{Department of Physics and Astronomy, UCL, UK}   
\author[9]{M.~Chithirasreemadam}
  \affil[9] {UFR Sciences, Universite Paris Saclay, France}
\author[10]{M.~Davydov}
   \affil[10]{Department of Physics, Moscow State University, 119899 Moscow, Russia}
\author[3]{L.I.~Estevez~Banos}
\author[11]{K.~Moral~Figueroa}
    \affil[11]{School of Physics \& Astronomy, University of Edinburgh, UK}
\author[12]{A.~B.~Galv\'an}
  \affil[12]{Universitat Polit\`ecnica de Catalunya (UPC), Barcelona, Spain}
\author[8]{C.~G\"utschow}
\author[3]{H.~Jung}
\author[13]{S.~Kim}
\affil[13]{University of Oxford, UK}
\author[14]{K.~Koennonkok}
 \affil[14]{Department of Physics, Mahidol University, Thailand}
\author[15]{A.~León~Quirós}
  \affil[15]{Universidad Autónoma de Madrid, Madrid, Spain}
\author[16]{L.~Marsili}
  \affil[16]{ Department of Physics and Astronomy, University of Bologna, Italy}
  \author[3]{M.~Mendizabal}
\author[17]{S.~Pl\"atzer}
  \affil[17]{Institute of Physics, NAWI Graz, University of Graz, Universit\"atsplatz 5, A-8010 Graz, Austria and Particle Physics, Faculty of Physics, University of Vienna, Boltzmanngasse 5, A-1090 Wien, Austria}
\author[18]{N.~Rahimova}
  \affil[18]{Department of Physics, Yeditepe University, Turkey}
\author[3]{S.~Schmitt}
\author[8]{J.~Shannon}
\author[6]{S.~K.~Singh}
\author[19]{C.~S\"usl\"u}
  \affil[19]{Department of Physics, Ihsan Dogramaci Bilkent University, Turkey}
  \author[3]{S.~Taheri~Monfared}
\author[14]{N.~Trakulphorm}
\author[20]{P.~van Mechelen}
   \affil[20]{Elementary Particle Physics, University of Antwerp, Belgium}
\author[21]{A.~Verbytskyi}
   \affil[21]{Max-Planck Institut f\"ur Physik, Munich, Germany}
\author[3,22]{Q.~Wang}
   \affil[22]{School of Physics, Peking University, China} 
\author[23]{G.~Watt}
   \affil[23]{ IPPP, Department of Physics, Durham University, UK}
\author[24]{D.~Wilson}
  \affil[24]{ School of Physics, University of the Witwatersrand, South Africa}
\author[3,8]{M.~Wing}
\author[3,22]{H.~Yang}
\author[25]{W.~Zhang}
 \affil[25]{University of Cambridge, United Kingdom\vspace*{-22.5cm}}

\date{}
\begin{titlepage}
\maketitle
\begin{flushright}
DESY-21-222\\
\end{flushright}
\end{titlepage}

\begin{abstract}
During the DESY summer student program 2021, young scientists from more than 13 different countries worked together, connecting from remote, to provide computer codes within the \rivet framework for 19~\hera measurements. Most of these measurements were originally available within the \hztool package, but no longer accessible for modern analysis packages such as \rivet.
The temporary \rivethztool interface was used to validate most of the new \rivet plugins.
\end{abstract}

\section{Introduction}
Since HERA times, a dedicated effort has been made to include published measurements into computer codes which can be used for comparison of the data with predictions from event generators. One of the first general packages was \hztool~\cite{hztoolv1,Waugh:2006ip}, started in 1990, written in Fortran. With the preparation for LHC, new developments started to apply modern C++ and Python techniques, and the Rivet~\cite{Bierlich:2019rhm} package is now a standard for comparison of experimental measurements with theoretical calculations, in the form of particle-level Monte Carlo event generators and partonic next-to-leading order (NLO) calculations.

Since \rivet was developed after many of the \hera measurements were completed, a dedicated effort was needed to make \hera measurements available and accessible in \rivet. One of the complications was, that the original \hera publications contain comparison of measurements with predictions which are no longer accessible, thus making the validation of the new codes difficult. A few years ago, the \rivethztool~\cite{Rivet-HZTOOL-Vienna-2019,Rivet-HZTool-DESY-2019}  interface between \hztool and \rivet was developed, which allowed the Fortran code of \hztool to be called from inside \rivet. While in principle \rivethztool was functional, it had severe limitations, since all the histogramming was based on CERN HBOOK~\cite{Brun:1987vv}, and new developments were needed to port \rivethztool, originally developed for the \rivet 2.7.X series, to the newer \rivet~3 series. Instead of porting \rivethztool to \rivet~3, a dedicated effort was started to code \hera analyses directly into \rivet and to use \rivethztool as a validation tool only.

During the DESY summer student program 2021, young scientists from  13 different countries joined together, connecting from remote, working in virtual meetings together twice a day, to provide the tools for future comparison of HERA measurements with modern computational predictions. 

\section{\rivet and \hztool }
The  \rivethztool~\cite{Rivet-HZTOOL-Vienna-2019,Rivet-HZTool-DESY-2019} project was initiated several years ago, also as part of the DESY summer student program\footnote{\rivethztool is available upon request from simon.plaetzer@uni-graz.at or hannes.jung@desy.de.}. The idea was to use as much as possible the Fortran code of \hztool and build an interface to \rivet.  While \hztool was built upon the CERN HBOOK histogramming package,  in \rivethztool the Rivet histogramming package YODA was used, and all calls to HBOOK were replaced or wrapped. This attempt was successful, but some of the features of HBOOK were difficult to port into the \rivet histogramming package. In addition  several features of \rivet could not be easily ported back to \hztool, an  example is  the usage of event-multiweights in \rivet~3.

With the further development of \rivet, it was decided to concentrate the effort in coding HERA results entirely in \rivet. However, coding older analyses faces challenges: in most of the publications old and no longer supported Monte Carlo event generators were used, and sometimes the detailed parameter settings for the theoretical calculations were not properly quoted. In other cases it turned out that the data points stored on HEPData~\cite{Maguire:2017ypu} were either incomplete or sometimes missing, while they were available inside the \hztool codes. In a common effort between the summer-student code developers and the representatives from the H1 and ZEUS collaborations and HEPData, missing information was provided, approved and uploaded to HEPData to make it available for future analyses as well as for \rivet.

The working environment of \rivethztool, although only in \rivet 2.7.X, was used to validate newly coded \rivet plugins, applying exactly the same Monte Carlo event generator predictions.

\section{\rivet for HERA }
In this section a brief description of the H1 and ZEUS HERA analyses now available in \rivet is given, together with validation figures obtained from the corresponding \rivethztool code, where available. In all cases the \RAPGAP~\cite{Jung:1993gf,RAPGAP33} MC event generator was used, with CTEQ6L1~\cite{Pumplin:2002vw} and HERAPDF2.0 LO~\cite{Abramowicz:2015mha} parton densities. In order to study the effect of parton shower in addition to the DGLAP type parton shower (PS), the Color Dipole Model as implemented in \ARIADNE~\cite{Lonnblad:1999cx,Lonnblad:1992tz} was used.  In one case, the \HERWIG 7 MC event generator was applied~\cite{Bellm:2015jjp}.

In $ep$ scattering often different frames are used, the hadronic center-of-mass frame and the Breit frame. It is common, to define the hadronic center-of-mass frame such, that the virtual photon direction is along the positive $z$-axis (while the laboratory frame has usually the initial proton along the positive $z$-axis). In the Breit frame,  the photon is always along the negative $z$-axis. Some HERA analyses use a different definition of the hadronic center-of-mass frame, with the photon along the negative $z$-axis, and when distributions of rapidity or even cuts on rapidity are used, this has to be treated with special care.

The \rivet implementation of the deep-inelastic scattering (DIS) observables follows its established approach of agnosticism to the originating MC generator, and resulting strong preference to compute observables portably from final-state (or certainly post-hadronization) particles rather than from particular event-graph structures in and around the matrix-element representation. This approach post-dates and is subtly distinct from the more generator-specific ``truth level'' observable definitions used at the time of original measurement. In particular, the scattered lepton is identified not by event-graph connections, but by selection of the direct (not from hadron-decay) lepton with the largest $p_z$, $\eta$, or $E$ value (as specified by the analysis-routine author). The kinematic variables $x$, $Q^2$ and $y$ were calculated using the four-momentum of that lepton and the momenta of the unambiguously identified beam particles. Most of the time these definitions coincide with the HERA ones, as there are rarely ambiguities in direct-lepton choice, but it should be noted that this introduces a slight mismatch with the original definition, limiting the precision of phenomenology that can be conducted using these routines. The robustness of the fiducial approach for observable definition, as evidenced by the LHC measurement programme's adoption of the same method, argues for this approach to be taken by future $ep$ collider experiments.

The jet measurements at HERA were performed using the jet algorithms implemented in the Fortran \ktclus package, which implemented the longitudinally invariant \kt  clustering and inclusive jet reconstruction algorithms~\cite{Catani:1993hr,Ellis:1993tq,Catani:1992zp} with various definitions of the jet distances, which were not available in the 
 \fastjet~\cite{Cacciari:2011ma} package used in \rivet. The necessary additions were included in the \rivet  plugins, where needed.

In the description of the \rivet plugins, we first cite parts of the abstracts from the original publication, followed by a description of the validation of the \rivet plugin and some further information on the analyses. In the titles of the following subsections  both the \rivet and the \hztool identifiers are given.

The data points displayed in the following might differ between \rivet and \rivethztool , sometimes because only statistical uncertainties are used in \rivethztool , sometimes also the central values are slightly different.  We rely on the values which are stored on HEPData, and did not make any effort to update the data points in \rivethztool .

\subsection{Measurement of multiplicity and momentum spectra in the current fragmentation region of the Breit frame at HERA (ZEUS) (ZEUS\_1995\_I392386, HZ95007)}
\begin{figure}[htbp]
\begin{center}
\includegraphics[width=0.5\linewidth]{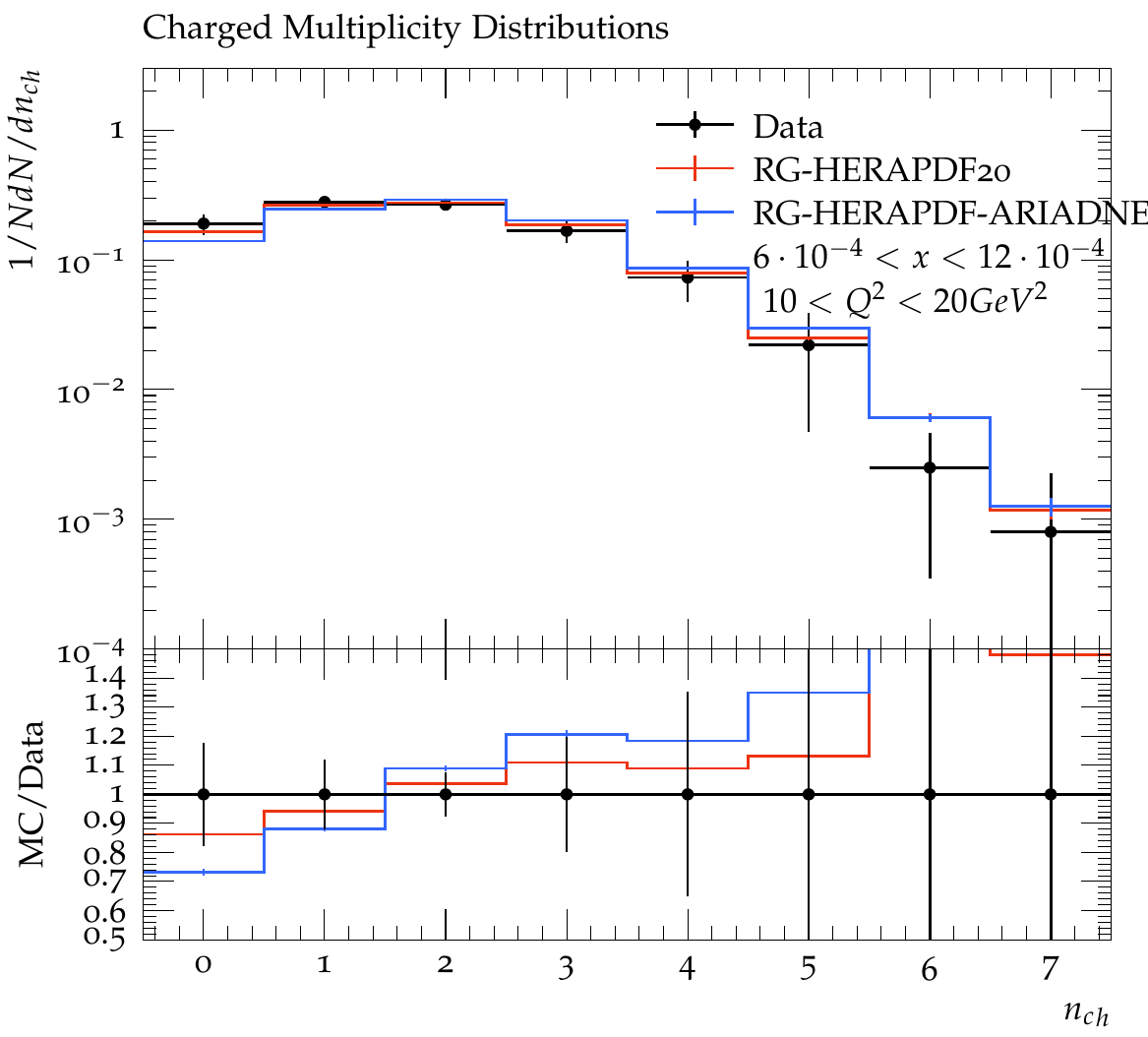}\includegraphics[width=0.5\linewidth]{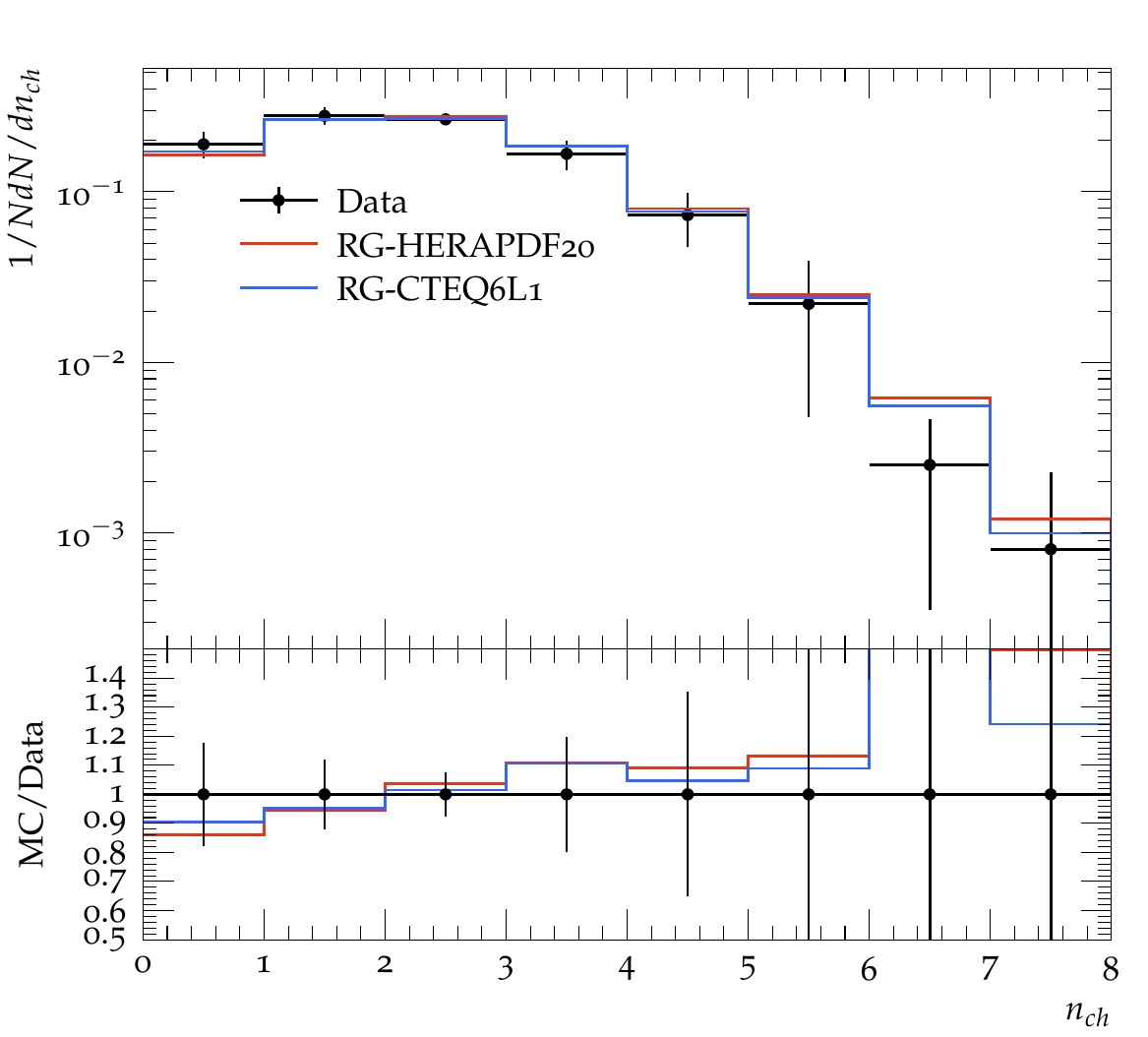}
\caption{Comparison of results of charged multiplicity distribution for   $6 \times 10^{-4} < x_{bj} <1.2 \times 10^{-3}$ and  $ 10 < Q^2 < 20 $ \GeV$^2$, using HERAPDF20 and CTEQ6L, obtained from \rivet (left) and the one from \rivethztool (right).
Please note the different definitions of bins for multiplicities between \rivet and \rivethztool .}
\label{fig:H1_1995_I392386_a}
\end{center}
\end{figure}

\begin{figure}[htbp]
\begin{center}
\includegraphics[width=0.5\linewidth]{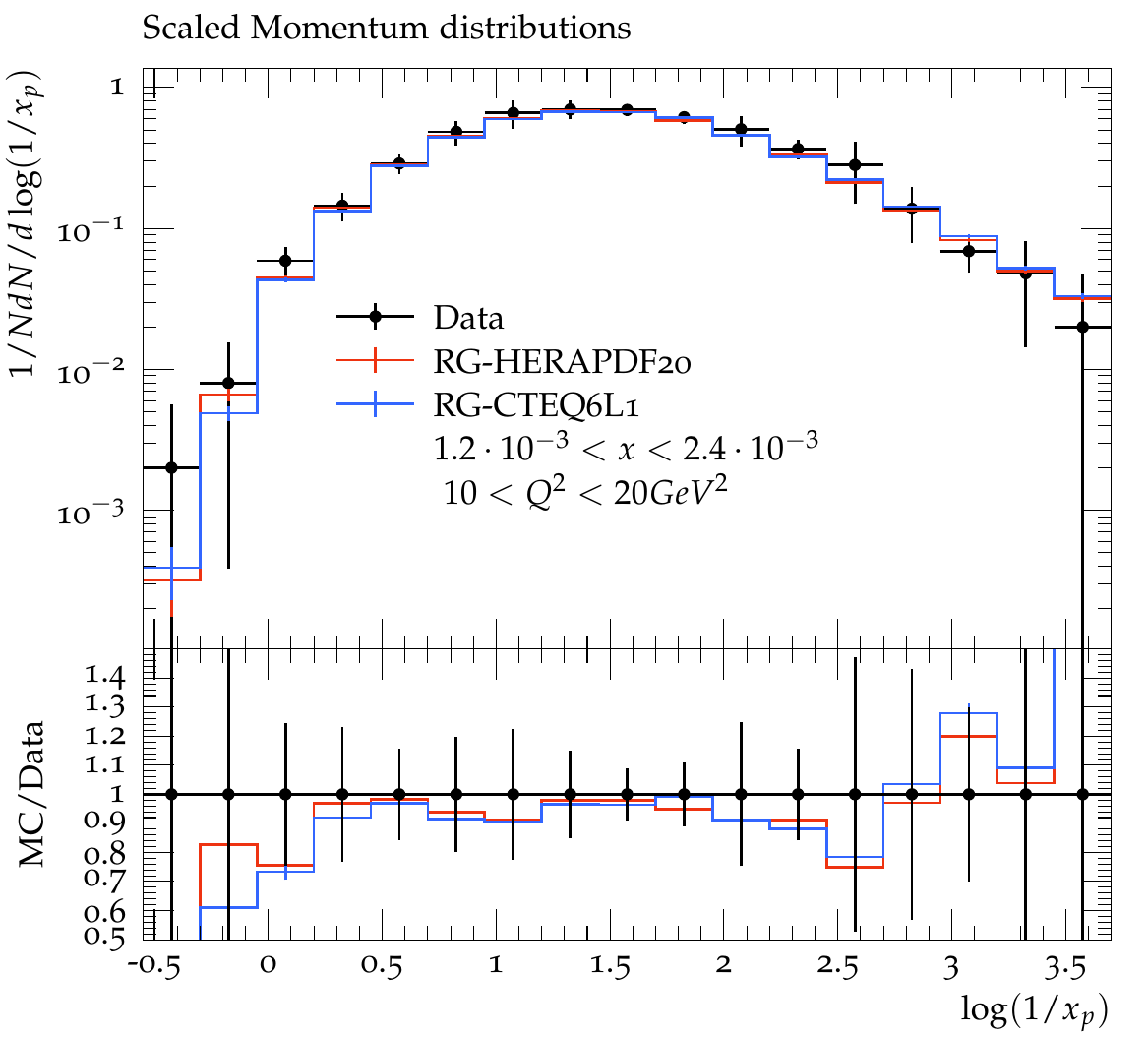}\includegraphics[width=0.5\linewidth]{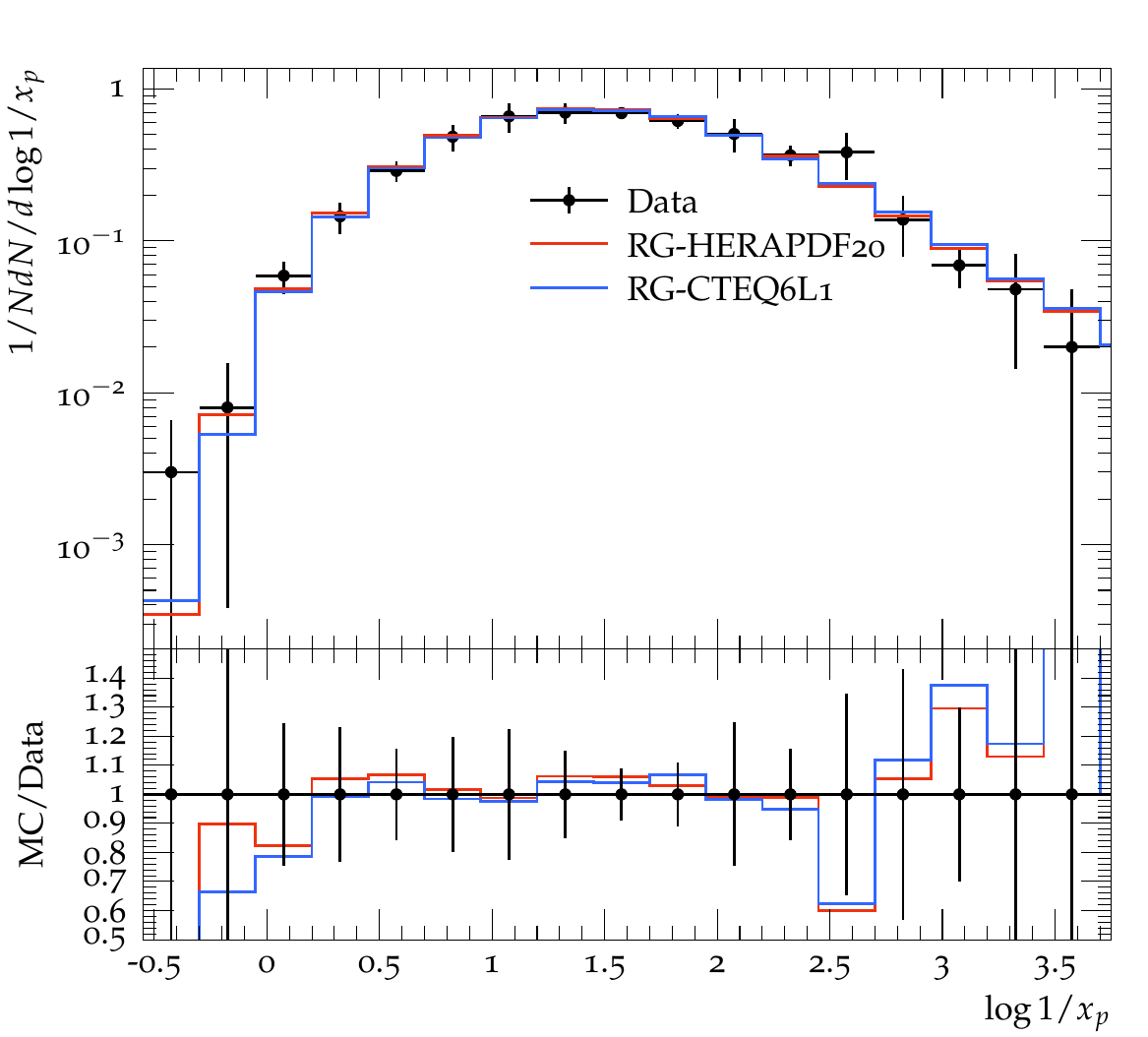}
\caption{Comparison of results of scaled momentum distribution $\log{1/x_p}$ for   $6 \times 10^{-4} < x_{bj} <1.2 \times 10^{-3}$ and  $ 10 < Q^2 < 20$ \GeV$^2$, using HERAPDF20 and CTEQ6L, obtained from \rivet (left) and the one from \rivethztool (right).}
\label{fig:H1_1995_I392386_b}
\end{center}
\end{figure}

\begin{figure}[htbp]
\begin{center}
\includegraphics[width=0.5\linewidth]{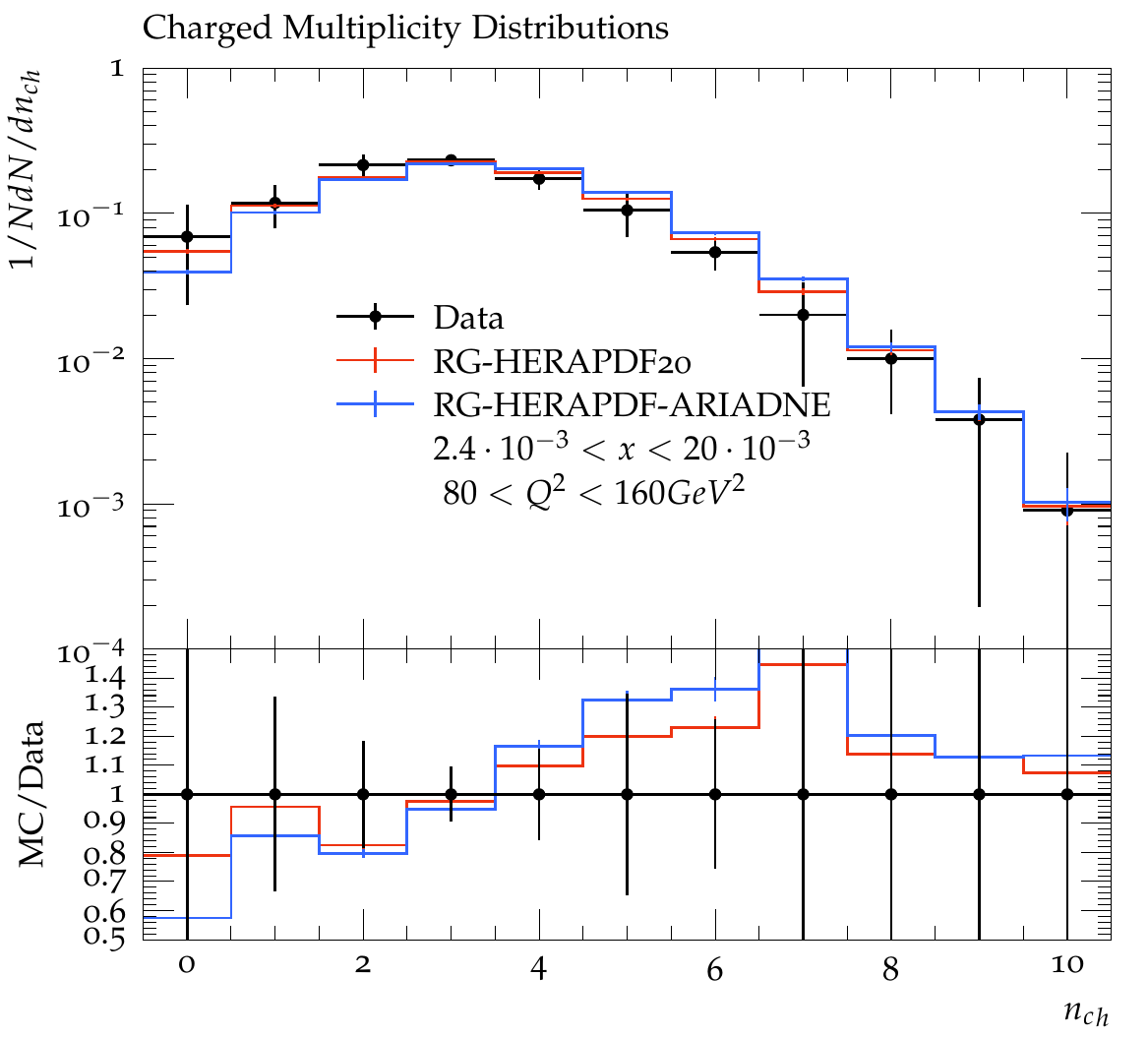}\includegraphics[width=0.5\linewidth]{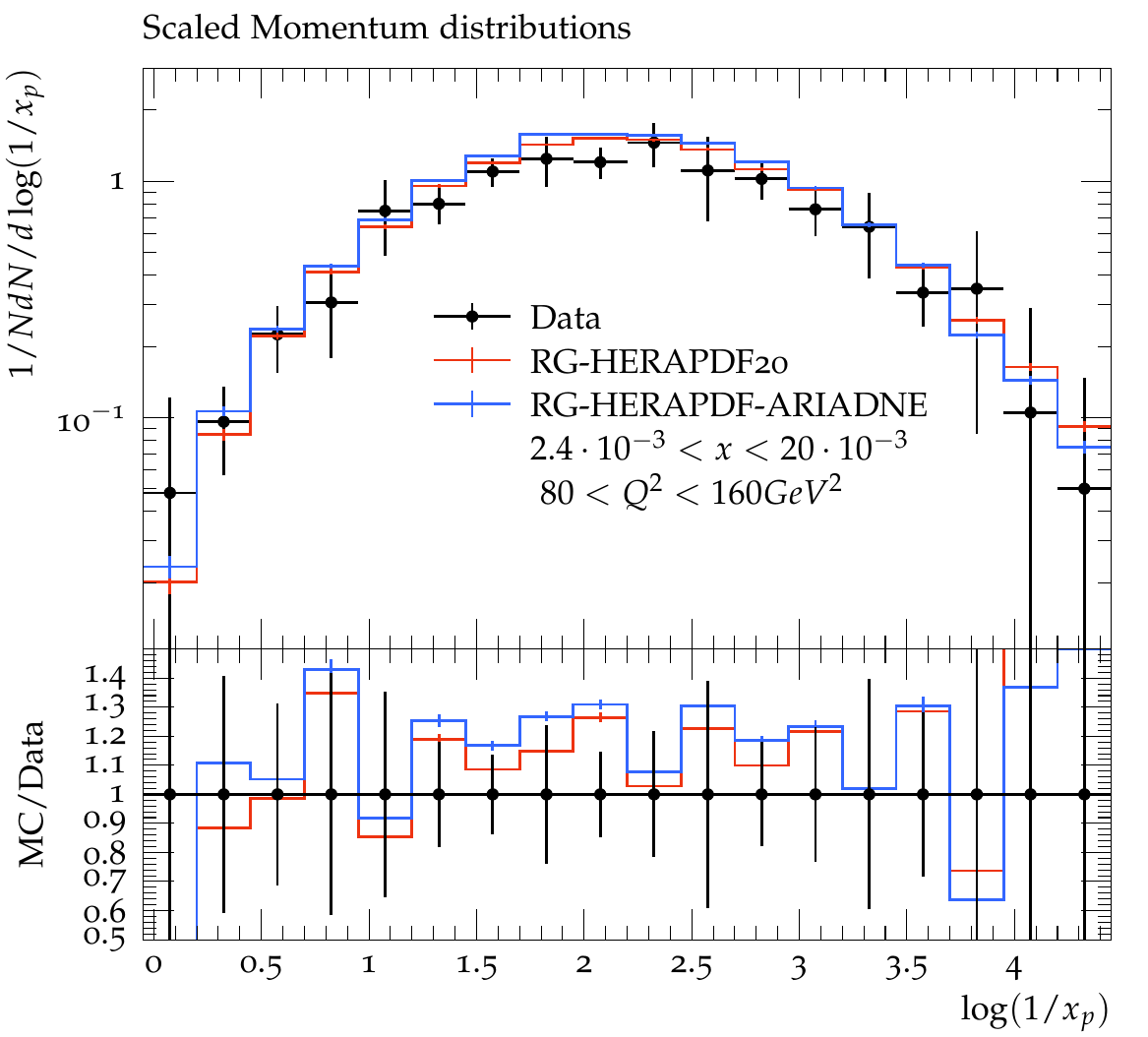}
\caption{Comparison of the charged particle multiplicity and the scaled momentum distribution obtained with DGLAP parton shower and with ARIADNE.}
\label{fig:H1_1995_I392386_c}
\end{center}
\end{figure}

\renewcommand{\thissection}{ZEUS\_1995\_I392386, HZ95007 }
\index{HZ95007 }
\index{ZEUS\_1995\_I392386}
\markboth{\thischapter}{\thissection}
{\bf Abstract} (cited from  Ref.~\cite{ZEUS:1995red}): "Charged particle production has been measured in Deep Inelastic Scattering (DIS) events using the ZEUS detector over a large range of $Q^2$ from 10 to 1280 GeV$^2$. The evolution with $Q$ of the charged multiplicity and scaled momentum has been investigated in the current fragmentation region of the Breit frame." \\
The results of the \rivet plugin\footnote{Author: Susie Kim } are compared with those from  \rivethztool  for the same kinematic range. Validation plots are shown in Fig.~\ref{fig:H1_1995_I392386_a} for the charged particle multiplicity and in Fig.~\ref{fig:H1_1995_I392386_b} for the scaled momentum distribution $\log{1/x_p}$ with $x_p=p_z/Q$ and $Q=\sqrt{Q^2}$.  
After the validation, the effect of a different parton shower, the Colour Dipole Model as implemented in ARIADNE, was studied.  As shown in Fig.~\ref{fig:H1_1995_I392386_c}, the distributions are not changed significantly.

\subsection{A Study of the fragmentation of quarks in \boldmath$ep$ collisions at HERA (H1) (H1\_1995\_I394793, HZ95072) }
\begin{figure}[htbp]
\begin{center}
\includegraphics[width=0.5\linewidth]{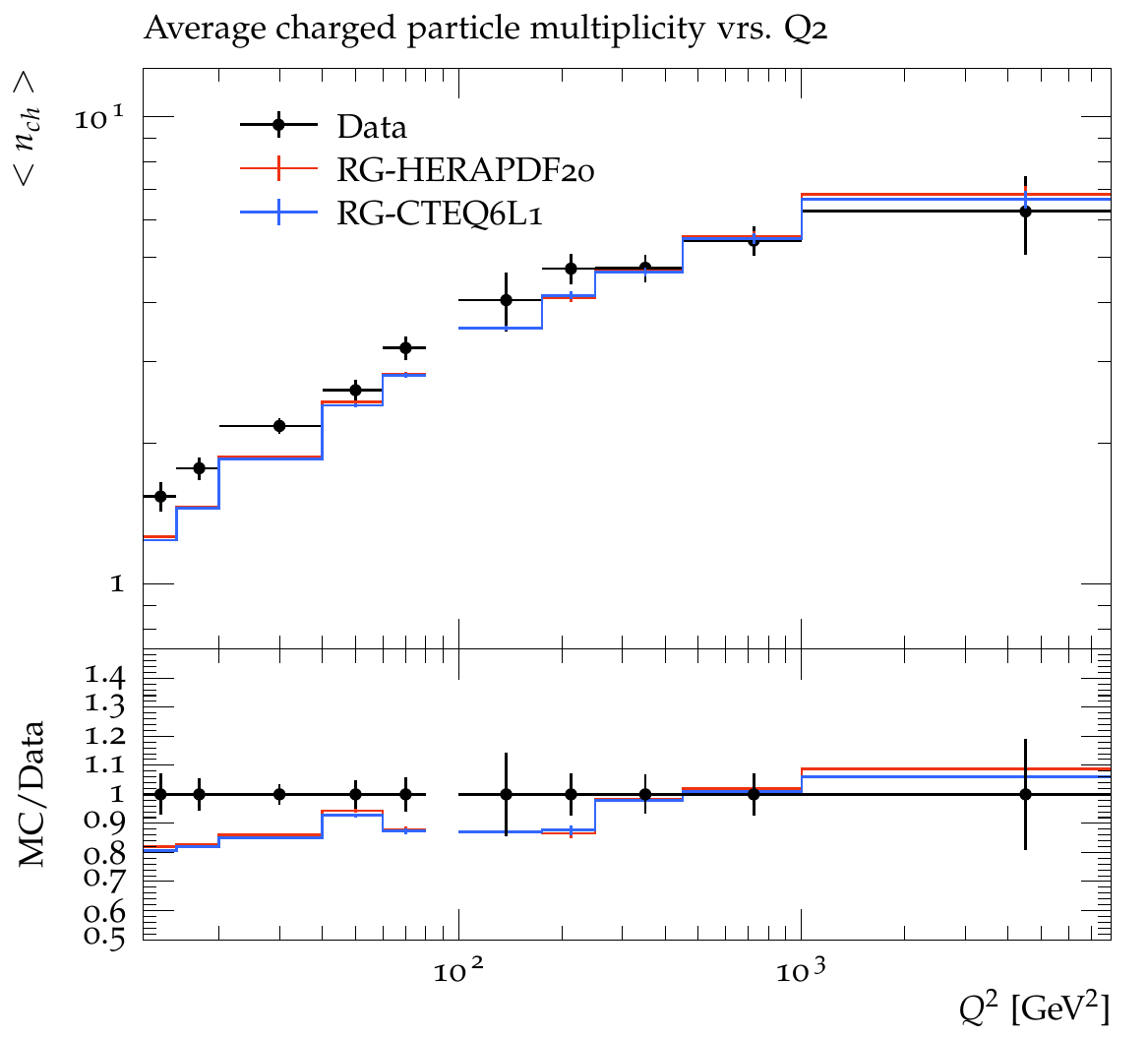}\includegraphics[width=0.5\linewidth]{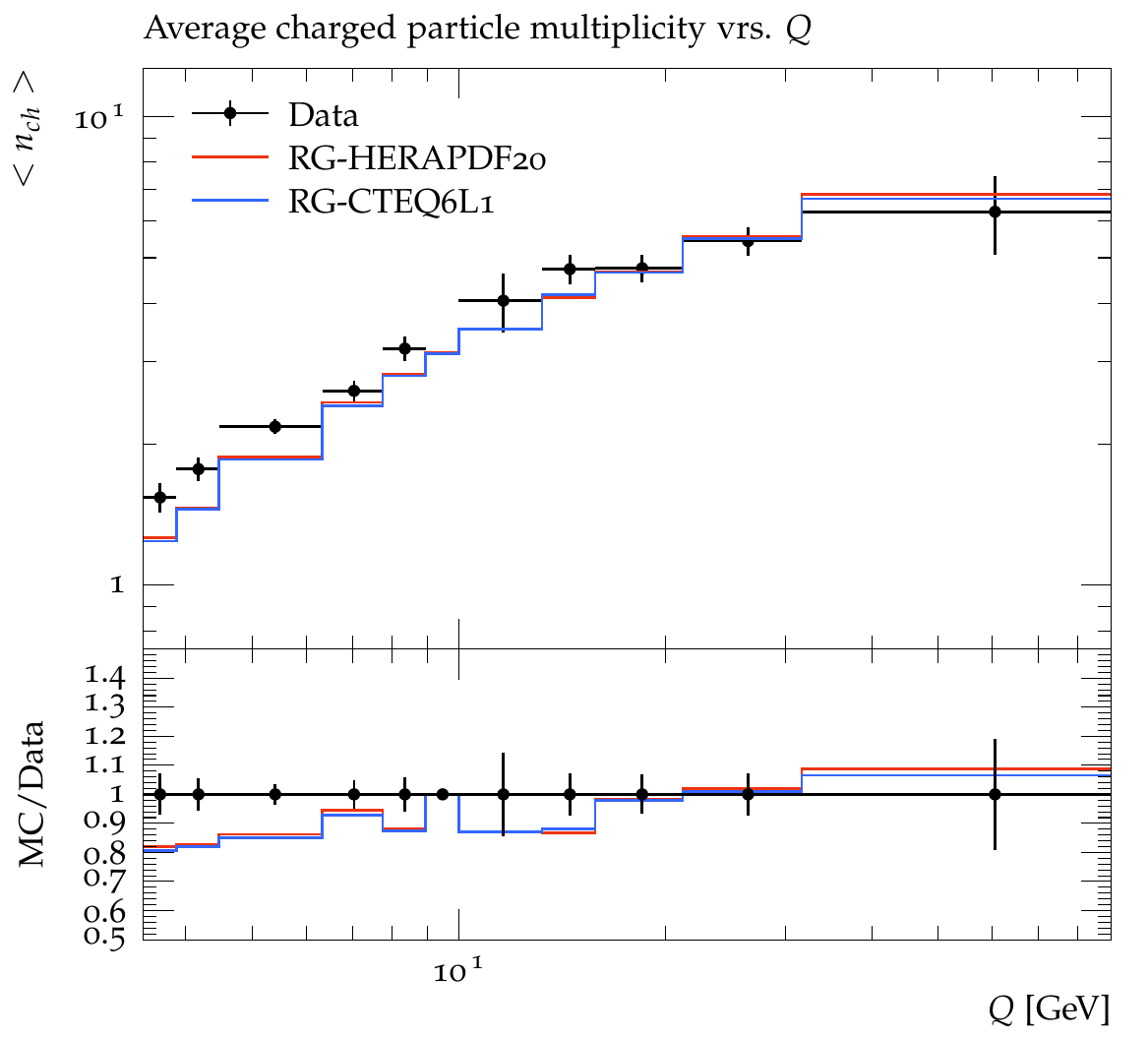}
\caption{Average charged particle multiplicity in the current region of the Breit frame as a function of $Q^2$ obtained from \rivet (left) and as a function of $Q$ from \rivethztool (right). Please note, that the $x$-axes are defined differently.}
\label{fig:H1_1995_I394793_a}
\end{center}
\end{figure}

\begin{figure}[htbp]
\begin{center}
\includegraphics[width=0.5\linewidth]{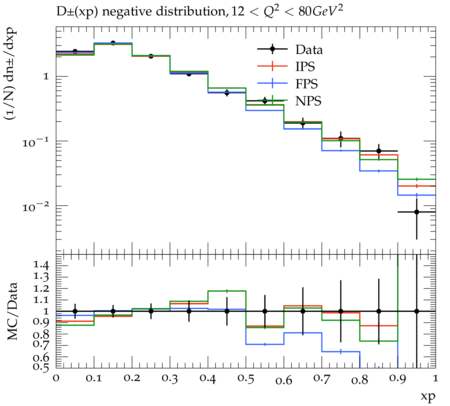}\includegraphics[width=0.5\linewidth]{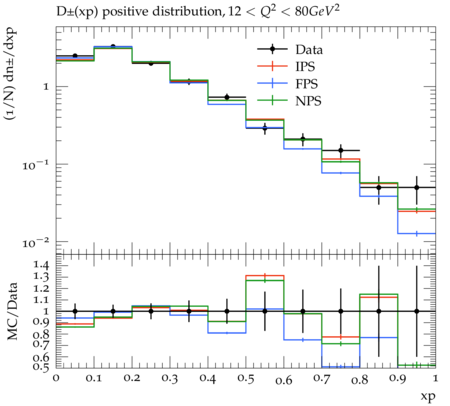}
\caption{The fragmentation functions, $D^\pm(x_p)$, for the current hemisphere of the Breit frame shown separately for positive and negative tracks results obtained from \rivet.
Shown are predictions with initial state (IPS), with final state (FPS) and without any parton shower (NPS).}
\label{fig:H1_1995_I394793_b}
\end{center}
\end{figure}

\renewcommand{\thissection}{H1\_1995\_I394793, HZ95072 }
\index{HZ95072 }
\index{H1\_1995\_I394793}
\markboth{\thischapter}{\thissection}
{\bf Abstract} (cited from  Ref.~\cite{H1:1995cqf}): "Deep inelastic scattering (DIS) events, selected from 1993 data taken by the H1 experiment at HERA, are studied in the Breit frame of reference. It is shown that certain aspects of the quarks emerging from within the proton in $ep$ interactions are essentially the same as those of quarks pair-created from the vacuum in $e^+e^-$ annihilation."
\\
The results of the \rivet plugin\footnote{Author: Narmin Rahimova} are compared with those from \rivethztool  for the same kinematic range. 
Validation plots are shown 
in Fig.~\ref{fig:H1_1995_I394793_a} for  the average charged particle multiplicity as a function of $Q^2$ (or $Q$ for \rivethztool ) obtained with different parton densities HERAPDF20 and CTEQ6L1.
In Fig.~\ref{fig:H1_1995_I394793_b} the fragmentation function $D^\pm(x_p)$ is shown for positive and negative tracks. The predictions were obtained with different selections on the parton shower: with initial state-, with final state- and without any parton shower.

\subsection{Neutral strange particle production in deep inelastic scattering at HERA (ZEUS)  (ZEUS\_1995\_I395196, HZ95084)}
\begin{figure}[htbp]
\begin{center}
\includegraphics[width=0.5\linewidth]{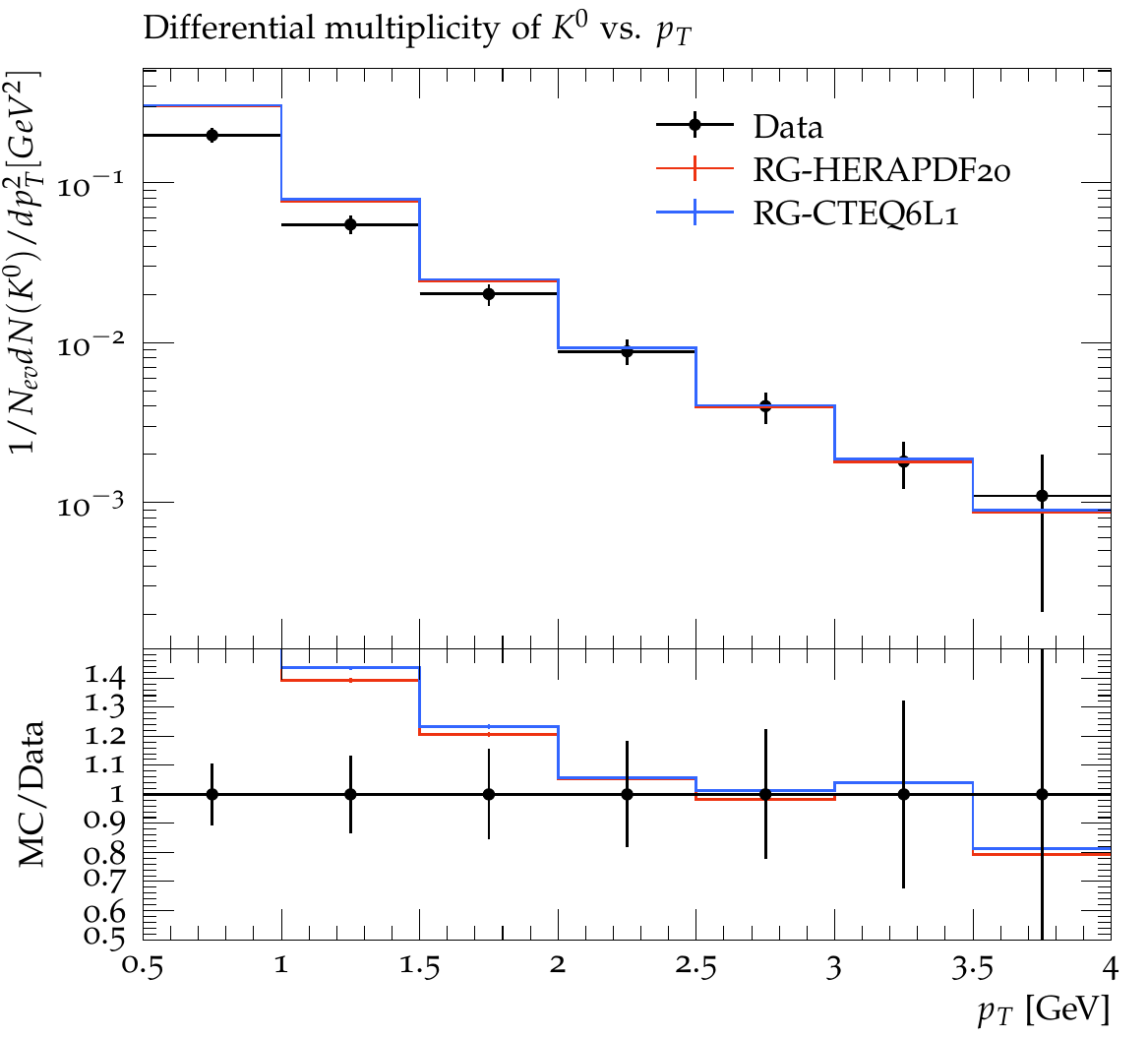}\includegraphics[width=0.5\linewidth]{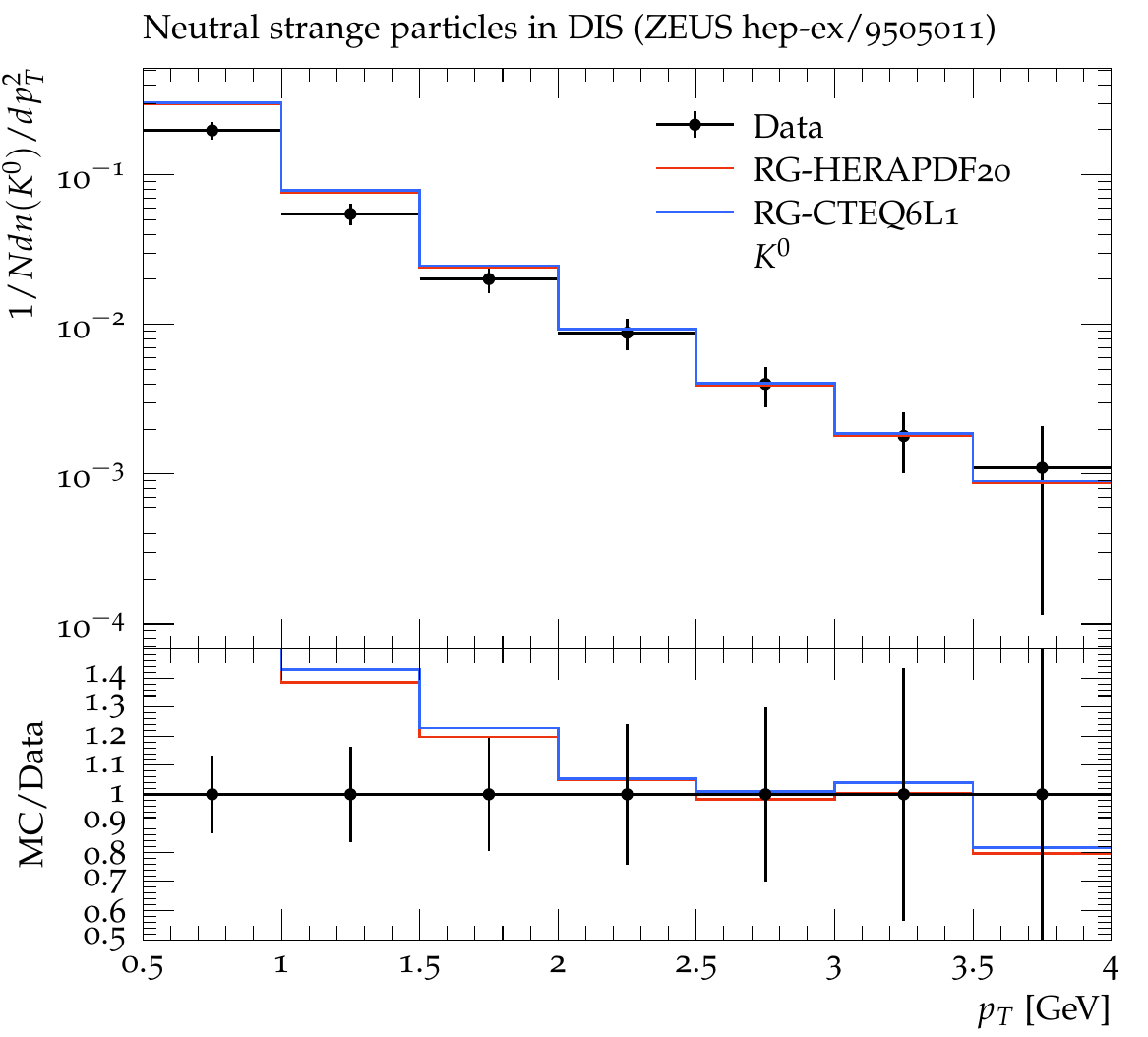}
\caption{Comparison of the differential multiplicity of $K^0$ as a function of $\pt $  obtained from \rivet using HERAPDF20 and CTEQ6L1 (left), and the corresponding one from \rivethztool (right).}
\label{fig:ZEUS_1995_I395196_1}
\end{center}
\end{figure}

\renewcommand{\thissection}{ZEUS\_1995\_I395196, HZ95084 }
\index{HZ95084 }
\index{ZEUS\_1995\_I395196}
\markboth{\thischapter}{\thissection}
{\bf Abstract} (cited from  Ref.~\cite{ZEUS:1995how}): 
"This paper presents measurements of  $K^0$ and $\Lambda$ production in neutral current, deep inelastic scattering of 26.7 \GeV\ electrons and 820 \GeV\ protons in the kinematic range $10<Q^2<640$\,\GeV$^2$,  $0.0003<x<0.01$, and $y>0.04$. Average multiplicities for $K^0$ and $\Lambda$ production are determined for transverse momenta $p_T >0.5$ \GeV\ and pseudorapidities $|\eta| < 1.3 $. The production properties of $K^0$'s in events with and without a large rapidity gap with respect to the proton direction are compared. The ratio of neutral $K^0$'s to charged particles per event in the measured kinematic range is, within the present statistics, the same in both samples."\\ 
The results of the \rivet plugin\footnote{Author: Can S\"usl\"u } are compared with those from  \rivethztool  for the same kinematic range. Particles are obtained by their MC IDs, without reconstruction. It is observed that the changes of value of the strange quark suppression factor $P_s/P_u$ affects the distribution distribution, as mentioned in Ref.~\cite{ZEUS:1995how} (a value $P_s/P_u=0.3$ is used here).
Validation plots are shown in Fig.~\ref{fig:ZEUS_1995_I395196_1}.

\subsection{Inclusive $D^0$ and $D^{*\pm}$ production in neutral current deep inelastic $ep$ scattering at HERA (H1) (H1\_1996\_I421105)}
\begin{figure}[htbp]
\begin{center}
\includegraphics[width=0.5\linewidth]{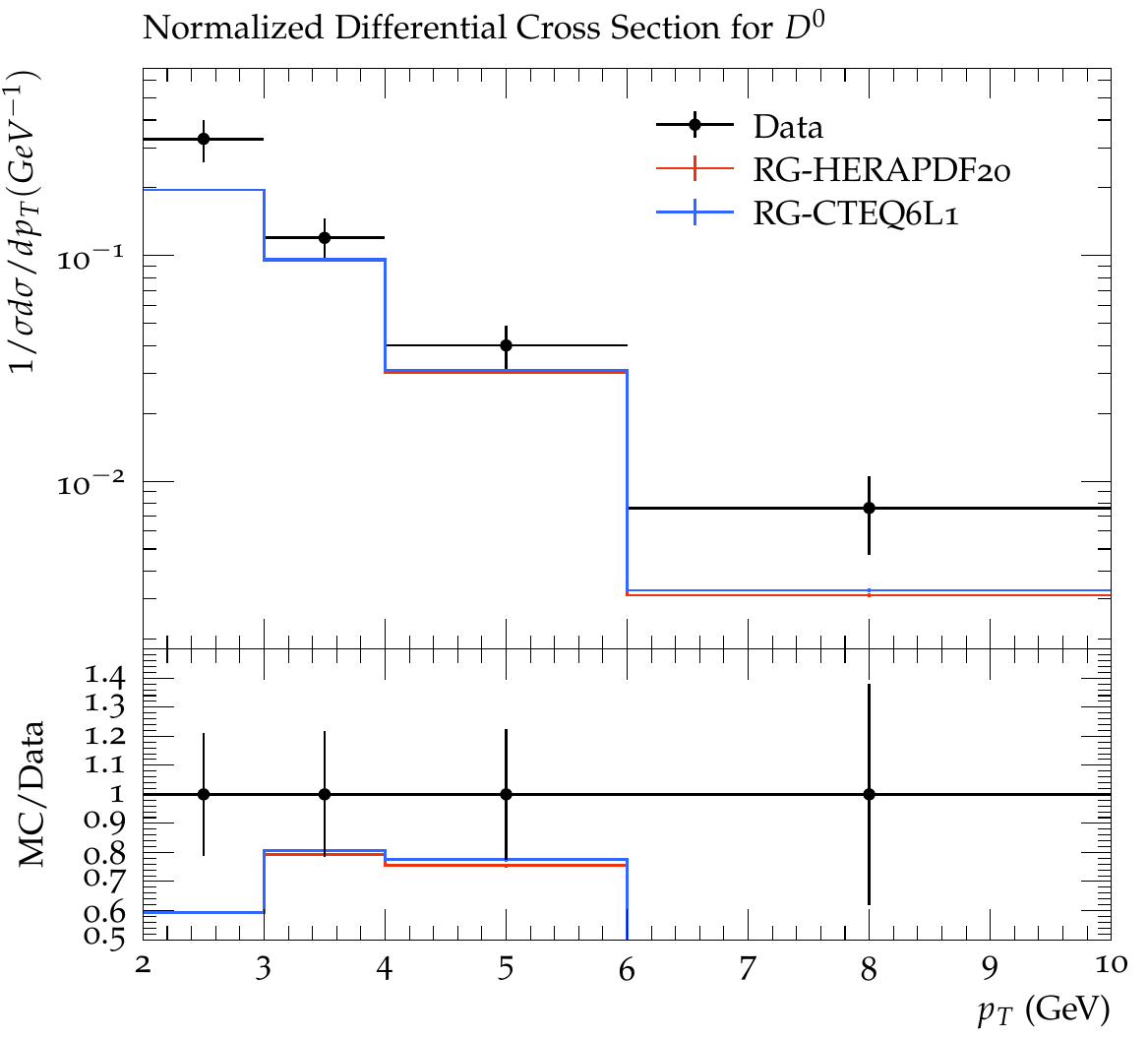}\includegraphics[width=0.5\linewidth]{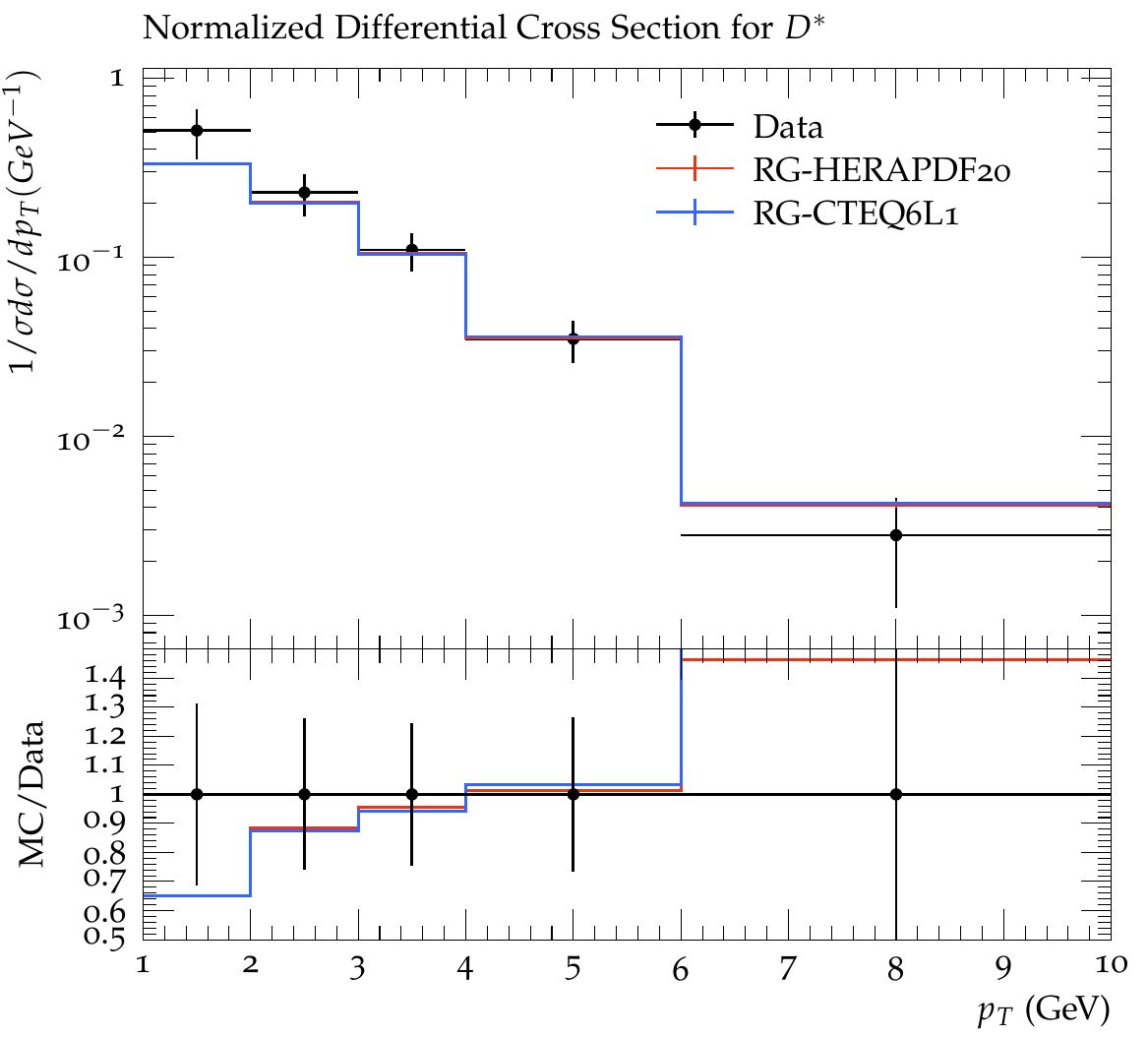}
\caption{Normalised cross section as a function of the transverse momentum of the $D^0$ (left) and the $D^{*}$ (right) mesons}
\label{fig:H1_1996_I421105_Pt}
\end{center}
\end{figure}

\begin{figure}[htbp]
\begin{center}
\includegraphics[width=0.5\linewidth]{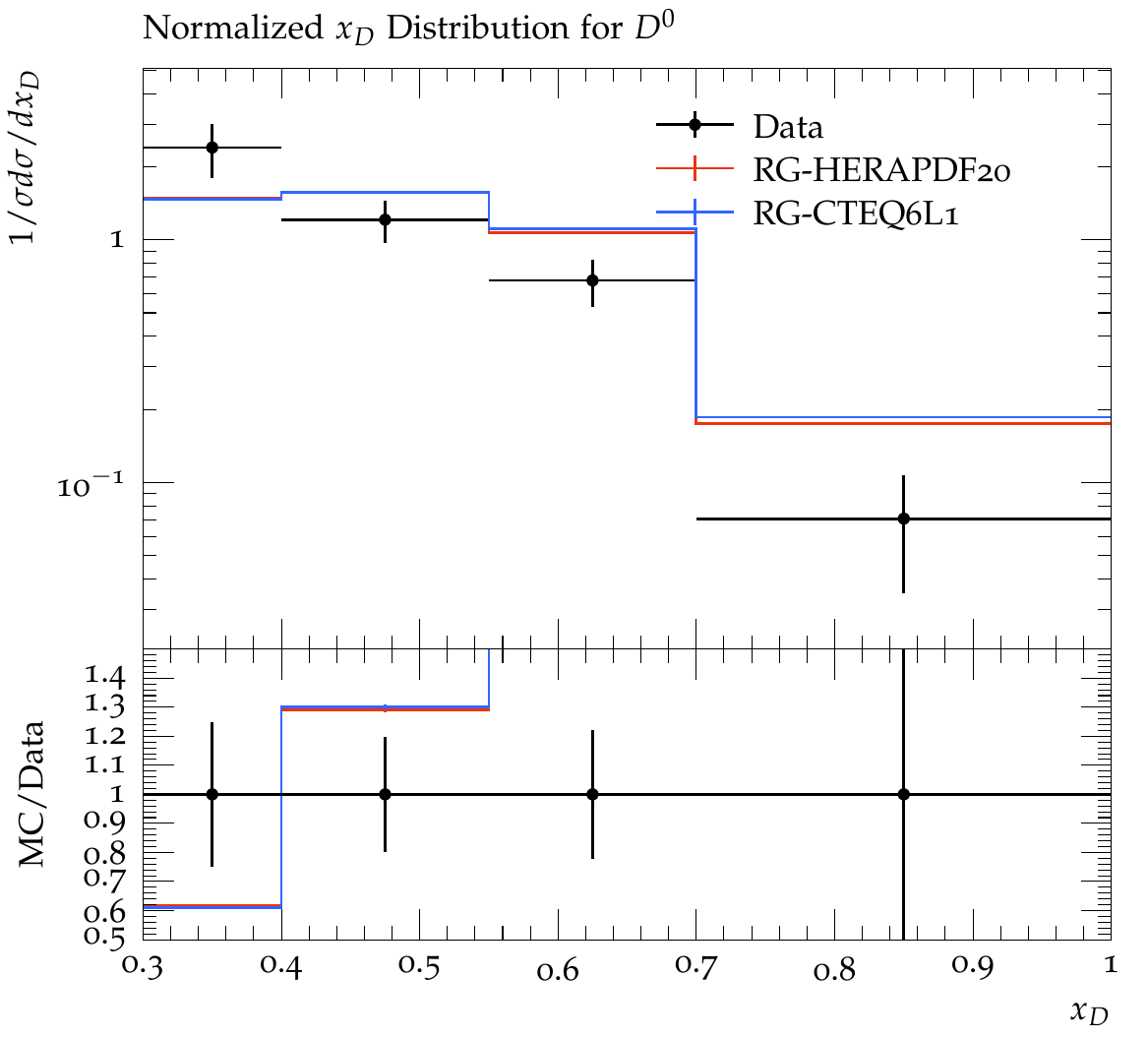}\includegraphics[width=0.5\linewidth]{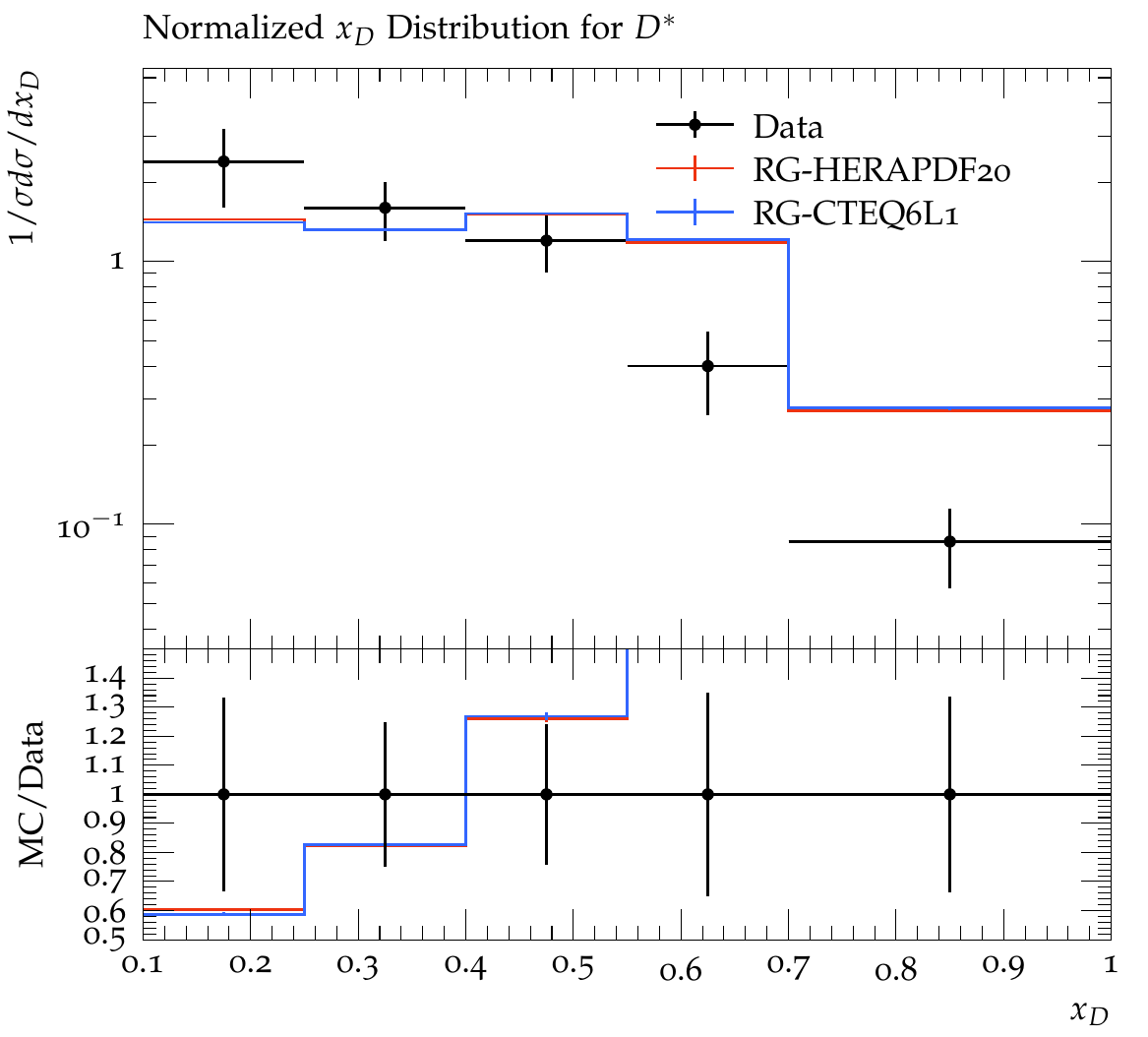}
\caption{Normalised distribution as a function of $x_D$  for $D^0$ (left) and the $D^{*\pm}$ (right) mesons.}
\label{fig:H1_1996_I421105_Xd}
\end{center}
\end{figure}

\renewcommand{\thissection}{H1\_1996\_I421105, HZ96138 }
\index{HZ96138 }
\index{H1\_1996\_I421105}
\markboth{\thischapter}{\thissection}
{\bf Abstract} (cited from  Ref.~\cite{H1:1996naa}): "First results on inclusive $D^0$ and $D^*$ production in deep inelastic $ep$ scattering are reported using data collected by the H1 experiment at HERA in 1994. " \\
The results of the \rivet plugin\footnote{Author: Muhammad Ibrahim Abdulhamid } are shown in Fig.~\ref{fig:H1_1996_I421105_Pt} for the normalized cross section of $D^0$ and $D^*$ mesons as a function of the transverse momentum.
In Fig.~\ref{fig:H1_1996_I421105_Xd} the normalized cross section is shown for $D^0$ and $D^*$ mesons as a function of $x_D= 2 |\vec{p}_{D^0}^*|/W$, where $\vec{p}_{D^0}^*$ is defined for the $D^0$ in the hadronic center-of-mass frame for both decay channels.  

Since no corresponding \rivethztool results were available, the \rivet plugin was validated with the results from the original publication.


\subsection{Charged particle multiplicities in deep inelastic scattering at HERA (H1) (H1\_1996\_I422230, HZ96160) }
\renewcommand{\thissection}{H1\_1996\_I422230, HZ96160 }
\index{HZ96160 }
\index{H1\_1996\_I422230}
\markboth{\thischapter}{\thissection}
{\bf Abstract} (cited from  Ref.~\cite{H1:1996ovs}): "Using the H1 detector at HERA, charged particle multiplicity distributions in deep inelastic $ep$ scattering have been measured over a large kinematical region. The evolution with $W$ and $Q^2$ of the multiplicity distribution and of the multiplicity moments in pseudorapidity domains of varying size is studied in the current fragmentation region of the hadronic centre-of-mass frame." \\

\begin{figure}[htbp]
\begin{center}
\includegraphics[width=0.5\linewidth]{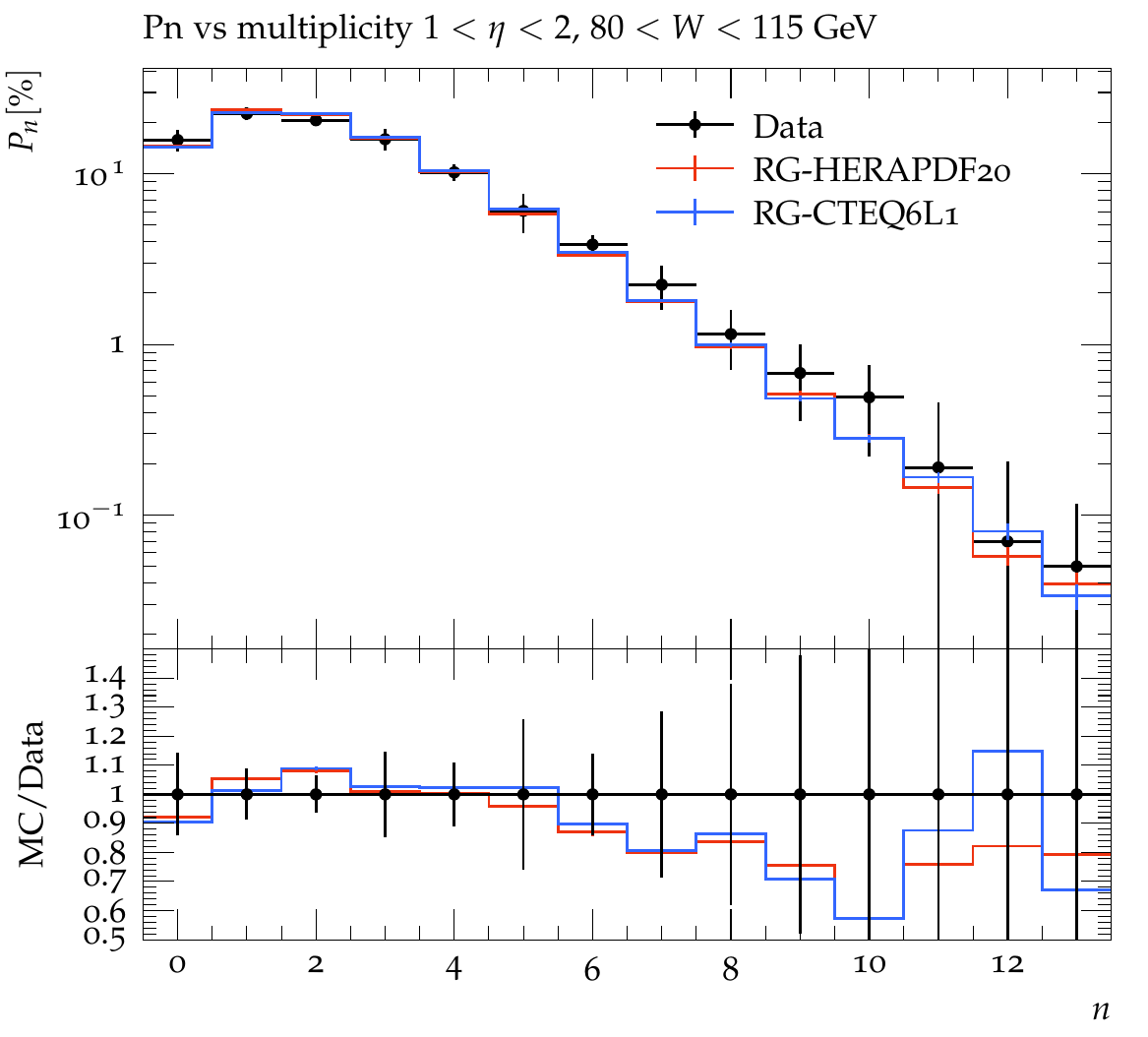}\includegraphics[width=0.5\linewidth]{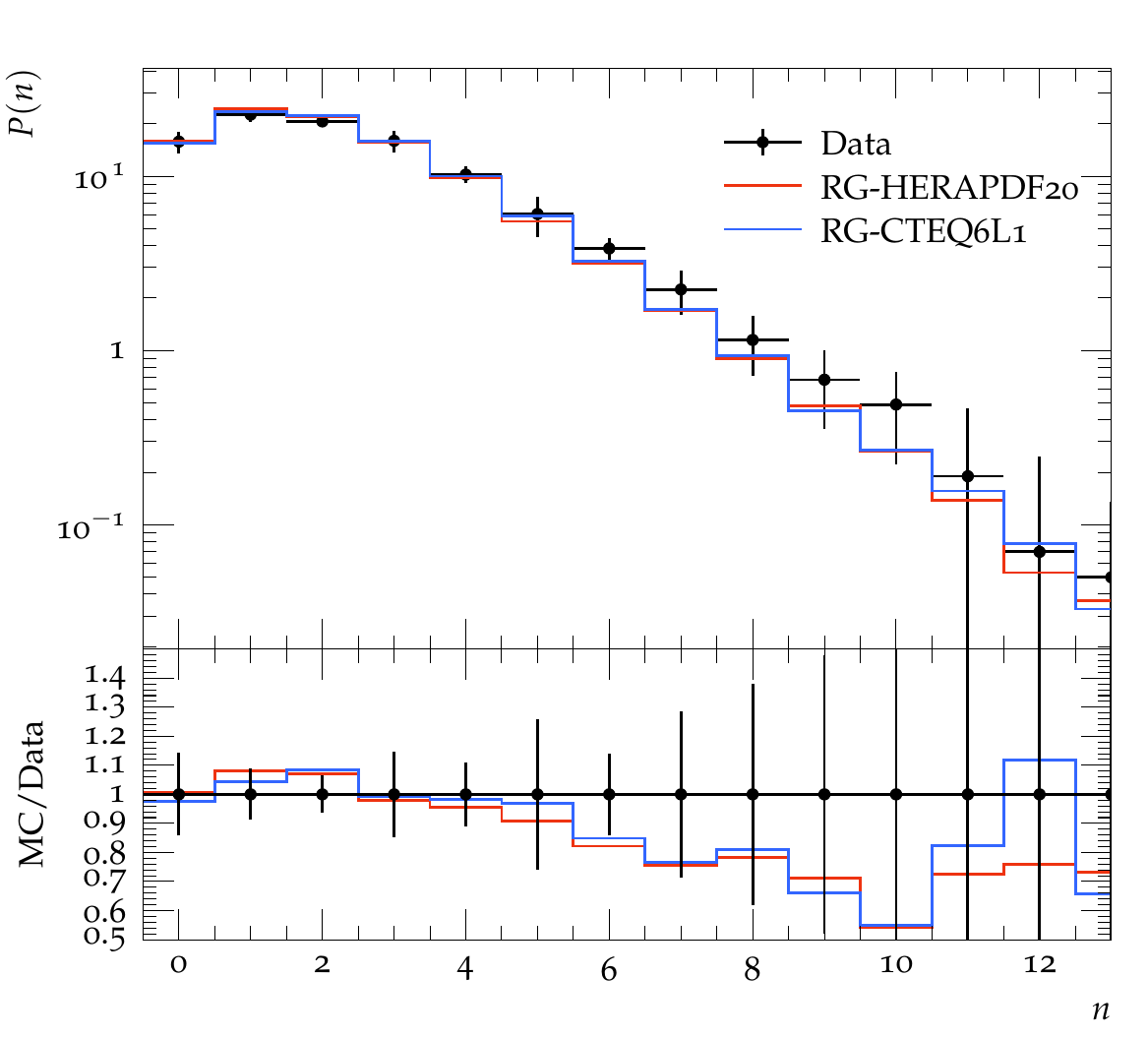}
\caption{The probability distribution of the charged particle multiplicity in the intervals $80<W<115$ \GeV\  and $1<\eta^*<2$. Compared are results obtained from \rivet (left) and from \rivethztool (right).}
\label{fig:H1_1996_I422230_1}
\end{center}
\end{figure}
\begin{figure}[htbp]
\begin{center}
\includegraphics[width=0.5\linewidth]{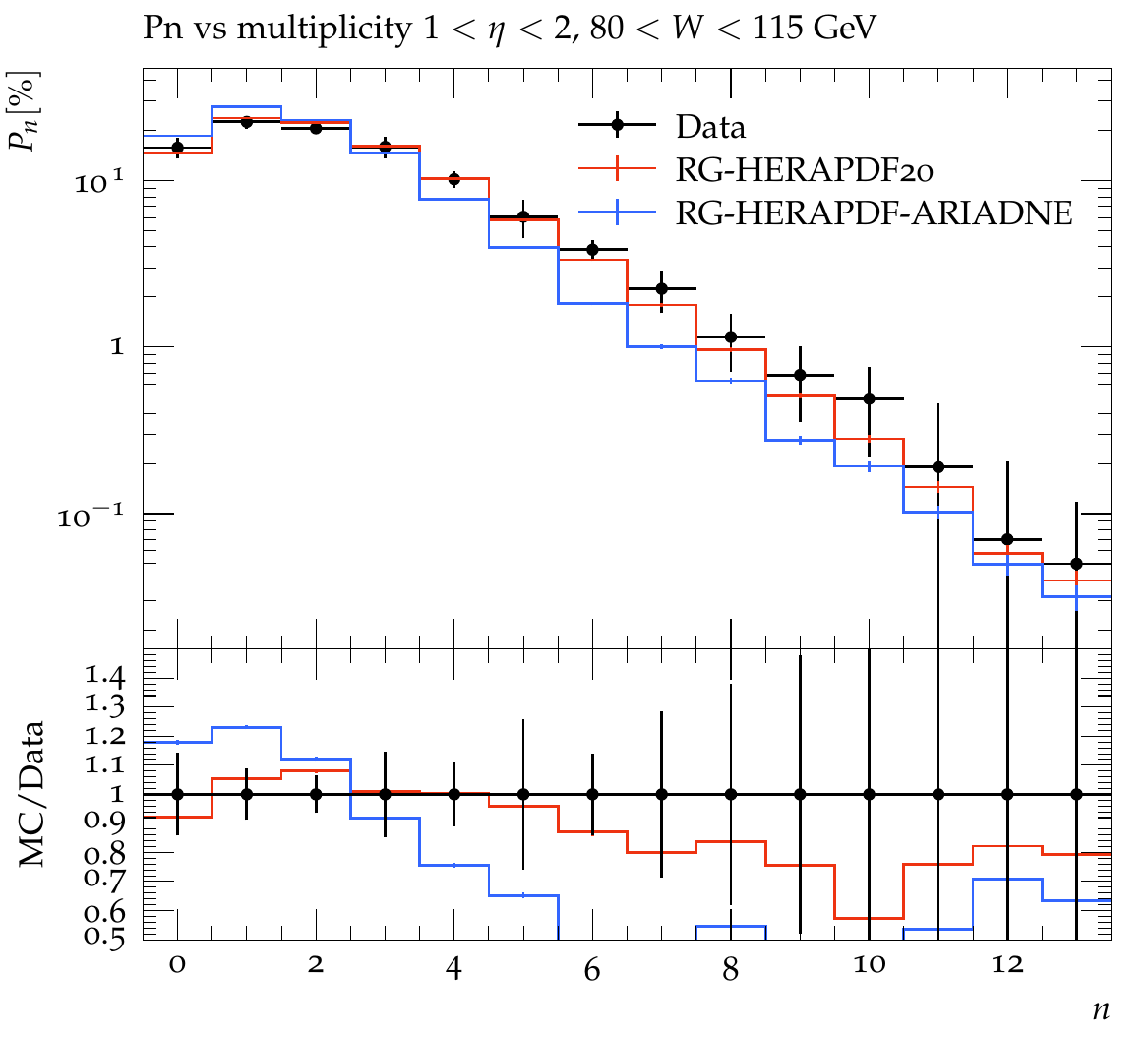}\includegraphics[width=0.5\linewidth]{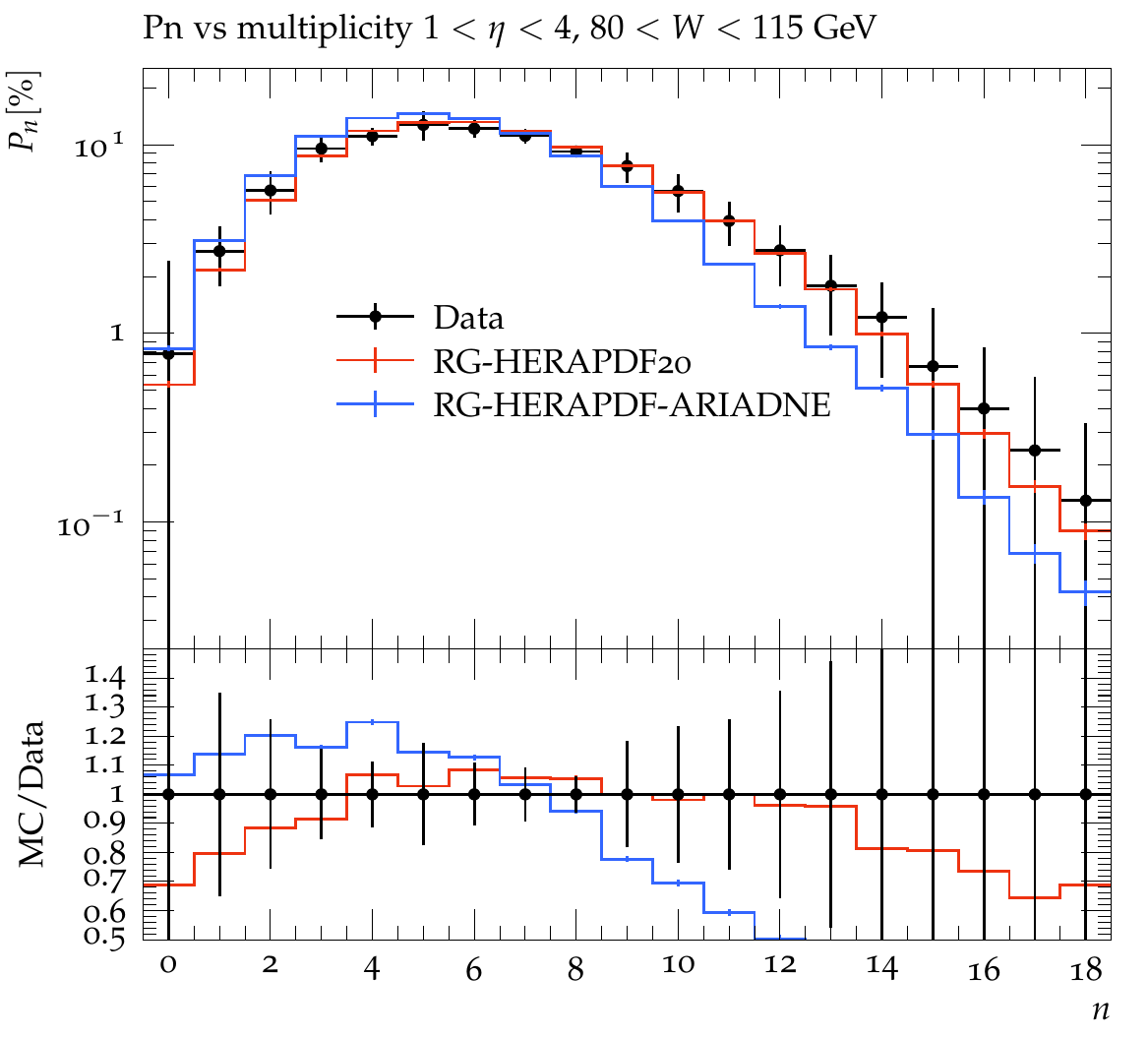}
\caption{Charged particle multiplicity in  $80<W<115$ GeV and $1<\eta^*<2$ (left) and $1<\eta^*<4$ (right)  obtained with DGLAP parton shower and with ARIADNE.}
\label{fig:H1_1996_I422230}
\end{center}
\end{figure}

The results of the \rivet plugin\footnote{Author: Ariadna Le\'on }  for the charged particle multiplicities in a fixed range of $\eta ^*$ (in the hadronic center of mass) are compared with those from \rivethztool  for the same kinematic range. 
Validation plots are shown in Fig.~\ref{fig:H1_1996_I422230_1}.
The comparison of results obtained with DGLAP parton shower and  with the ARIADNE model is presented in Fig.~\ref{fig:H1_1996_I422230} (left), and for different parton density options (right).
\begin{figure}[htbp]
\begin{center}
\includegraphics[width=0.4\linewidth]{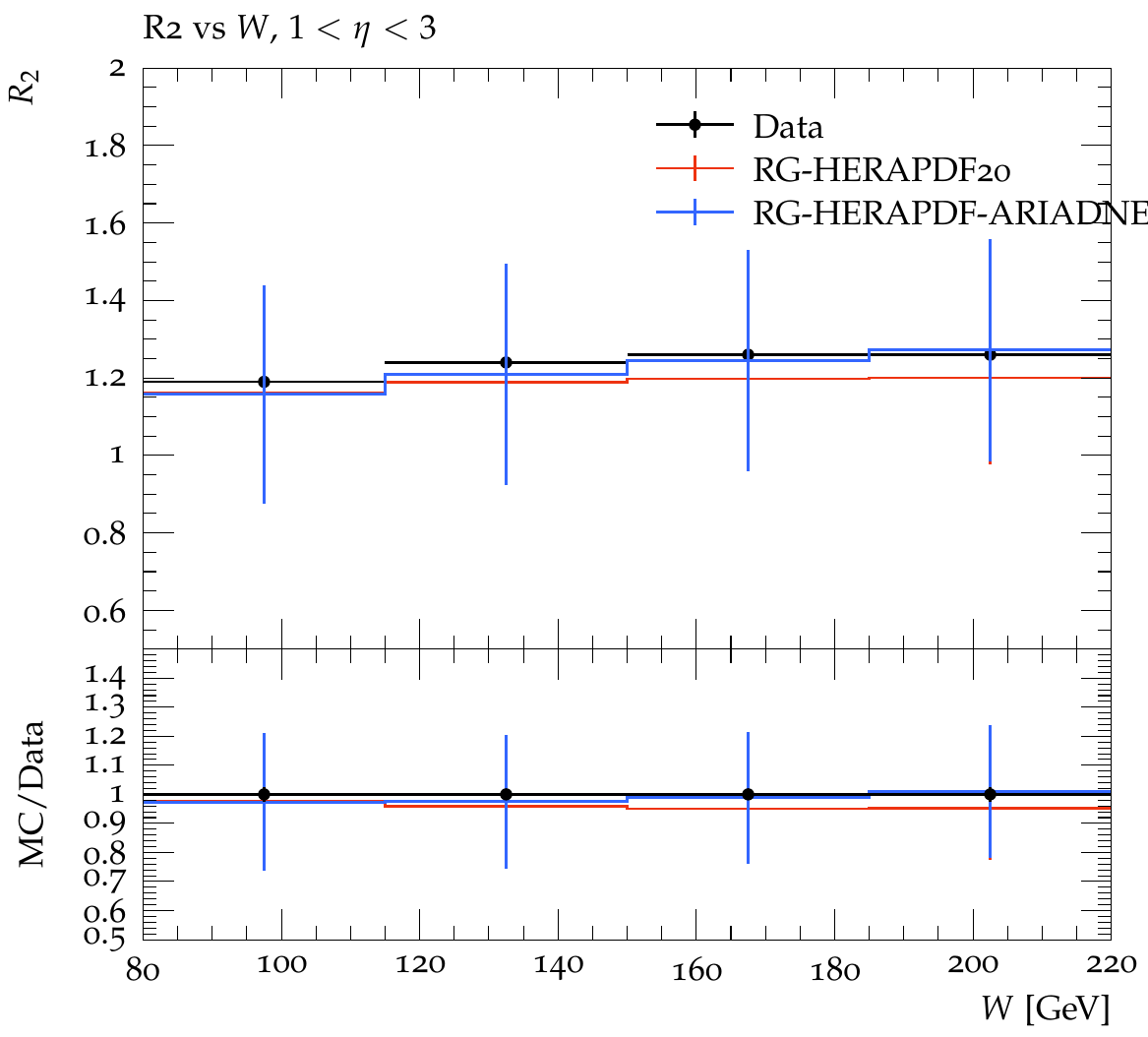}\includegraphics[width=0.2\linewidth]{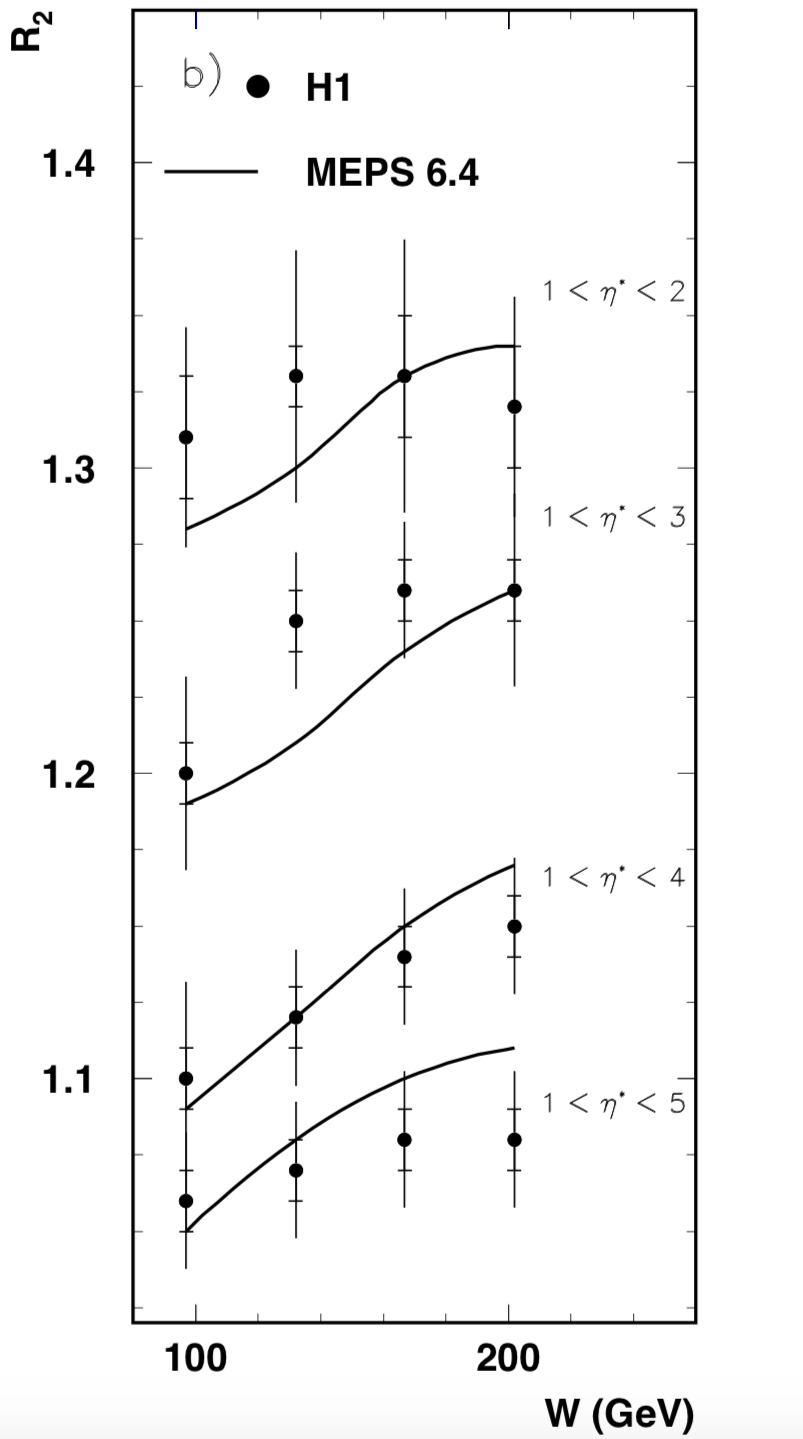}\includegraphics[width=0.4\linewidth]{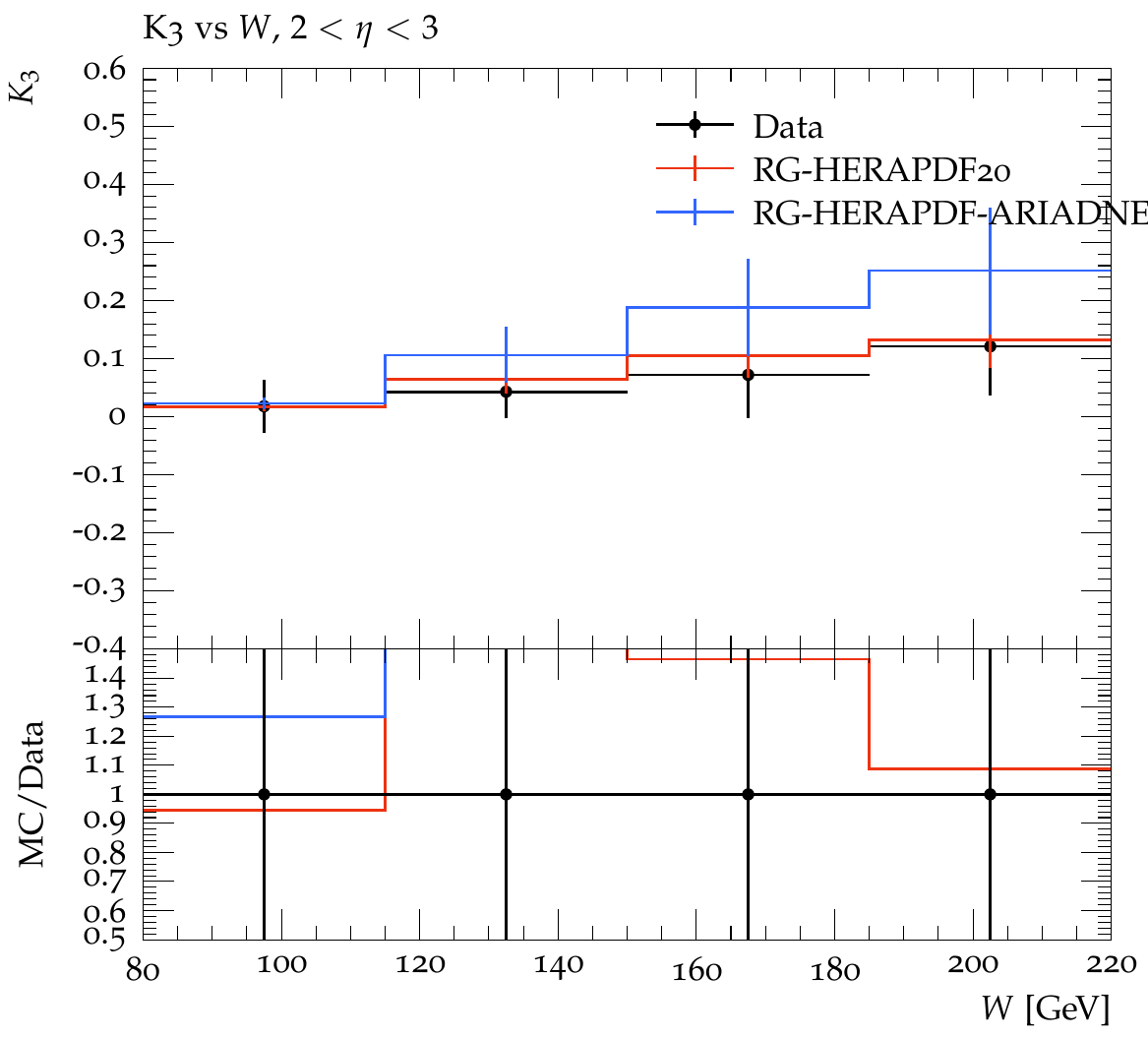}
\caption{Comparison of results obtained from \rivet (left) and from the original paper (middle) for the $R_2$ variable in the interval $1<\eta^*<3$. 
(Right) The distribution of $K_3$ for the interval $2<\eta^*<3$.}
\label{fig:H1_1996_I422230_2}
\end{center}
\end{figure}
The calculation of the various moments of the distributions were not provided by \rivethztool, therefore the new \rivet plugin had to be validated with the distributions shown in the original reference~\cite{H1:1996ovs}. In Fig.~\ref{fig:H1_1996_I422230_2} (left) the distribution of $R_2$ is compared with that  (middle) of Ref.~\cite{H1:1996ovs}.

The moments allow a deeper statistical study of the multiplicity. 
{\it Correlated production} occurs when the presence of a particle in the system affects the probability to generate new particles. This probability is represented by the Muller moments. A positive Muller moment $K_3$ means that the presence of $n-1$ particles enhances the production of a new particle.
In Fig.~\ref{fig:H1_1996_I422230_2} (right) $K_3$ in one of the pseudorapidity intervals is shown. 

\subsection{Measurement of charged particle transverse momentum spectra in deep inelastic scattering (H1) (H1\_1996\_I424463, HZ96215)}
\begin{figure}[htbp]
\begin{center}
\includegraphics[width=0.5\linewidth]{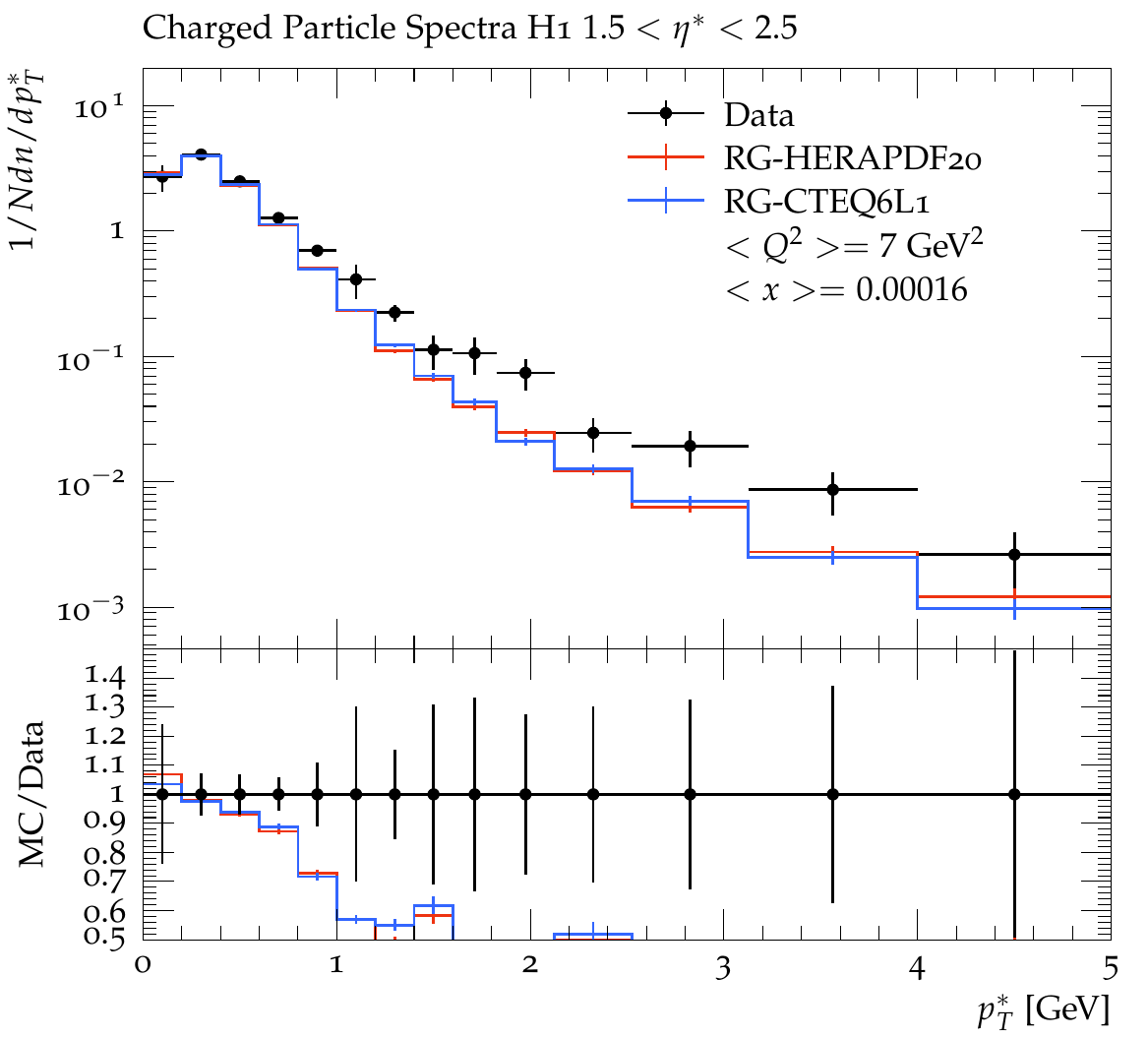}\includegraphics[width=0.5\linewidth]{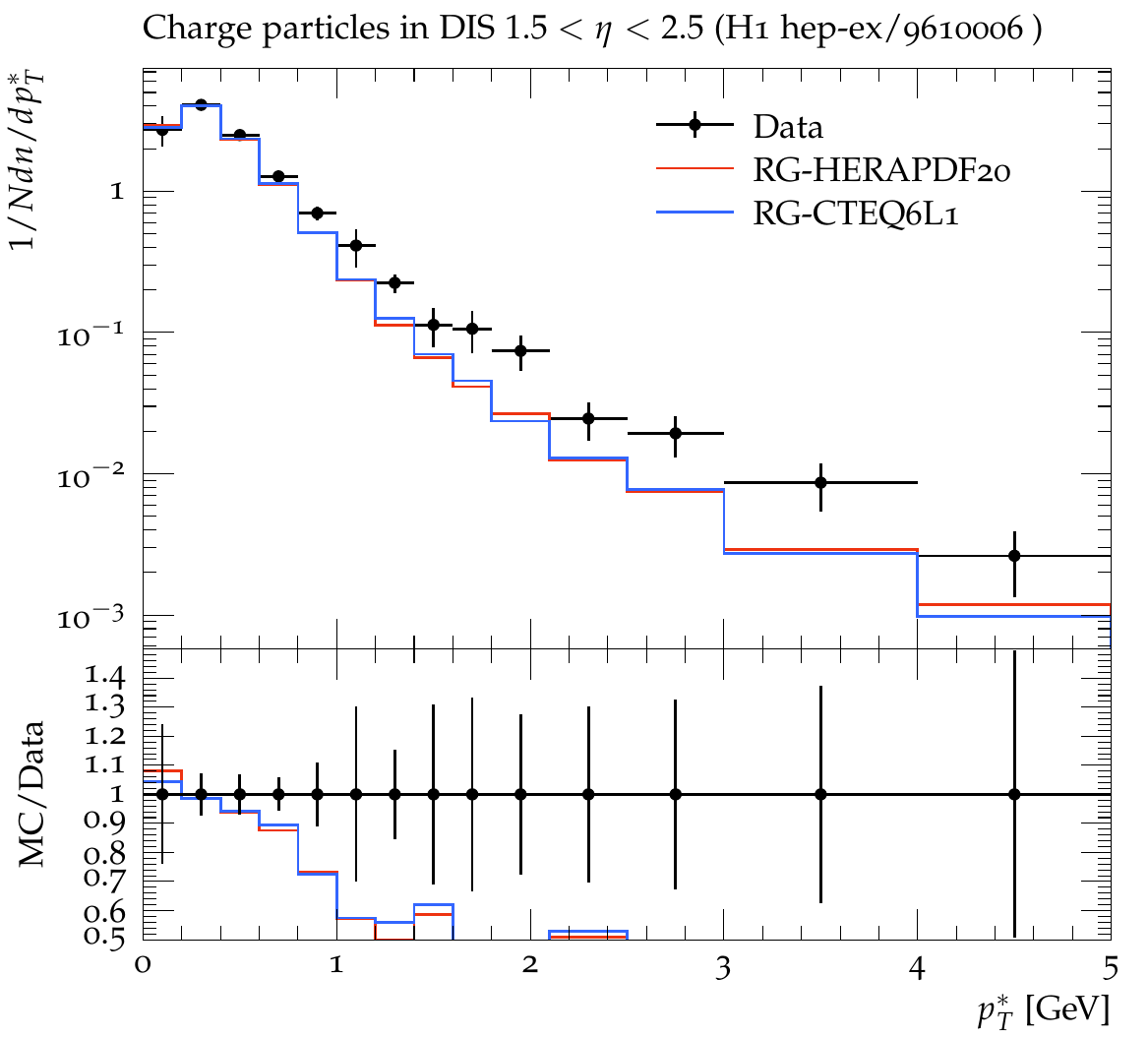}
\caption{Charged particle multiplicity as a function of $p_T$ in $1.5 < \eta^* < 2.5$, obtained from \rivet (left)  and from \rivethztool (right).}
\label{fig:H1_1996_I42446}
\end{center}
\end{figure}

\renewcommand{\thissection}{H1\_1996\_I424463, HZ96215}
\index{HZ96215}
\index{H1\_1996\_I424463}
\markboth{\thischapter}{\thissection}
\newcommand{\dif}{\mbox{\rm d}}
{\bf Abstract} (cited from  Ref.~\cite{Adloff:1996dy}): "Transverse momentum spectra of charged particles produced in deep inelastic scattering are measured as a function of the kinematic variables $x_B$ and $Q^2$ using the H1 detector at the $ep$ collider HERA. The data are compared to different parton emission models, either with or without ordering of the emissions in transverse momentum. The data provide evidence for a relatively large amount of parton radiation between the current and the remnant systems." \\
The results of the \rivet plugin\footnote{Author: Suraj Kumar Singh} are compared with those from \rivethztool  for the same kinematic range. 
Validation plots for the charged multiplicity as a function of the transverse momentum in the hadronic center-of-mass frame are shown in Fig.~\ref{fig:H1_1996_I42446}.

\newpage

\subsection{Evolution of \boldmath$e p$ fragmentation and multiplicity distributions in the Breit frame (H1) (H1\_1997\_I445116 - HZ97108)}
\renewcommand{\thissection}{H1\_1997\_I445116, HZ97108 }
\index{HZ97108 }
\index{H1\_1997\_I445116}
\markboth{\thischapter}{\thissection}
{\bf Abstract} (cited from  Ref.~\cite{H1:1997mpq}): "Low $x$ deep-inelastic $ep$ scattering data, taken in 1994 at the H1 detector at HERA, are analysed in the Breit frame of reference. "\\

\begin{figure}[htbp]
\begin{center}
\includegraphics[width=0.5\linewidth]{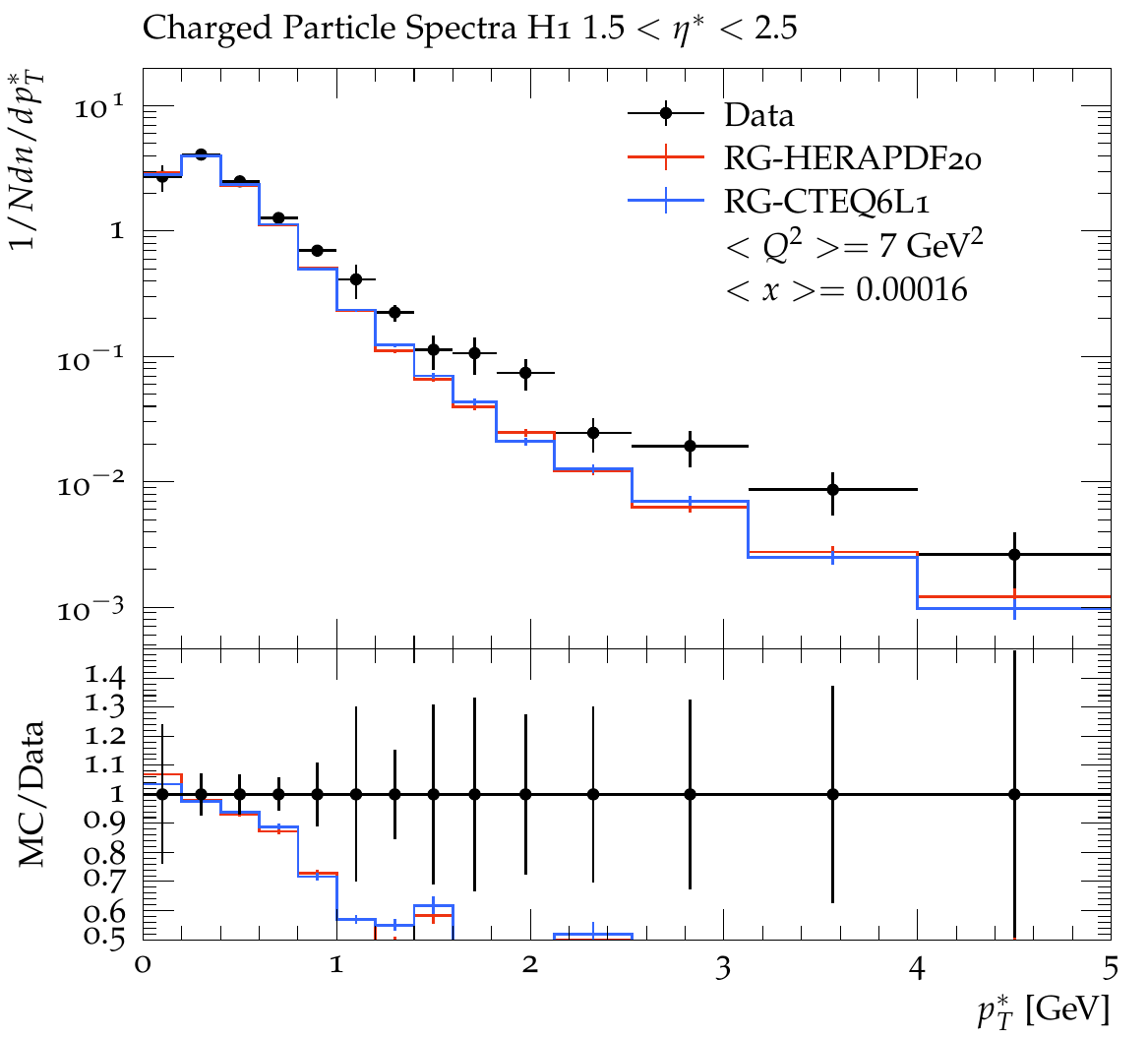}\includegraphics[width=0.5\linewidth]{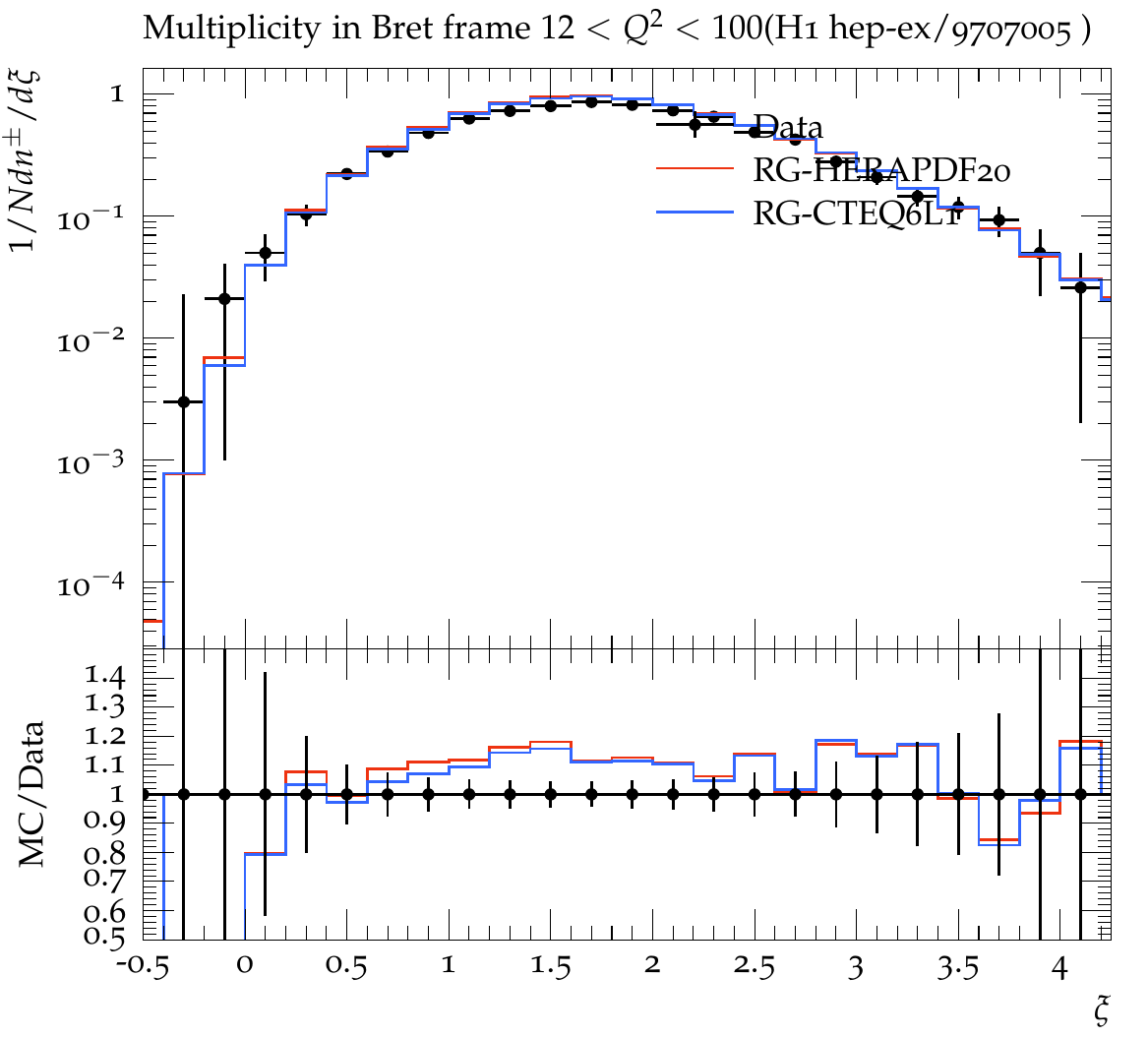}
\caption{Distribution $\xi = \log(1/x_p)$ obtained from \rivet (left)  and from \rivethztool (right) for  low $Q^2$ data.}
\label{fig:H1_1997_I445116_3}
\end{center}
\end{figure}

\begin{figure}[htbp]
\begin{center}
\includegraphics[width=0.5\linewidth]{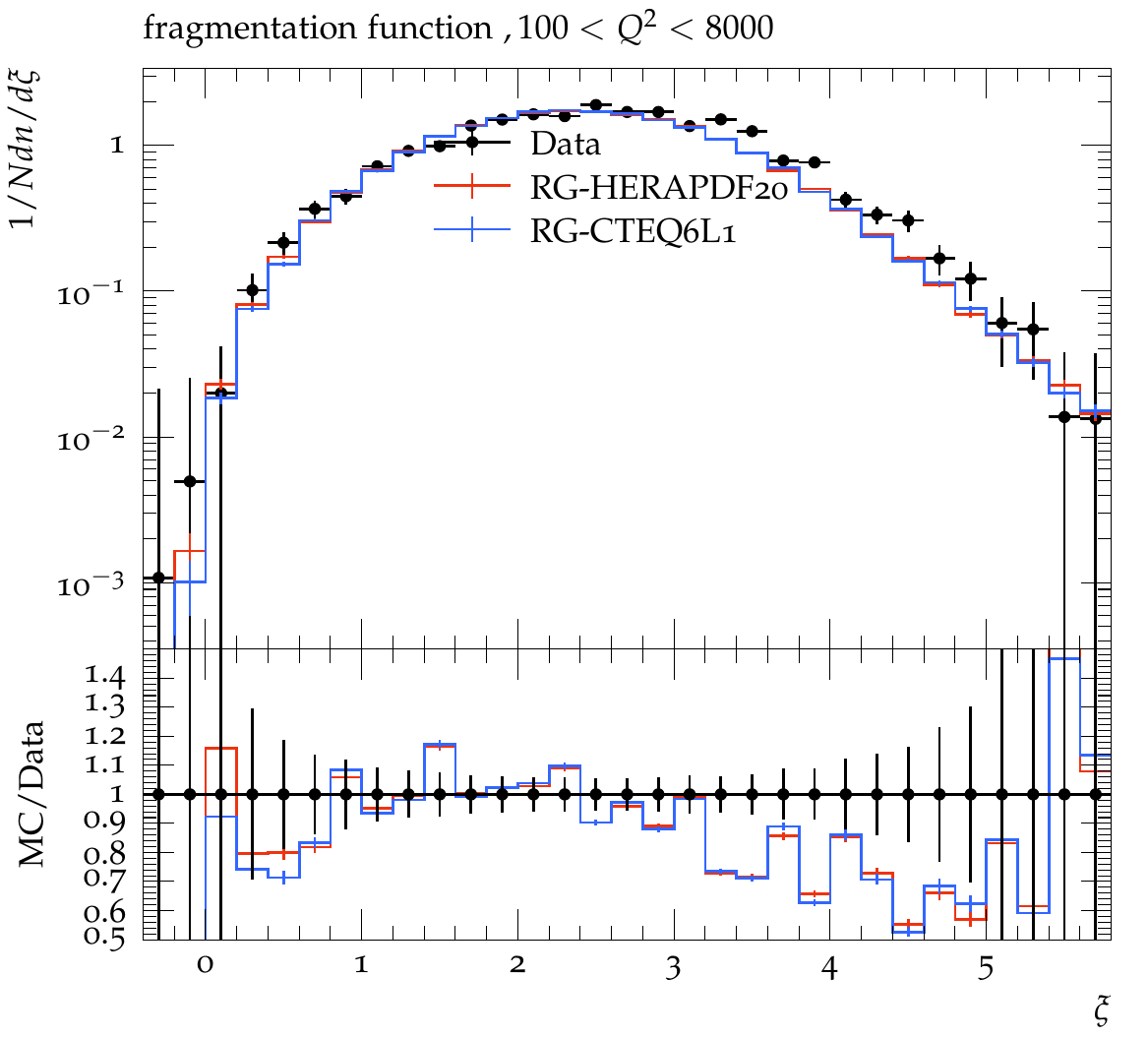}\includegraphics[width=0.5\linewidth]{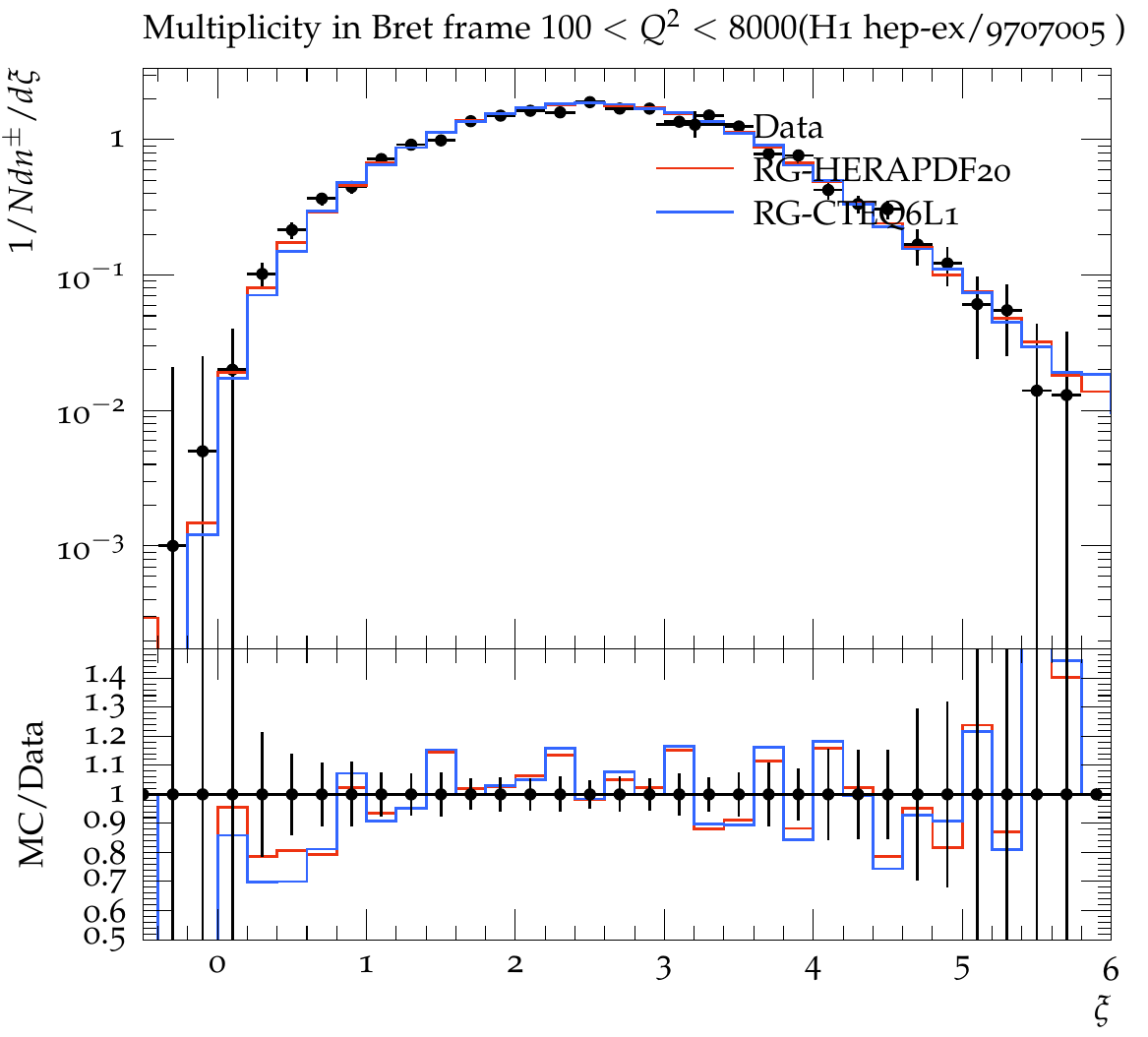}
\caption{Distribution $\xi = \log(1/x_p)$ obtained from \rivet (left)  and the corresponding one from \rivethztool (right) for high $Q^2$ data.}
\label{fig:H1_1997_I445116_4}
\end{center}
\end{figure}

The results of the \rivet plugin\footnote{Author: Kritsanon Koennonkok} are compared with those from \rivethztool  for the same kinematic range. 
Distributions for $\xi=\log(1/x_p)$ are shown in Fig.~\ref{fig:H1_1997_I445116_3} for $12 < Q^2 < 100 $ GeV$^2$ and in Fig.~\ref{fig:H1_1997_I445116_4} for $100 < Q^2 < 8000$ GeV$^2$. 
\subsection{Low \boldmath$Q^2$ jet production at HERA and virtual photon structure (H1) \\(H1\_1997\_I448449 - HZ97179)}
\renewcommand{\thissection}{H1\_1997\_I448449, HZ97179 }
\index{HZ97179 }
\index{H1\_1997\_I448449}
\markboth{\thischapter}{\thissection}
{\bf Abstract} (cited from  Ref.~\cite{H1:1997lbr}): "The transition between photoproduction and deep-inelastic scattering is investigated in jet production at the HERA $ep$ collider, using data collected by the H1 experiment. Measurements of the differential inclusive jet cross-sections $d\sigma^{ep}/dE_{t}^*$ and $d\sigma^{ep}/d\eta^*$, where $E_t^*$ and $\eta^*$ are the transverse energy and the pseudorapidity of the jets in the virtual photon-proton centre of mass frame, are presented for $0 < Q^2 < 49$ GeV$^2$ and $0.3 < y < 0.6$. " 

\begin{figure}[htbp]
\begin{center}
\includegraphics[width=0.5\linewidth]{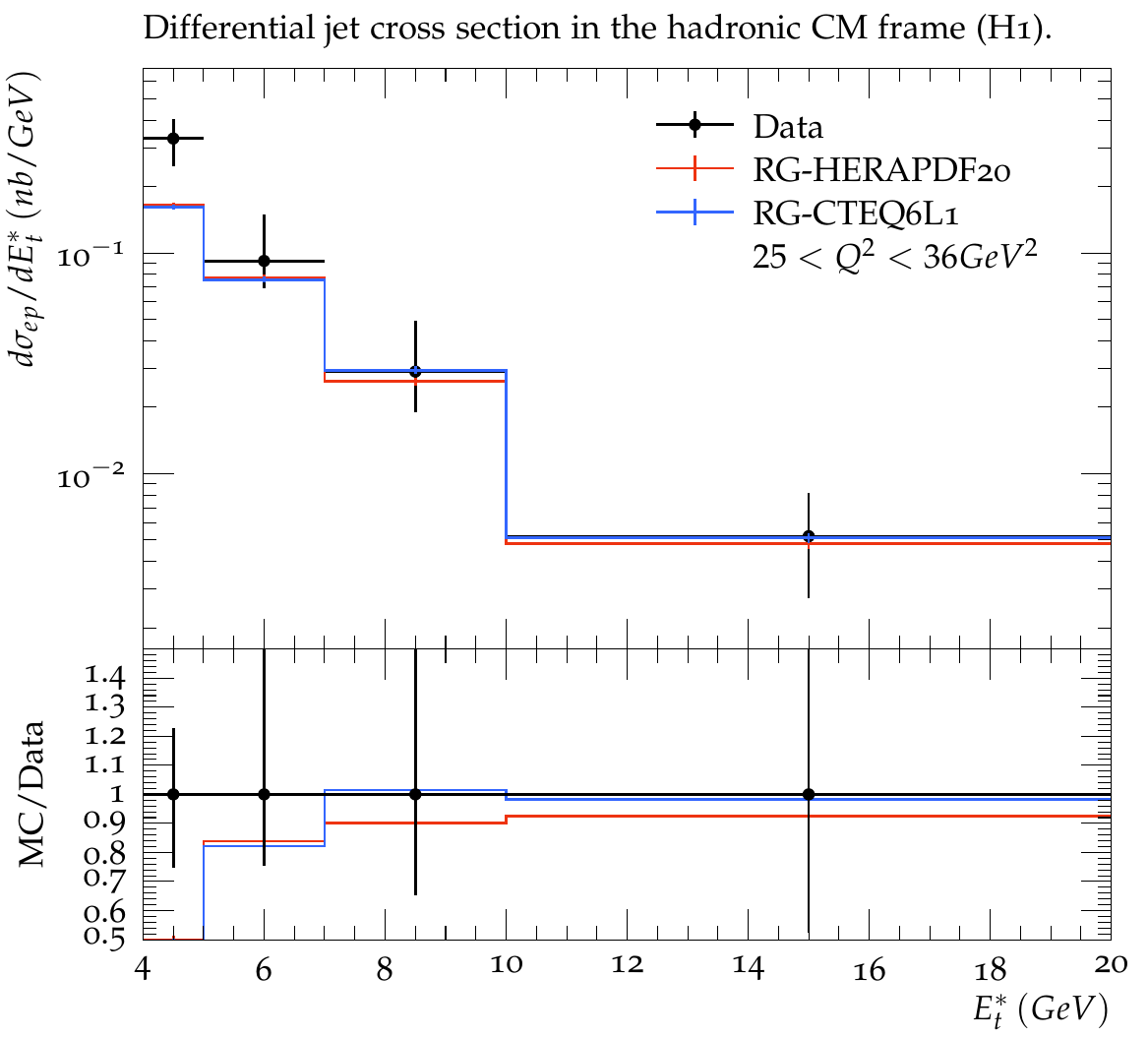}\includegraphics[width=0.5\linewidth]{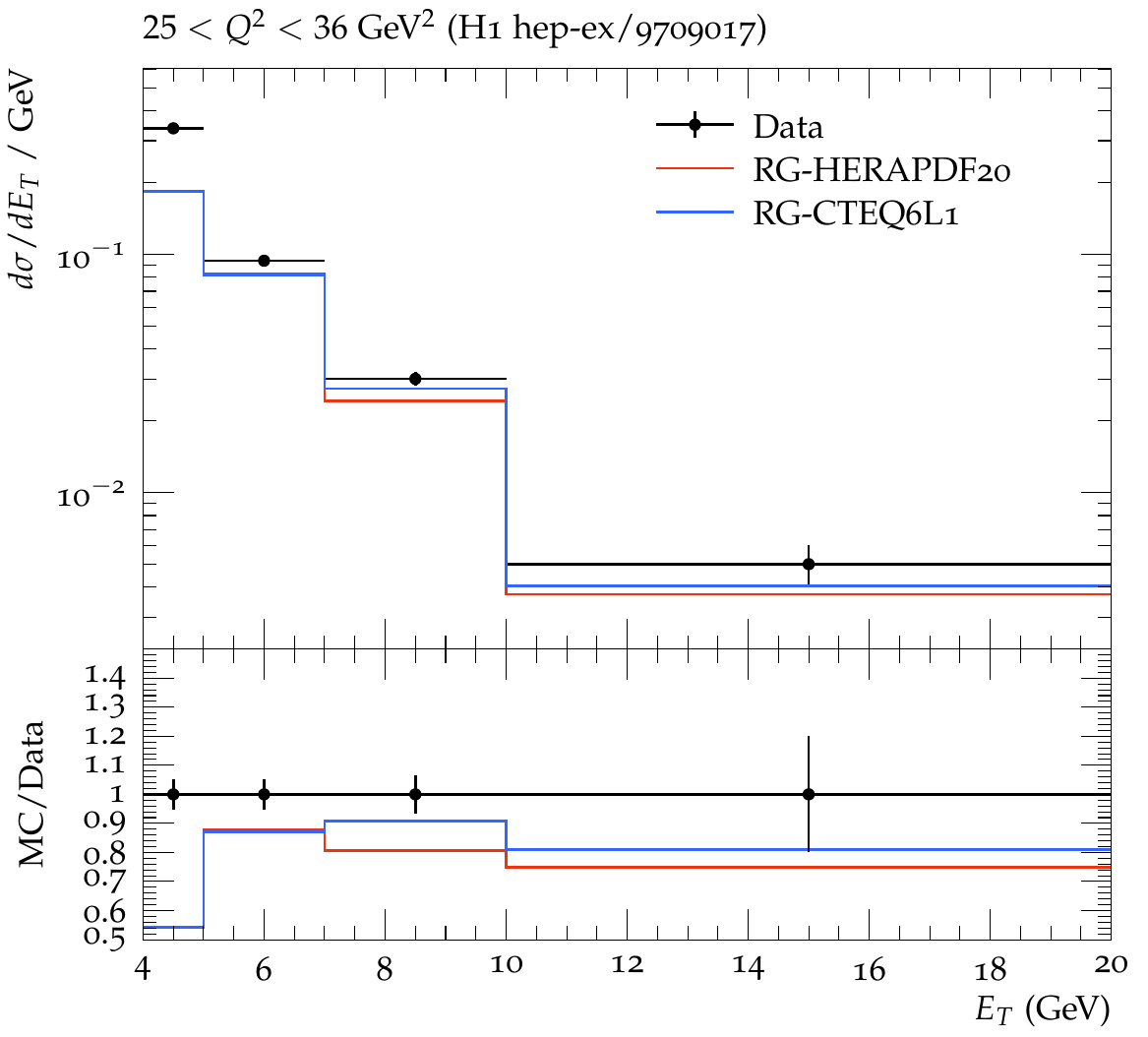}
\caption{Cross section as a function of $E^*_t$ as obtained from \rivet (left) and the corresponding one from \rivethztool (right),
showing only statistical uncertainties.}
\label{fig:H1_1997_I448449_et}
\end{center}
\end{figure}

Jets are reconstructed with the $k_t$-algorithm  used in the hadronic center-of-mass frame frame with $E^*_t > 4$ GeV and $-2.5 < \eta^* < -0.5$. 
The $k_t$-algorithm used in this analysis was not available in \fastjet and had to be implemented explicitly to be consistent with the experimental analysis.
The hadronic center-of-mass frame is defined with the photon along the $-p_z$, which is different from the standard definition.

\begin{figure}[htbp]
\begin{center}
\includegraphics[width=0.5\linewidth]{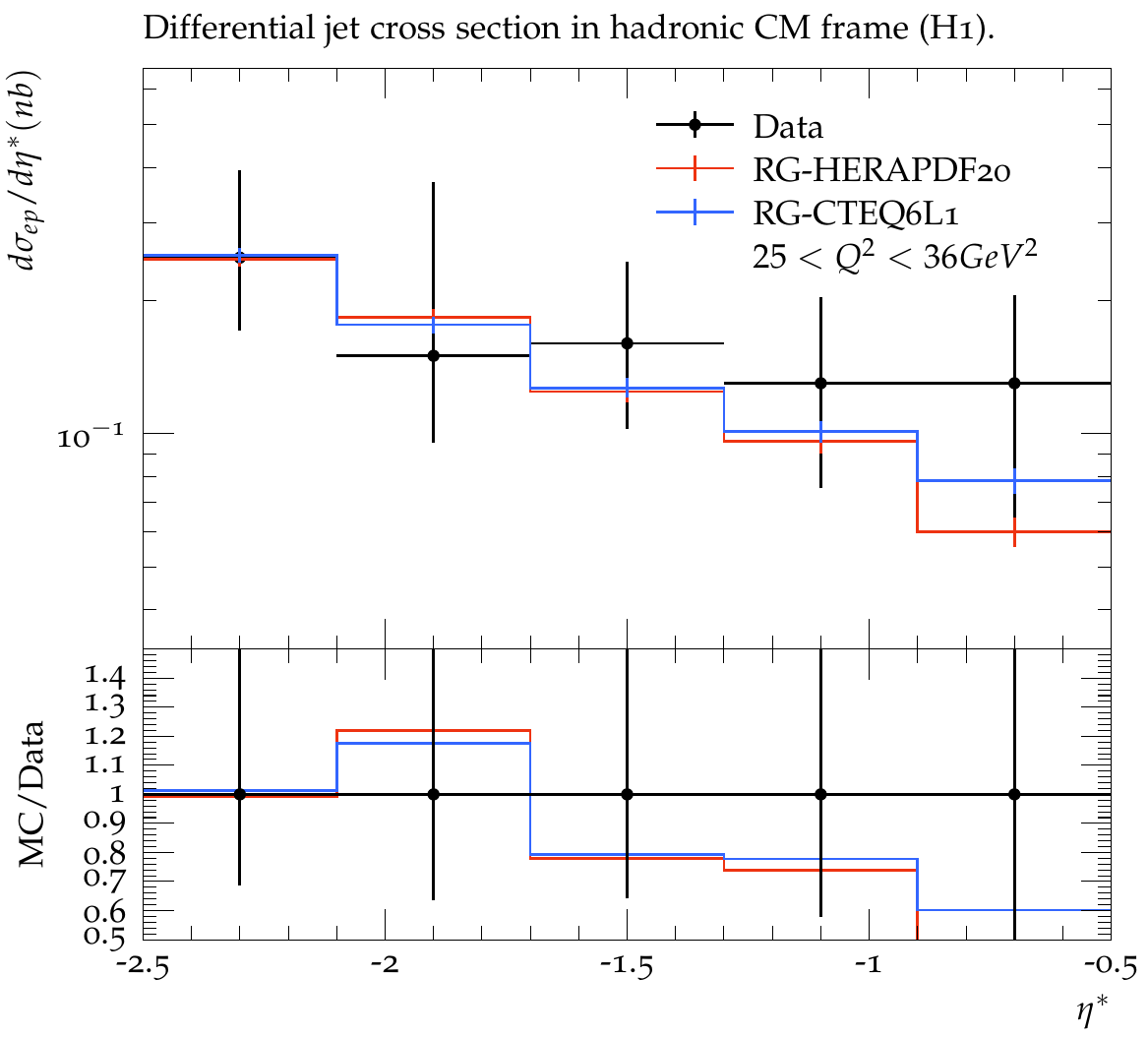}\includegraphics[width=0.5\linewidth]{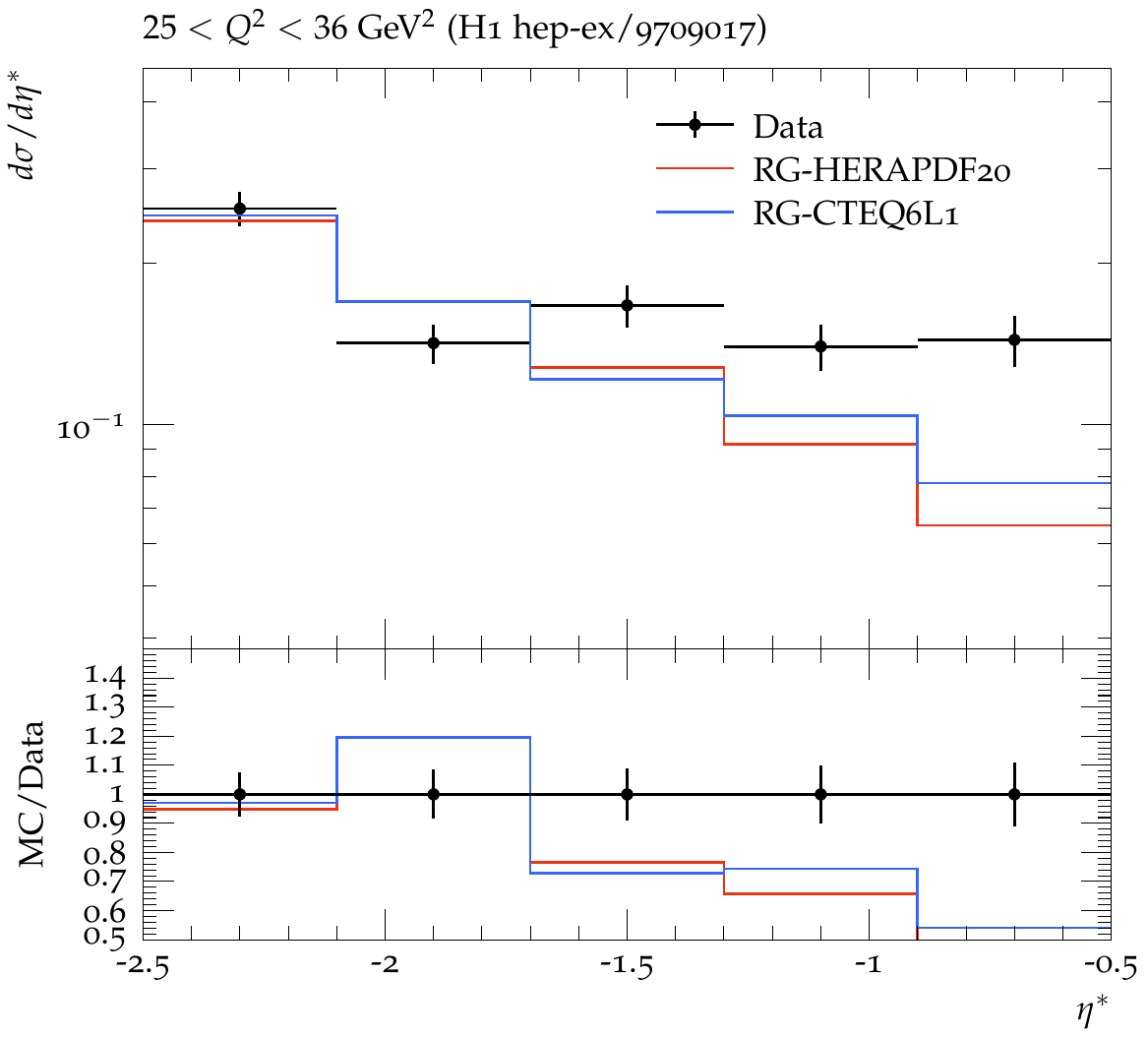}
\caption{Cross section as a function of $\eta^*$ as obtained from \rivet (left) and the corresponding one from \rivethztool (right),
showing only statistical uncertainties.}
\label{fig:H1_1997_I448449_eta}
\end{center}
\end{figure}

The results of the \rivet plugin\footnote{Author: Nattaporn Trakulphorm} are compared with those from \rivethztool  for the same kinematic range. 
Validation plots are shown for $d\sigma^{ep}/dE_{t}^*$  in Fig.~\ref{fig:H1_1997_I448449_et}, for $d\sigma^{ep}/d\eta^*$ in Fig.~\ref{fig:H1_1997_I448449_eta}.
The predictions use only direct photon processes, no resolved virtual photon processes are included.

\subsection{Forward jet production in deep inelastic scattering at HERA (ZEUS) \\(ZEUS\_1999\_I470499, ZEUS\_1999\_I508906, HZ98050) }
\begin{figure}[htpb]
\begin{center}
\includegraphics[width=0.5\linewidth]{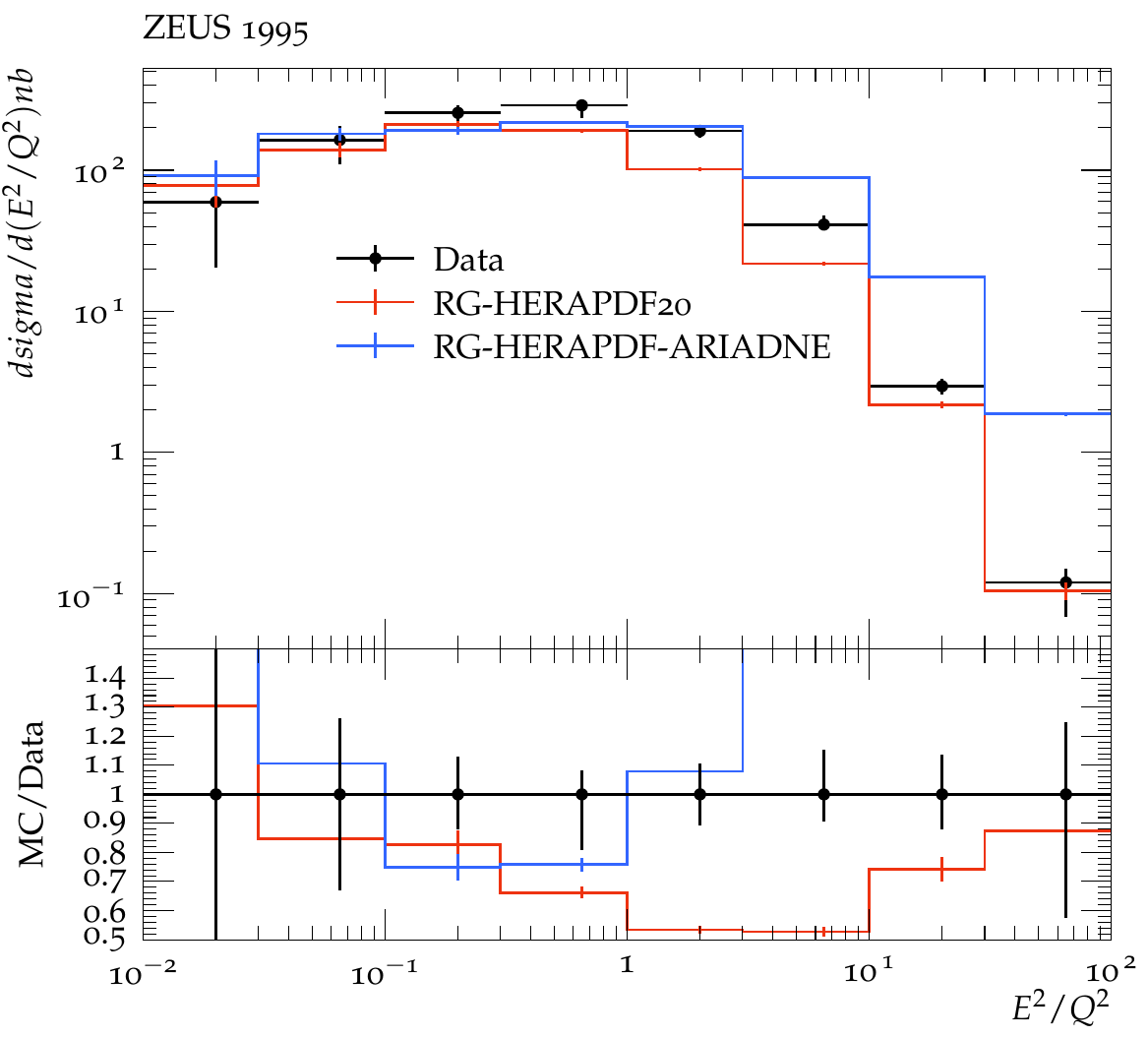}\includegraphics[width=0.5\linewidth]{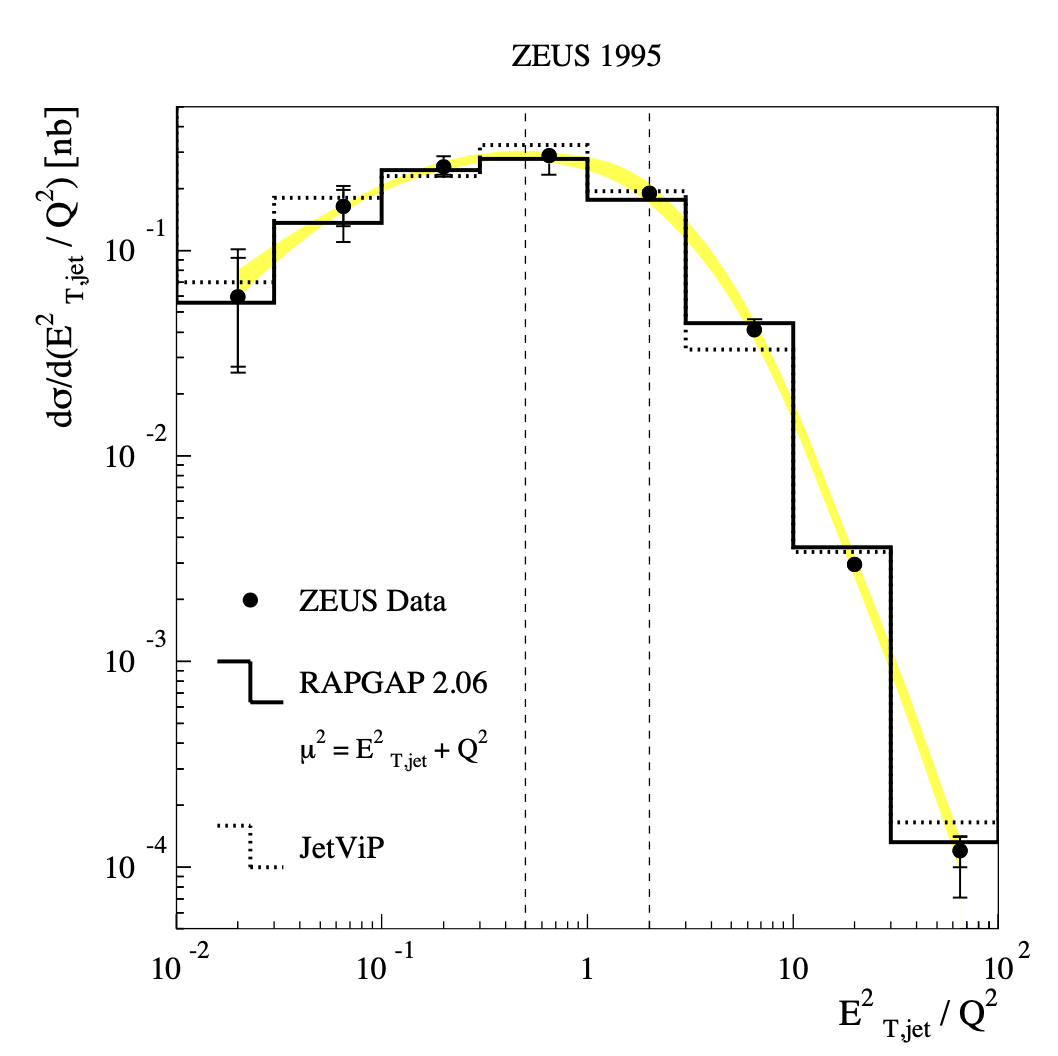}\\
\caption{Comparison of forward jet  cross section as a function of  $E_{T,jet} / Q^2 $. (Left) predictions obtained with \rivet plugin with DGLAP parton shower (red) and with \ARIADNE~(blue);  (right) from original publication.} 
\label{fig:ZEUS_1999_I470499_b}
\end{center}
\end{figure}

\renewcommand{\thissection}{ZEUS\_1999\_I470499, ZEUS\_1999\_I508906, HZ98050, }
\index{HZ98050 }
\index{ZEUS\_1999\_I470499}
\index{ZEUS\_1999\_I508906}
\markboth{\thischapter}{\thissection}
{\bf Abstract} (cited from  Ref.~\cite{Breitweg:1998ed,Breitweg:1999ss}): "The inclusive forward jet cross section in deep inelastic $e^+p$ scattering has been measured in the region of $x$-Bjorken, $4.5 \cdot 10^{-4}$ to $ 4.5 \cdot 10^{-2}$. This measurement is motivated by the search for effects of BFKL--like parton shower evolution.  The jet cross section is presented as a function of jet transverse energy squared, $E_{T,jet}^2$, and $Q^2$ in the kinematic ranges $10^{-2}<E_{T,jet}^2/Q^2<10^2$ and $2.5\cdot10^{-4}<x<8.0\cdot10^{-2}$.
"\\
\begin{figure}[htpb]
\begin{center}
\includegraphics[width=0.5\linewidth]{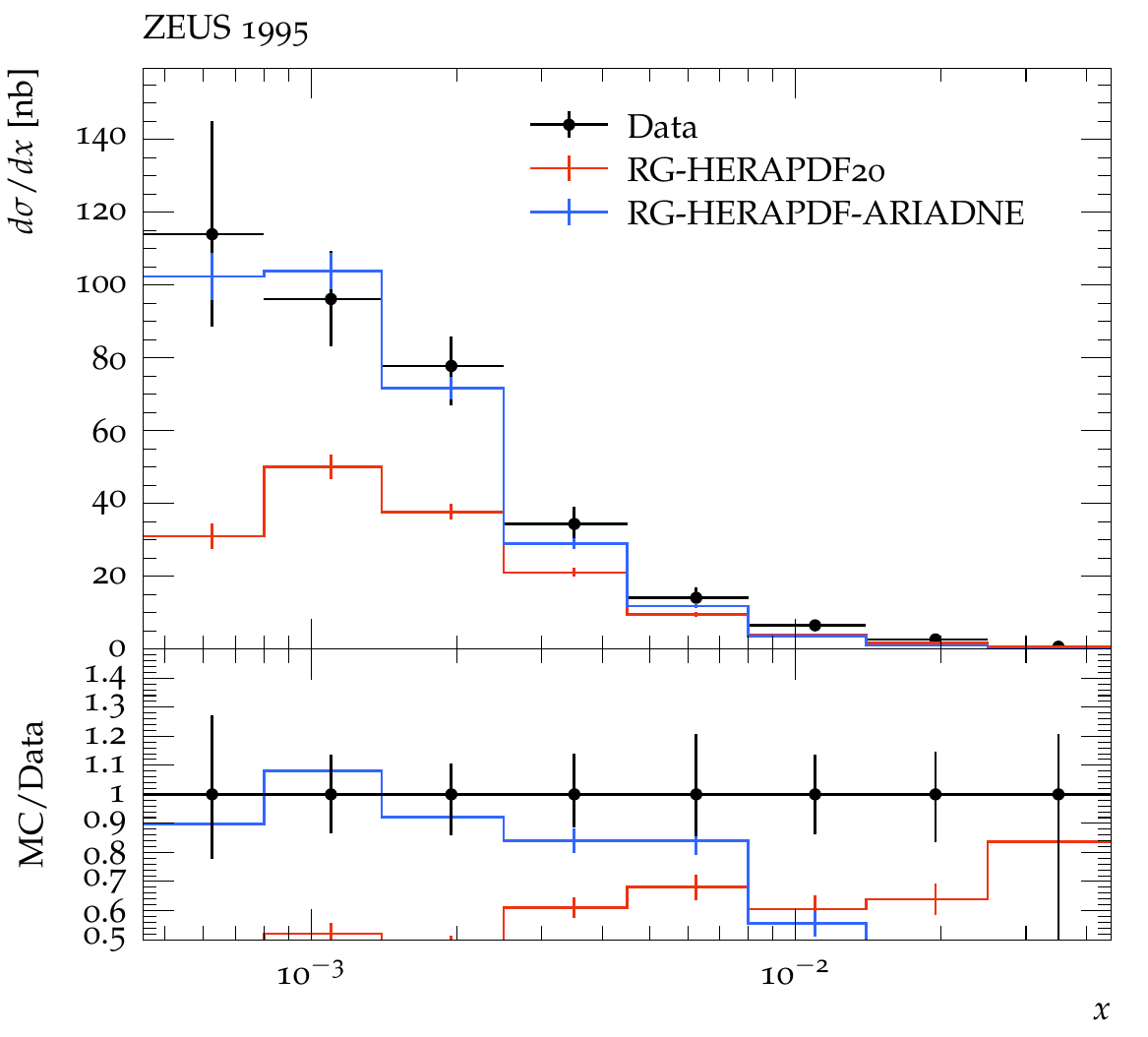}\includegraphics[width=0.5\linewidth]{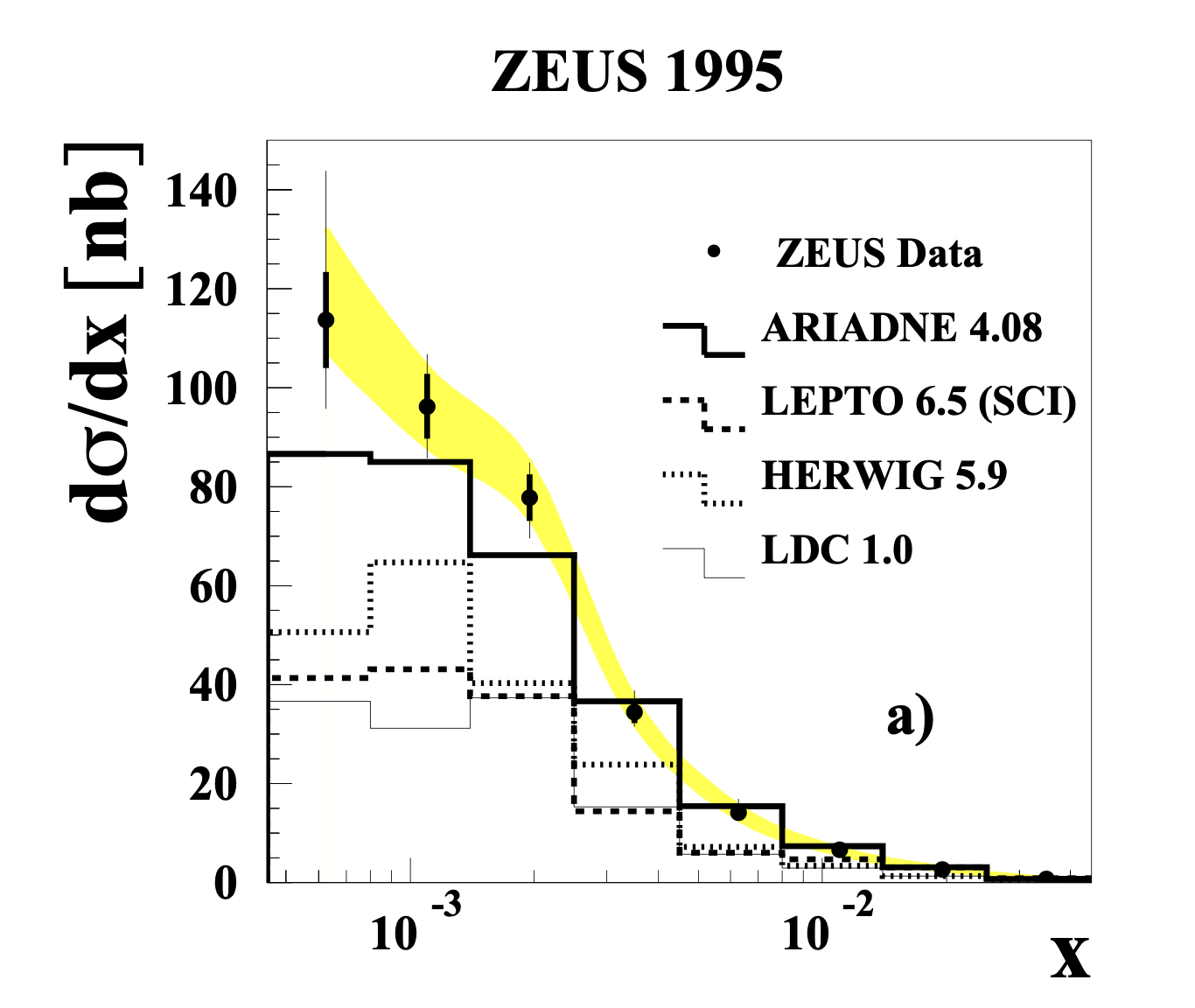}
\caption{Comparison of forward jet  cross section: (left) predictions obtained with \rivet plugin with DGLAP parton shower generated data(red) and with \ARIADNE~(blue), (right) from original publication.} 
\label{fig:ZEUS_1999_I470499_a}
\end{center}
\end{figure}
The results of the \rivet plugin\footnote{Author: Wenting Zhang} are compared with those from the original publication~\cite{Breitweg:1998ed,Breitweg:1999ss} in Fig.~\ref{fig:ZEUS_1999_I470499_a} for the forward jet cross section as a function of $x$ and in Fig.~\ref{fig:ZEUS_1999_I470499_b} for the forward jet cross section as a function of $E^2_{T,jet} / Q^2 $.

\subsection{Measurement of \boldmath$D^*$ meson cross sections at HERA and determination of the gluon density in the proton using NLO QCD \\(H1\_1999\_I481112, HZ98204)}
\renewcommand{\thissection}{H1\_1999\_I481112, HZ98204: Measurement of $D^*$ meson cross sections at HERA and determination of the gluon density in the proton using NLO QCD }
\index{HZ98204}
\index{H1\_1999\_I481112}
\markboth{\thischapter}{\thissection}
{\bf Abstract} (cited from  Ref.~\cite{H1:1998csb}): "With the H1 detector at the $ep$ collider HERA, $D^*$ meson production cross sections have been measured in deep inelastic scattering with four-momentum transfers $Q^2>2$ GeV$^2$ and in photoproduction at energies around $W_{\gamma p} \sim 88$ GeV and 194 GeV. Next-to-Leading Order QCD calculations are found to describe the differential cross sections within theoretical and experimental uncertainties. " \\
 
\begin{figure}[h!]
\begin{center}
\includegraphics[width=0.5\linewidth]{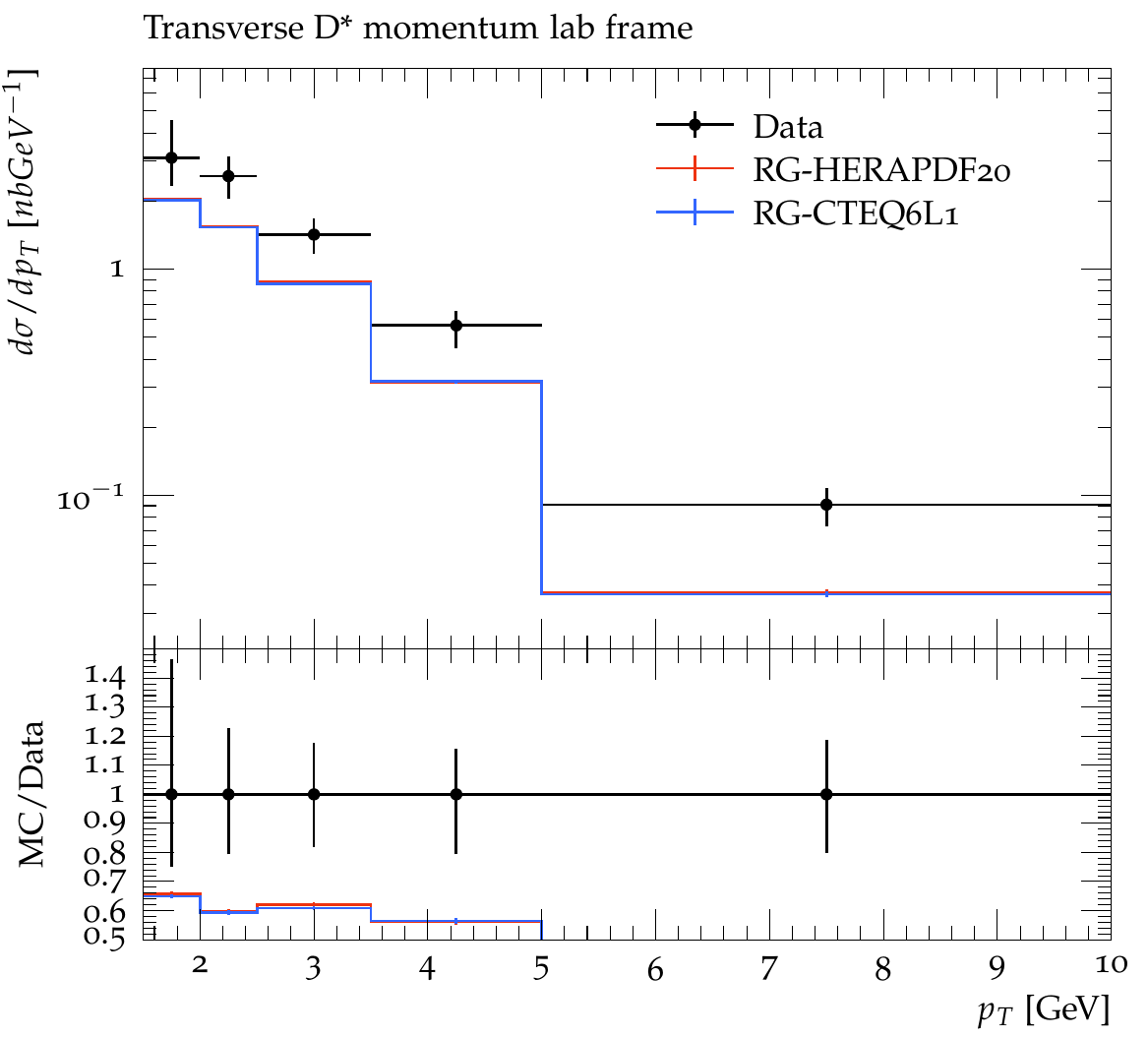}\includegraphics[width=0.5\linewidth]{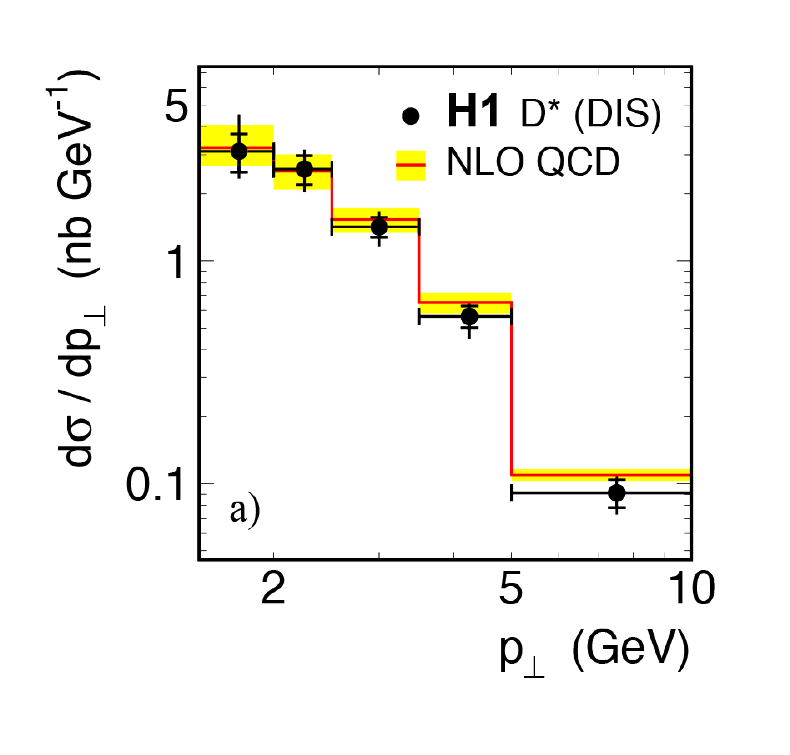}
\caption{Comparison of the transverse momentum of the $D^*$ meson in DIS as obtained from \rivet and the corresponding one from Ref.~\protect\cite{H1:1998csb}.}
\label{fig:H1_1999_I481112_a}
\end{center}
\end{figure}

 The results of the \rivet plugin\footnote{Author: Luca Marsili} are compared with those from the publication for the same kinematic range for DIS in Fig.~\ref{fig:H1_1999_I481112_a}.
 The MC simulation for photoproduction is based on the leading order subprocess $\gamma g \to c \bar{c}$,  using  CTEQ6L and HERAPDF20 as  parton density functions while in the original publication NLO calculations are used.

\subsection{Forward \boldmath$\pi^0$-meson production at HERA (H1) (H1\_1999\_I504022, HZ99094)}
\begin{figure}[htbp]
\begin{center}
\includegraphics[width=0.5\linewidth]{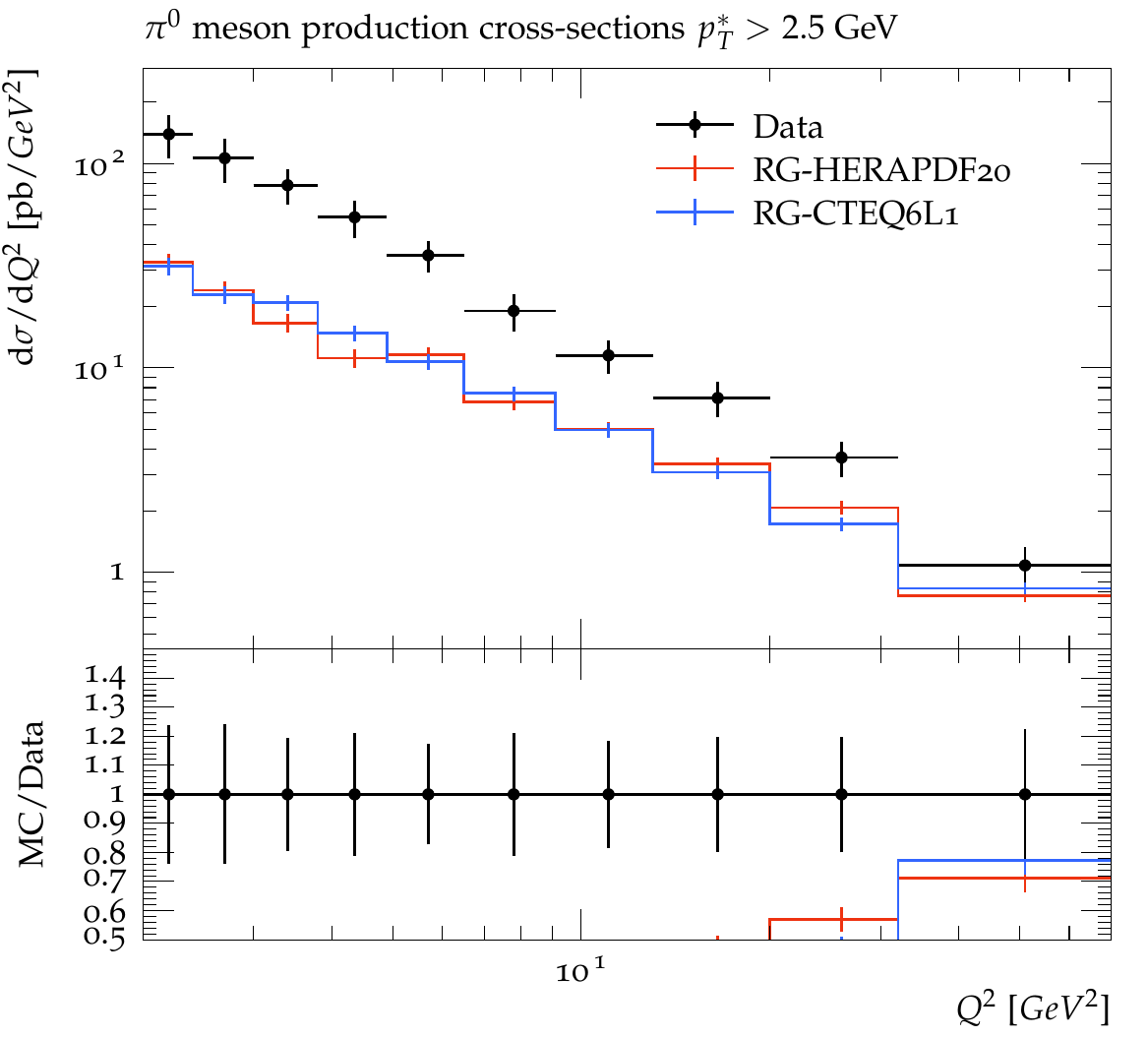}\includegraphics[width=0.5\linewidth]{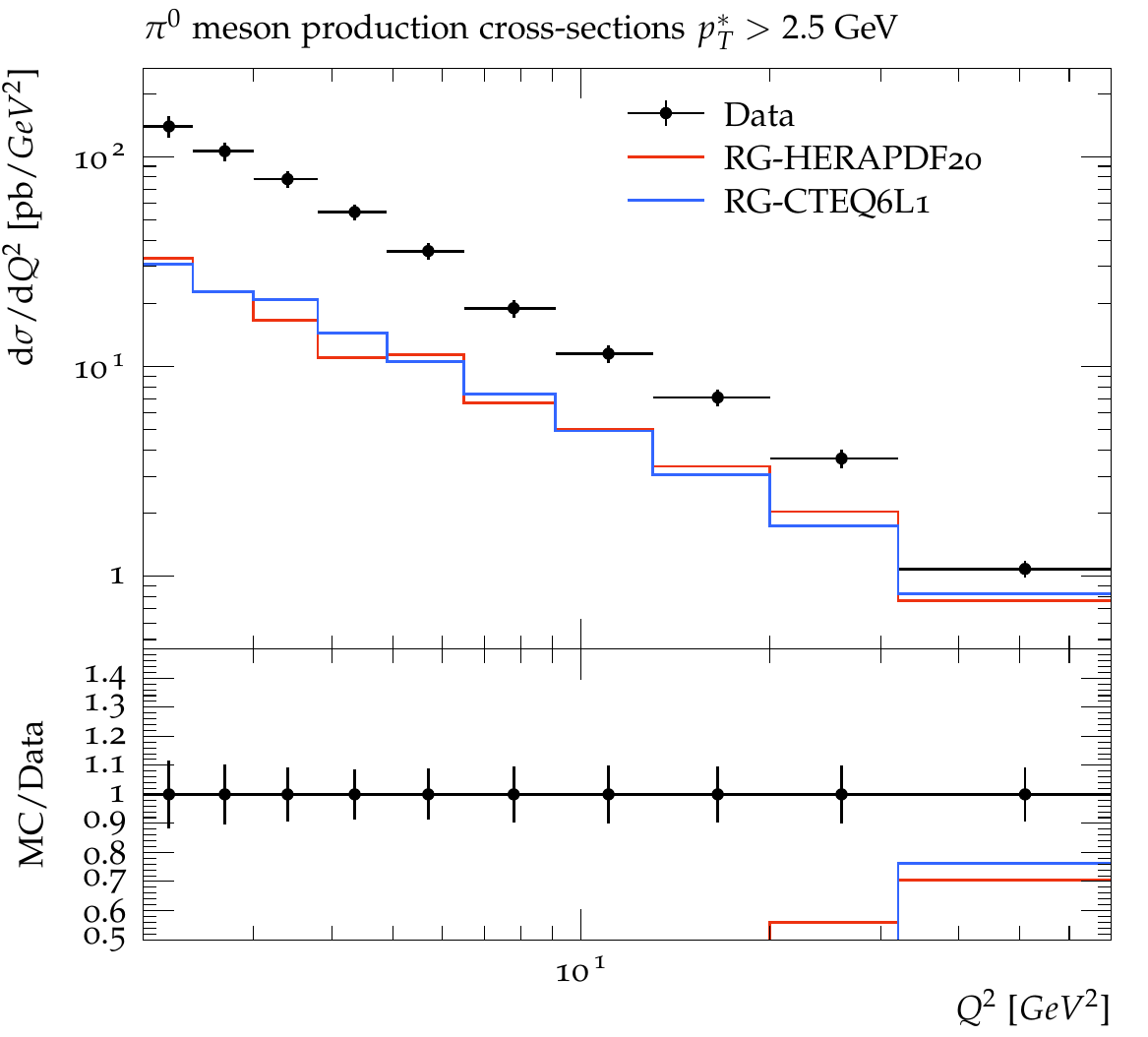}
\caption{Comparison of the forward $\pi^0$ cross section as a function of $Q^2$ obtained from \rivet (left) and the corresponding one from \rivethztool (right), showing only statistical data uncertainties. }
\label{fig:H1_1999_I504022}
\end{center}
\end{figure}

\renewcommand{\thissection}{H1\_1999\_I504022, HZ99094: Forward $\pi^0$-Meson Production at HERA (H1) }
\index{HZ99094}
\index{H1\_1999\_I504022}
\markboth{\thischapter}{\thissection}
{\bf Abstract} (cited from  Ref.~\cite{Adloff:1999zx}): "High transverse momentum $\pi^0$-mesons have been measured with the H1 detector at HERA in deep-inelastic $ep$ scattering events at low Bjorken-$x$, down to $x \leq 4 \times 10^{-5}$. The measurement is performed in a region of small angles with respect to the proton remnant in the laboratory frame of reference, namely the forward region, and corresponds to central rapidity in the centre-of-mass system of the virtual photon and proton. This region is expected to be particularly sensitive to QCD effects in hadronic final states. Differential cross sections for inclusive $\pi^0$-meson production are presented as a function of Bjorken-$x$ and the four-momentum transfer $Q^2$, and as a function of transverse momentum and pseudorapidity."\\
The results of the \rivet plugin\footnote{Author: Keila Moral Figueroa} are compared with those from \rivethztool  for the same kinematic range. 
Validation plots are shown in Fig.~\ref{fig:H1_1999_I504022}.

\subsection{Measurement of azimuthal asymmetries in deep inelastic scattering (ZEUS) (ZEUS\_2000\_I524911, HZ00040)}
\begin{figure}[htbp]
\begin{center}
\includegraphics[width=0.5\linewidth]{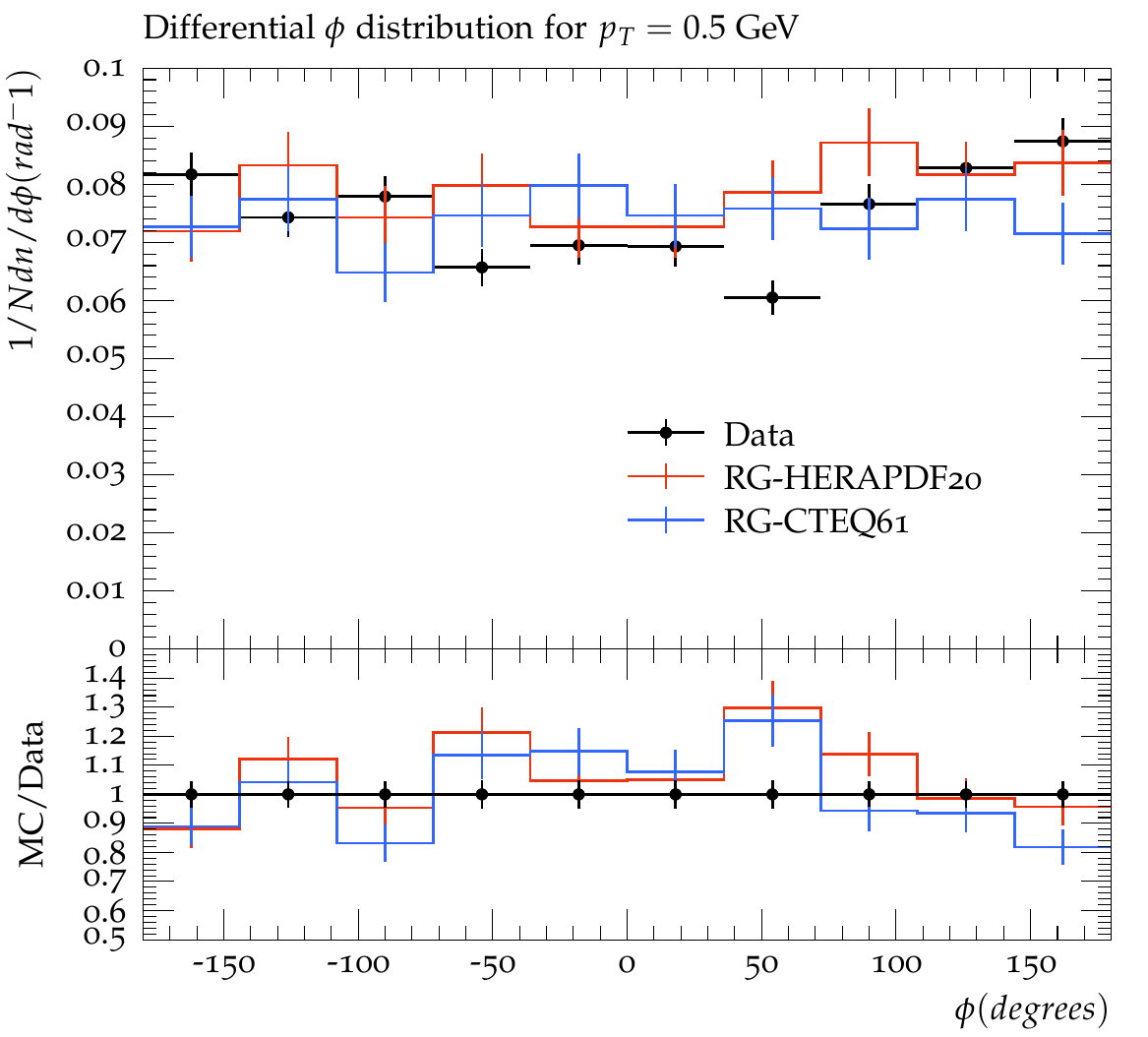}\includegraphics[width=0.5\linewidth]{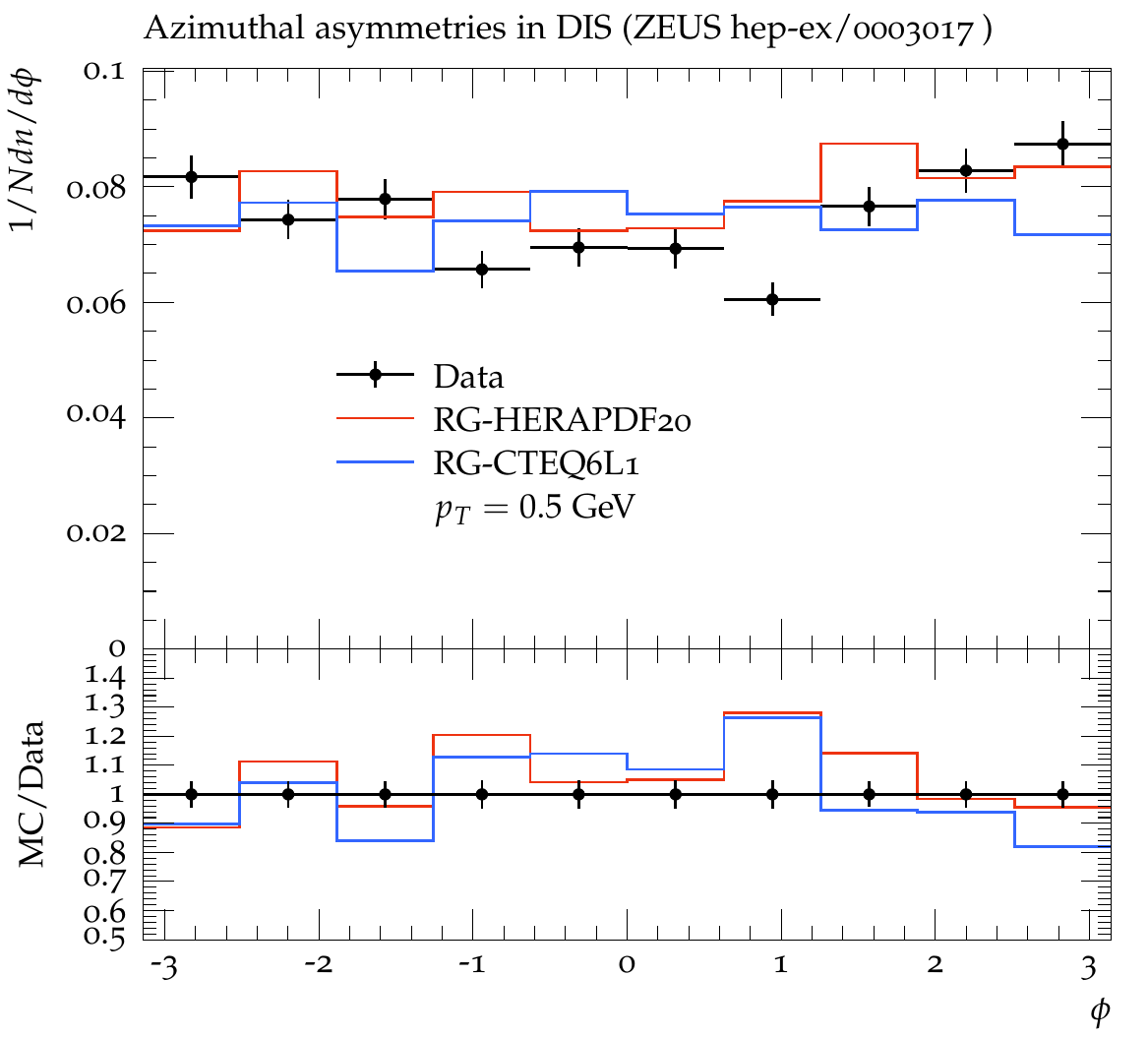}
\caption{Comparison of the azimuthal distribution of charged particles obtained from \rivet (left) and the corresponding one from \rivethztool (right).}
\label{fig:ZEUS_2000_I524911}
\end{center}
\end{figure}

\index{HZ00040}
\index{ZEUS\_2000\_I524911}
{\bf Abstract} (cited from  Ref.~\cite{ZEUS:2000esx}: "The distribution of the azimuthal angle for the charged hadrons has been studied in the hadronic centre-of-mass system for neutral current deep inelastic positron - proton scattering with the ZEUS detector at HERA. Measurements of the dependence of the moments of this distribution on the transverse momenta of the charged hadrons are presented. Asymmetries that can be unambiguously attributed to perturbative QCD processes have been observed for the first time."\\
The results of the \rivet plugin\footnote{Author: Aryan Borkar} are compared with those from \rivethztool  for the same kinematic range. 
Validation plots are shown in Fig.~\ref{fig:ZEUS_2000_I524911}.

\subsection{Measurement of \boldmath$D^{*\pm}$ Meson Production and \boldmath$F_2^c$ in deep inelastic scattering at HERA (H1) (H1\_2002\_I561885, HZ01100)}
\begin{figure}[htbp]
\begin{center}
\includegraphics[width=0.5\linewidth]{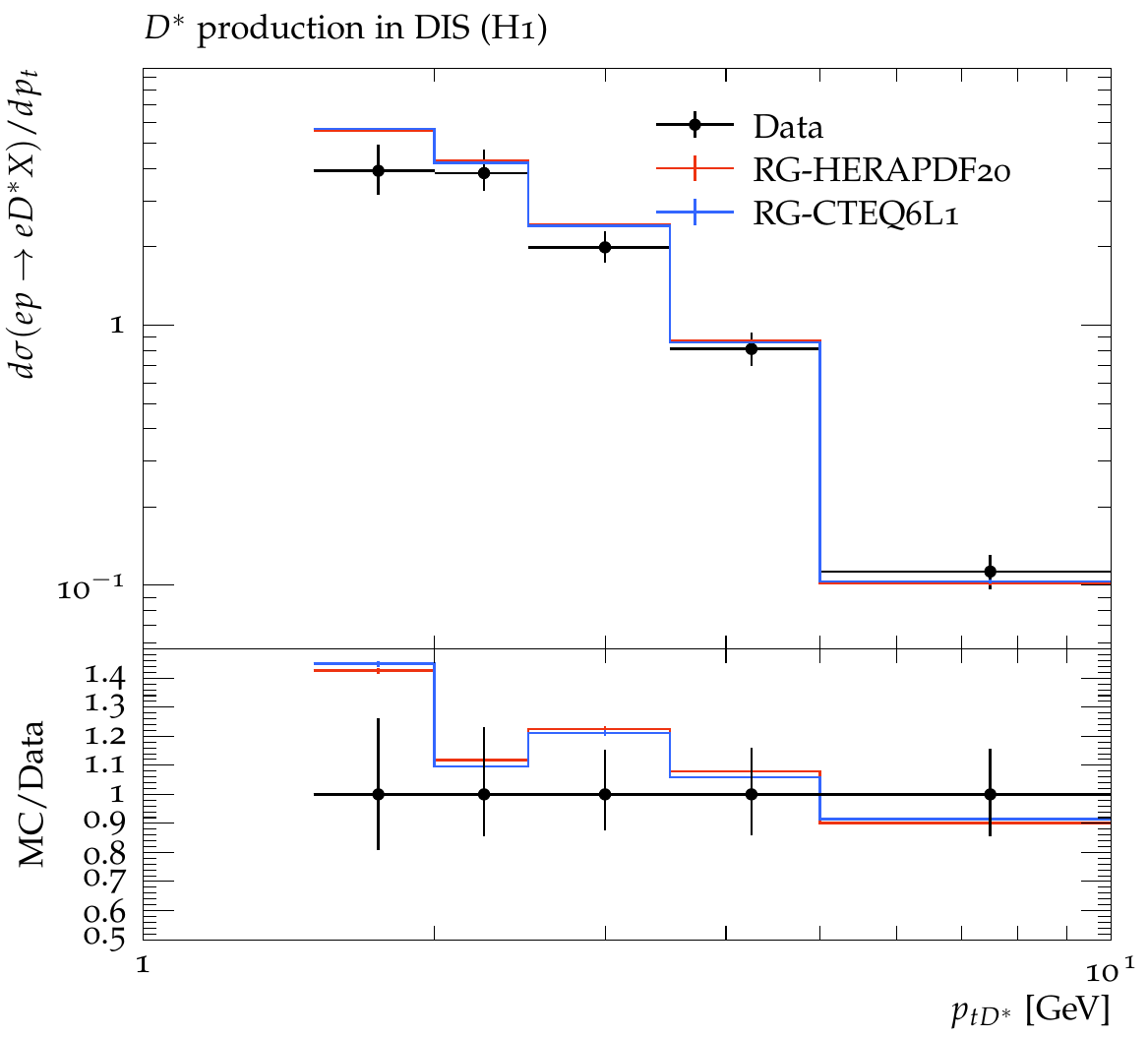}\includegraphics[width=0.5\linewidth]{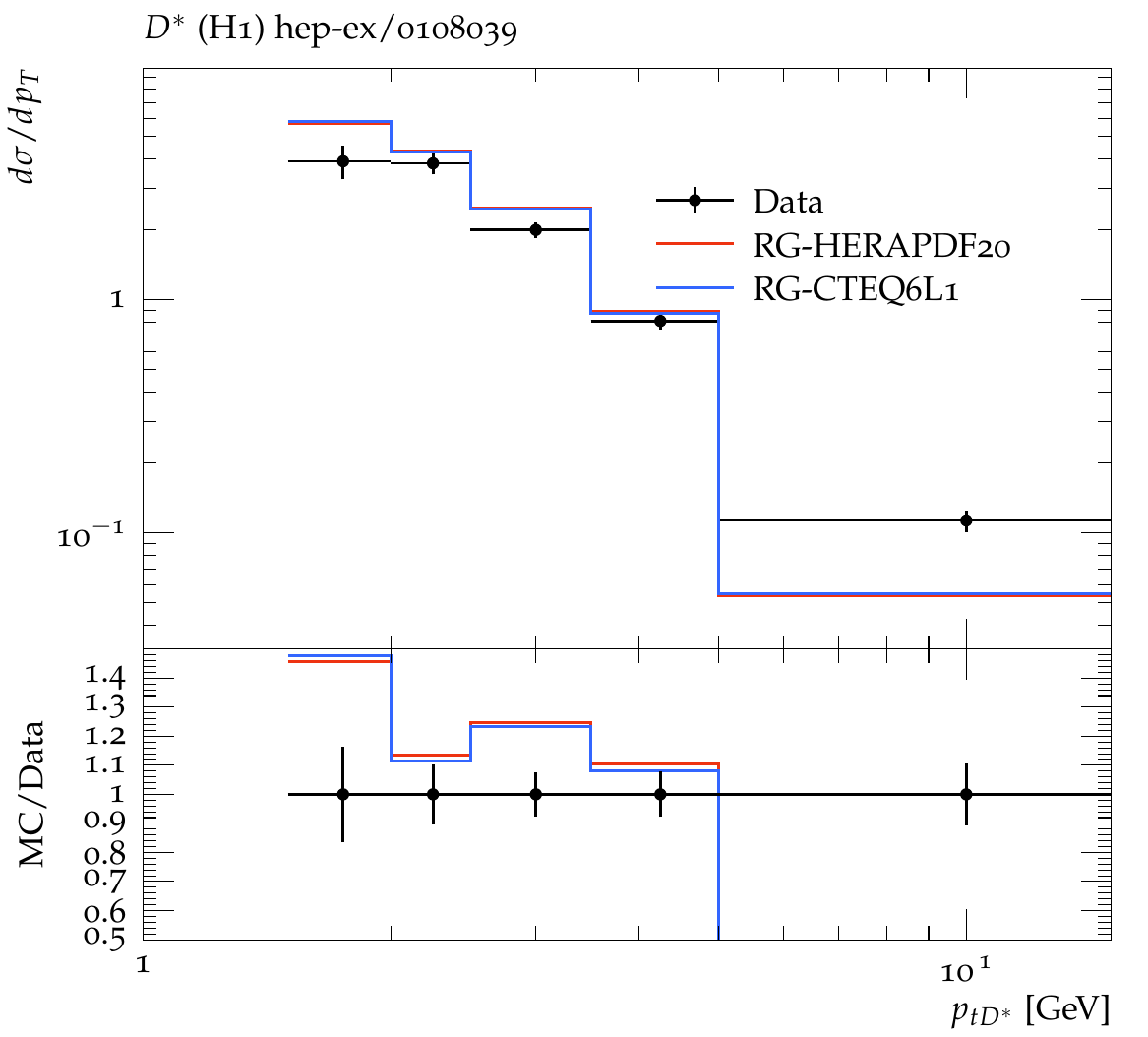}
\caption{Transverse momentum distribution of $D^{*\pm}(2010)$ mesons in DIS as obtained from \rivet and the corresponding one from \rivethztool.}
\label{fig:H1_2002_I561885}
\end{center}
\end{figure}

\renewcommand{\thissection}{H1\_2002\_I561885, HZ01100}
\index{HZ01100}
\index{H1\_2002\_I561885}
\markboth{\thischapter}{\thissection}
{\bf Abstract} (cited from  Ref.~\cite{H1:2001wps}: "The inclusive production of $D^{*\pm}(2010)$ mesons in deep-inelastic scattering is studied with the H1 detector at HERA. In the kinematic region $1< Q^2 <100$ GeV$^2$ and $0.05< y <0.7$ an $e^+ p$  cross section for inclusive $D^{*\pm}$ meson production of $8.50\pm0.42 {\rm (stat.)} ^{+1.21} _{-1.00} {\rm (syst.)}$\,nb is measured in the visible range $p_{t}(D^*) >1.5$  GeV and $|\eta (D^*) |<1.5$. Single and double differential inclusive $D^{*\pm}$ meson cross sections are compared to perturbative QCD calculations in two different evolution schemes." \\
The results of the \rivet plugin\footnote{Author: Madhav Chithirasreemadam} are compared with those from \rivethztool  for the same kinematic range. 
Validation plots are shown in Fig.~\ref{fig:H1_2002_I561885}.

\subsection{Measurement of beauty production at HERA using events with muons and jets (H1) (H1\_2005\_I676166, HZH0502010)}
\begin{figure}[htbp]
\begin{center}
\includegraphics[width=0.33\linewidth]{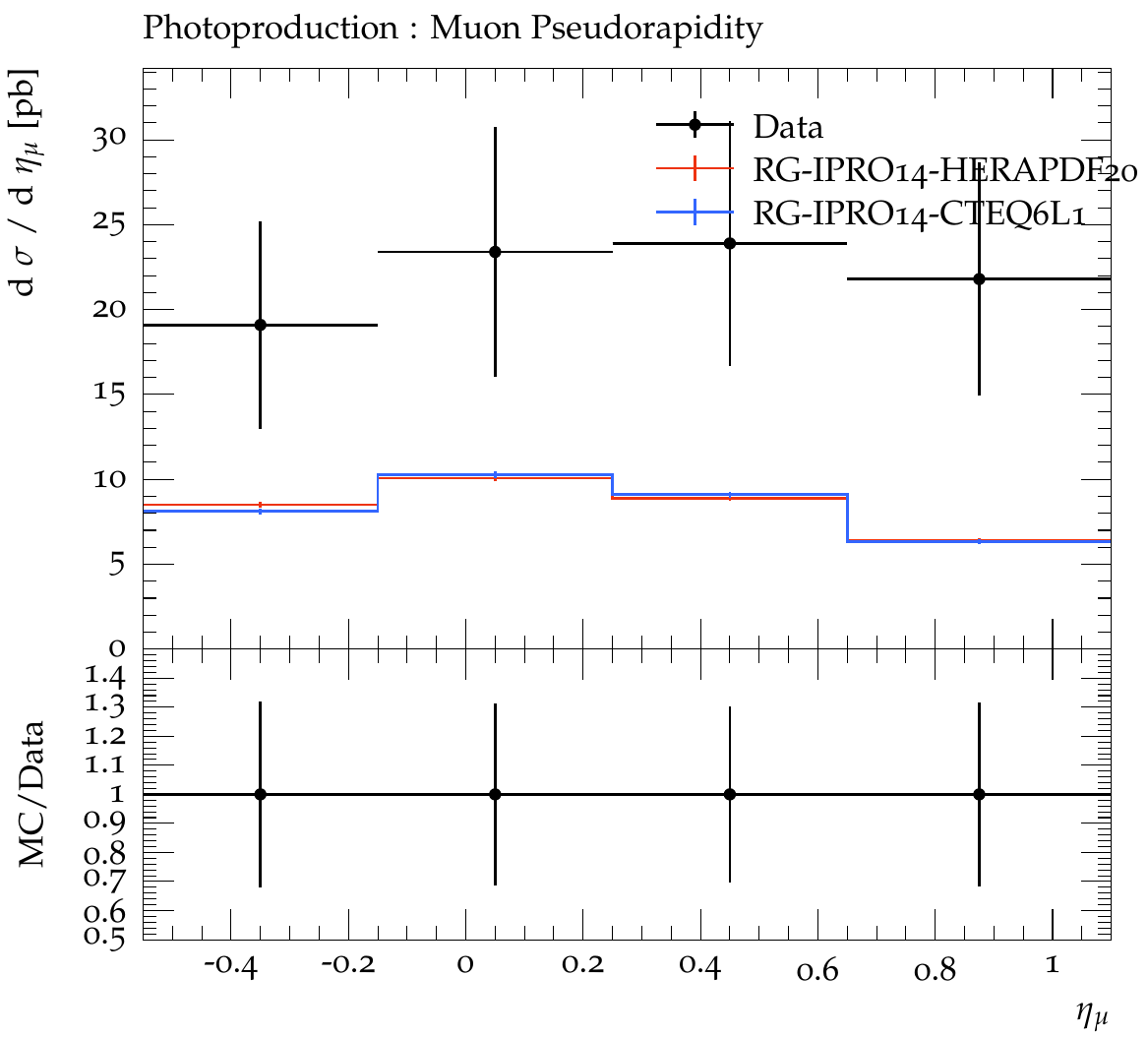}\includegraphics[width=0.33\linewidth]{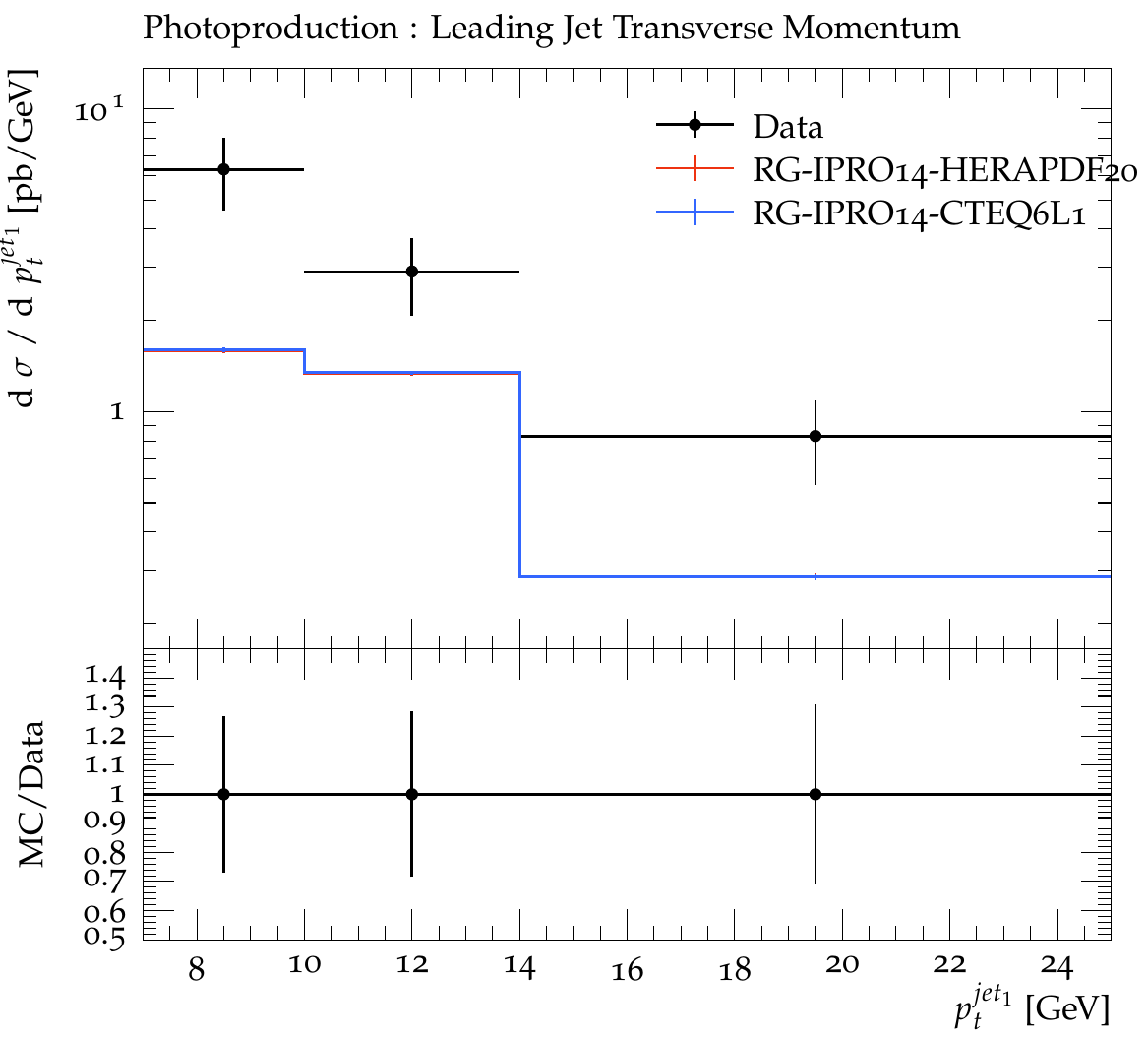}\includegraphics[width=0.33\linewidth]{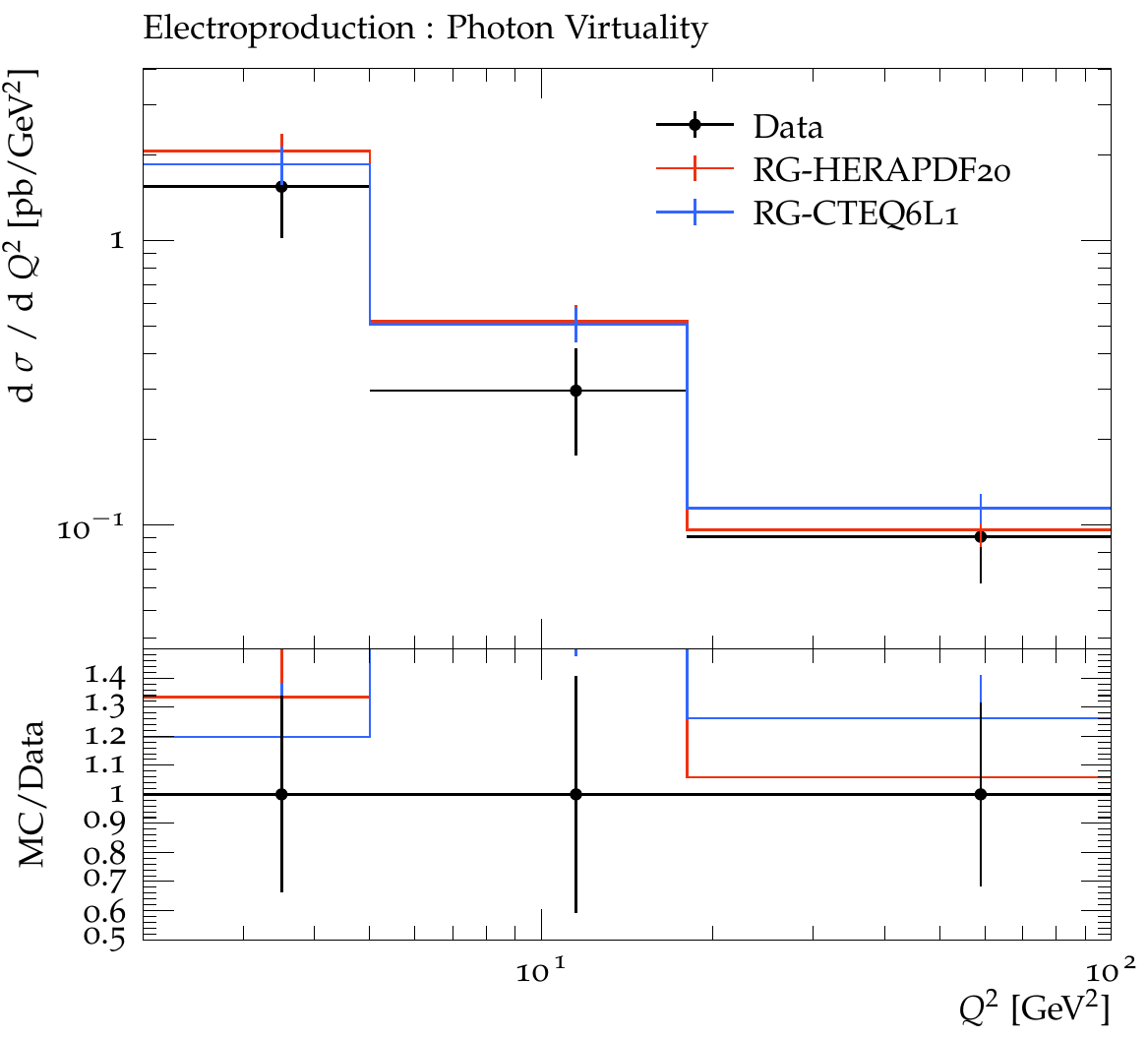}
\caption{The pseudorapidity spectrum of muons (left), the jet transverse momentum spectrum (middle) and the $Q^2$ distribution (right).}
\label{fig:H1_2005_I676166_NLO_11_14_18}
\end{center}
\end{figure}

\renewcommand{\thissection}{H1\_2005\_I676166, HZH0502010}
\index{HZH0502010}
\index{H1\_2005\_I676166}
\markboth{\thischapter}{\thissection}
\begin{figure}[htbp]
\begin{center}
\includegraphics[width=0.5\linewidth]{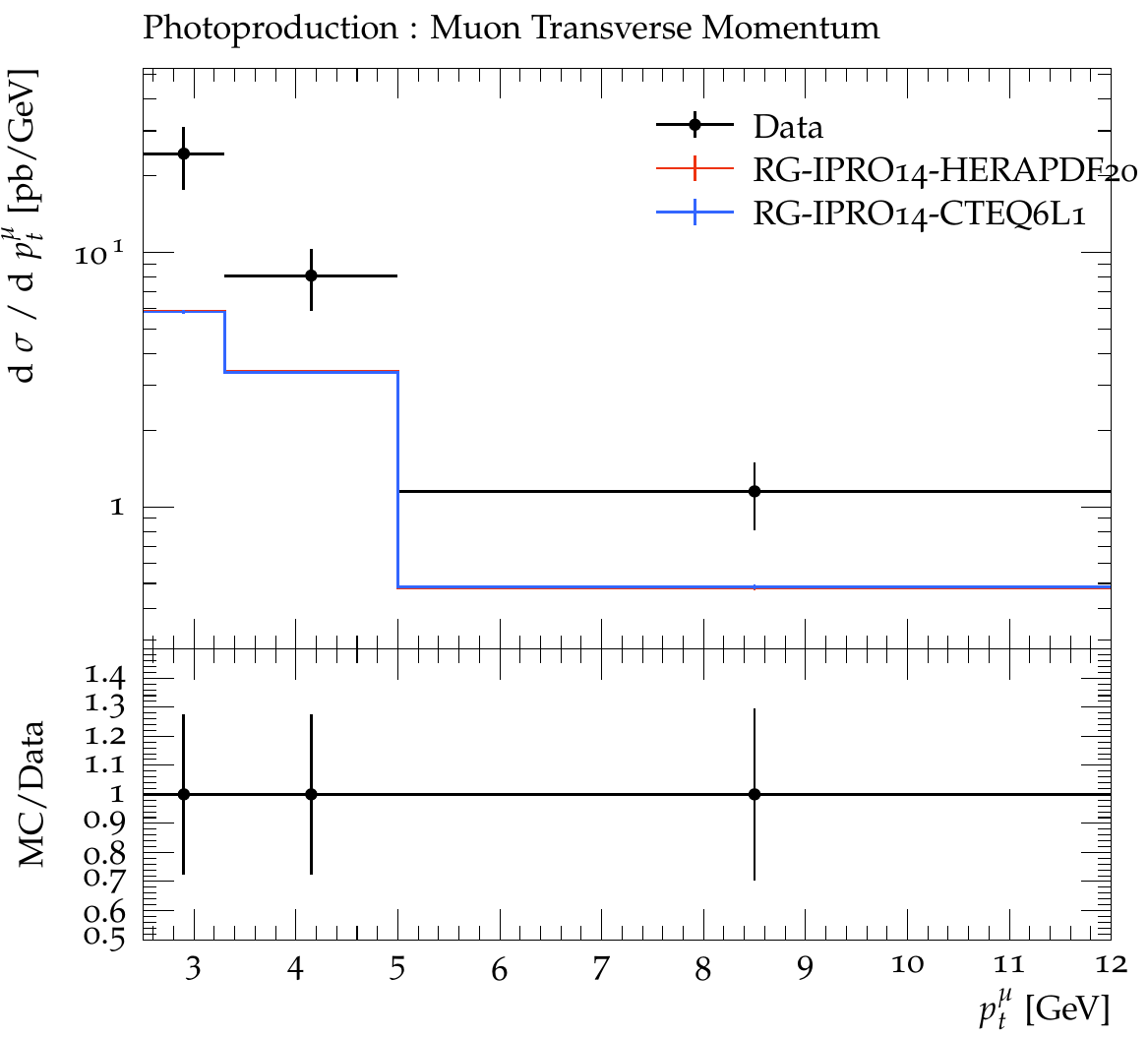}\includegraphics[width=0.5\linewidth]{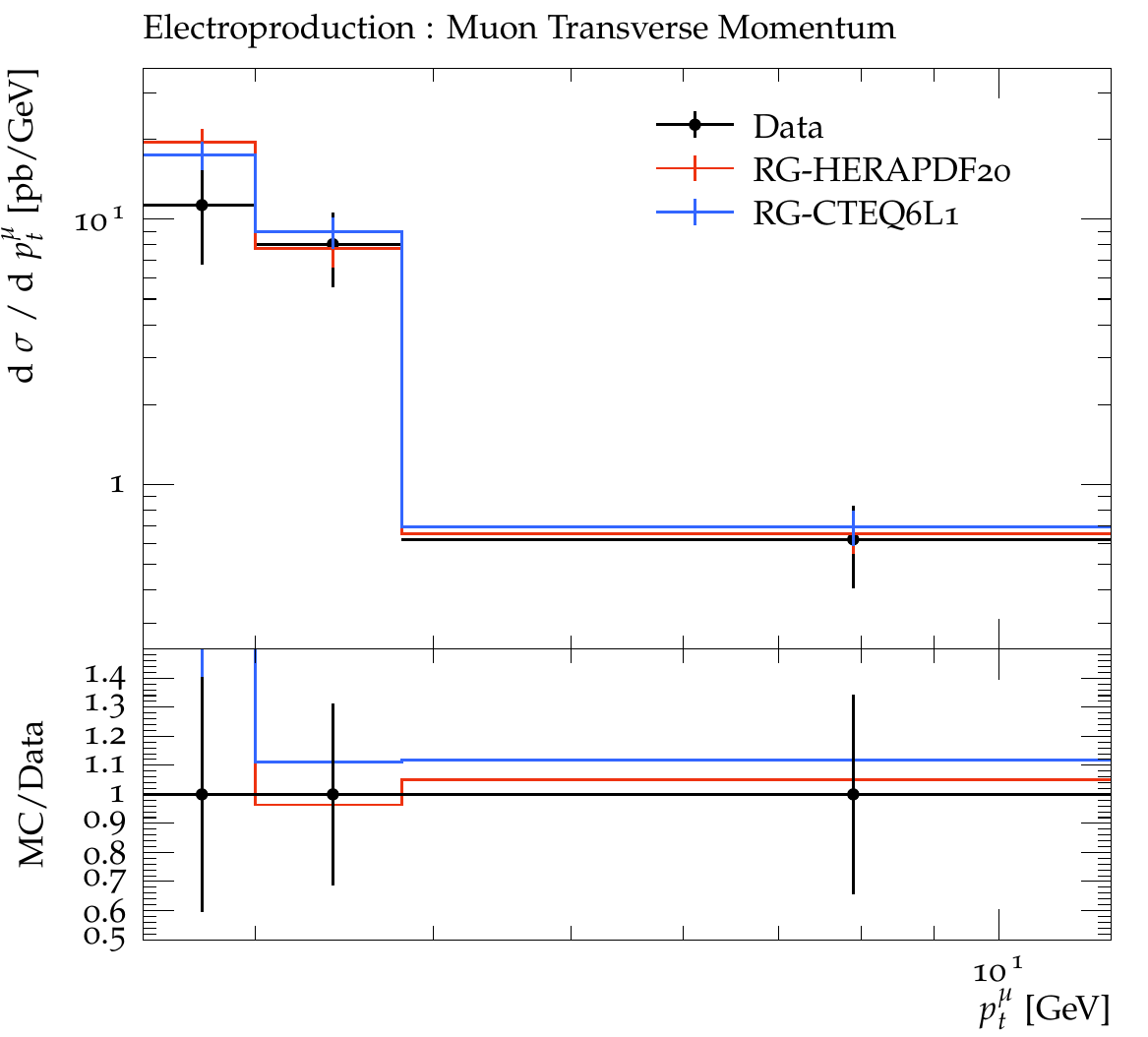}
\caption{Comparison of the transverse momentum spectrum of muons in photoproduction (left) and DIS (right) as obtained using \rivet. Photoproduction events are generated in the direct mode with  $e g \rightarrow b \bar{b}$ in massive mode (\texttt{IPRO=14}) while DIS events are generated inclusively with (\texttt{IPRO=12}).}
\label{fig:H1_2005_I676166_NLO_14}
\end{center}
\end{figure}
{\bf Abstract} (cited from Ref.~\cite{Aktas:2005zc}): "A measurement of the beauty production cross section in $ep$ collisions at a centre-of-mass energy of 319 GeV is presented. The data were collected with the H1 detector at the HERA collider in the years 1999-2000. Events are selected by requiring the presence of jets and muons in the final state. Both the long lifetime and the large mass of b-flavoured hadrons are exploited to identify events containing beauty quarks. Differential cross sections are measured in photoproduction, with photon virtualities $Q^2 < 1$ GeV$^2$, and in deep inelastic scattering, where $2 < Q^2 < 100$ GeV$^2$. " \\
The results of the \rivet plugin\footnote{Author: Danielle Wilson}  are shown in Fig.~\ref{fig:H1_2005_I676166_NLO_14} and Fig.~\ref{fig:H1_2005_I676166_NLO_11_14_18}.

\subsection{Forward Jet Production in Deep Inelastic Scattering at HERA \\ (H1\_2006\_I690939, HZH0508055)}
\begin{figure}[htbp]
\begin{center}
\includegraphics[width=0.5\linewidth]{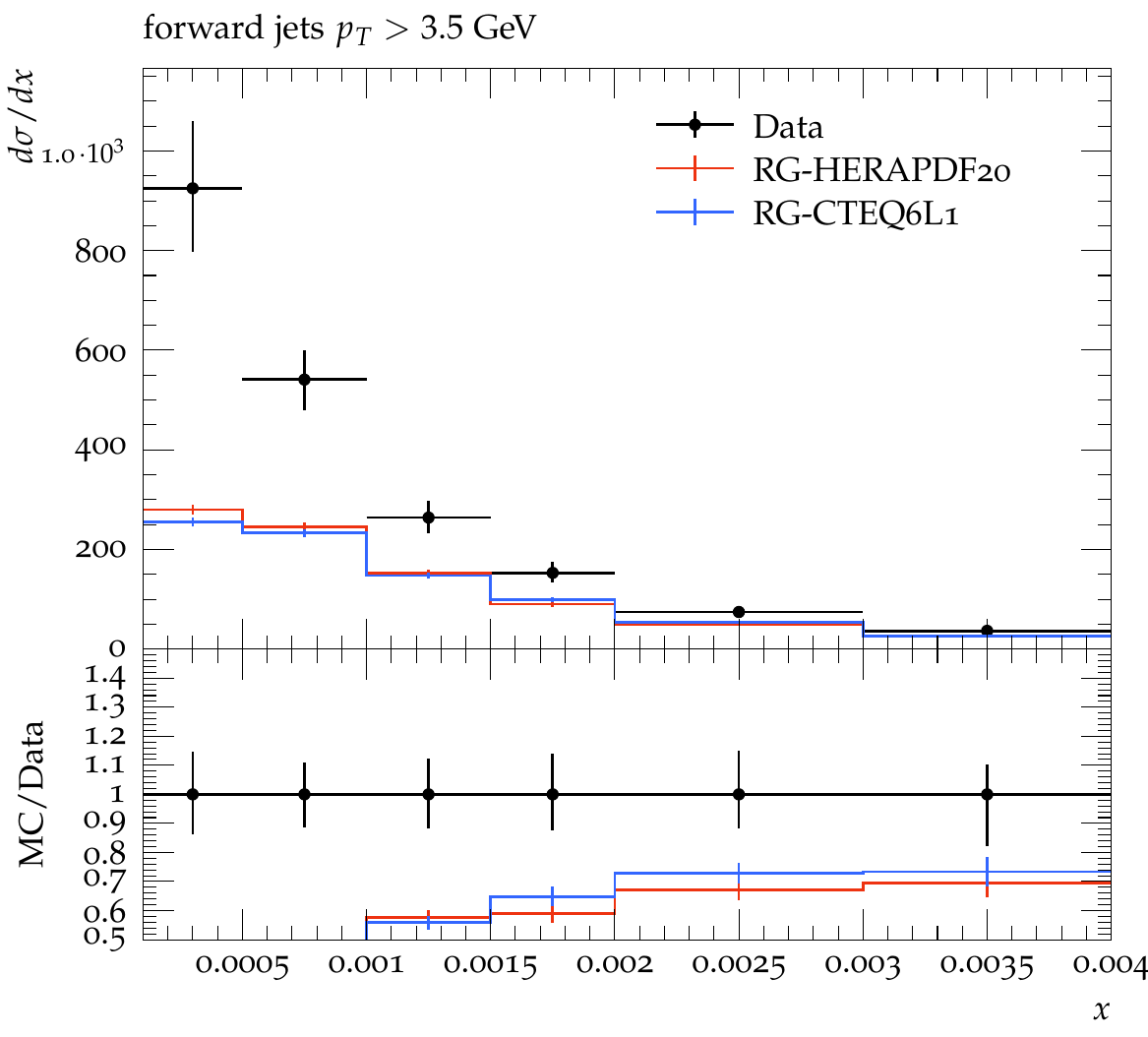}\includegraphics[width=0.5\linewidth]{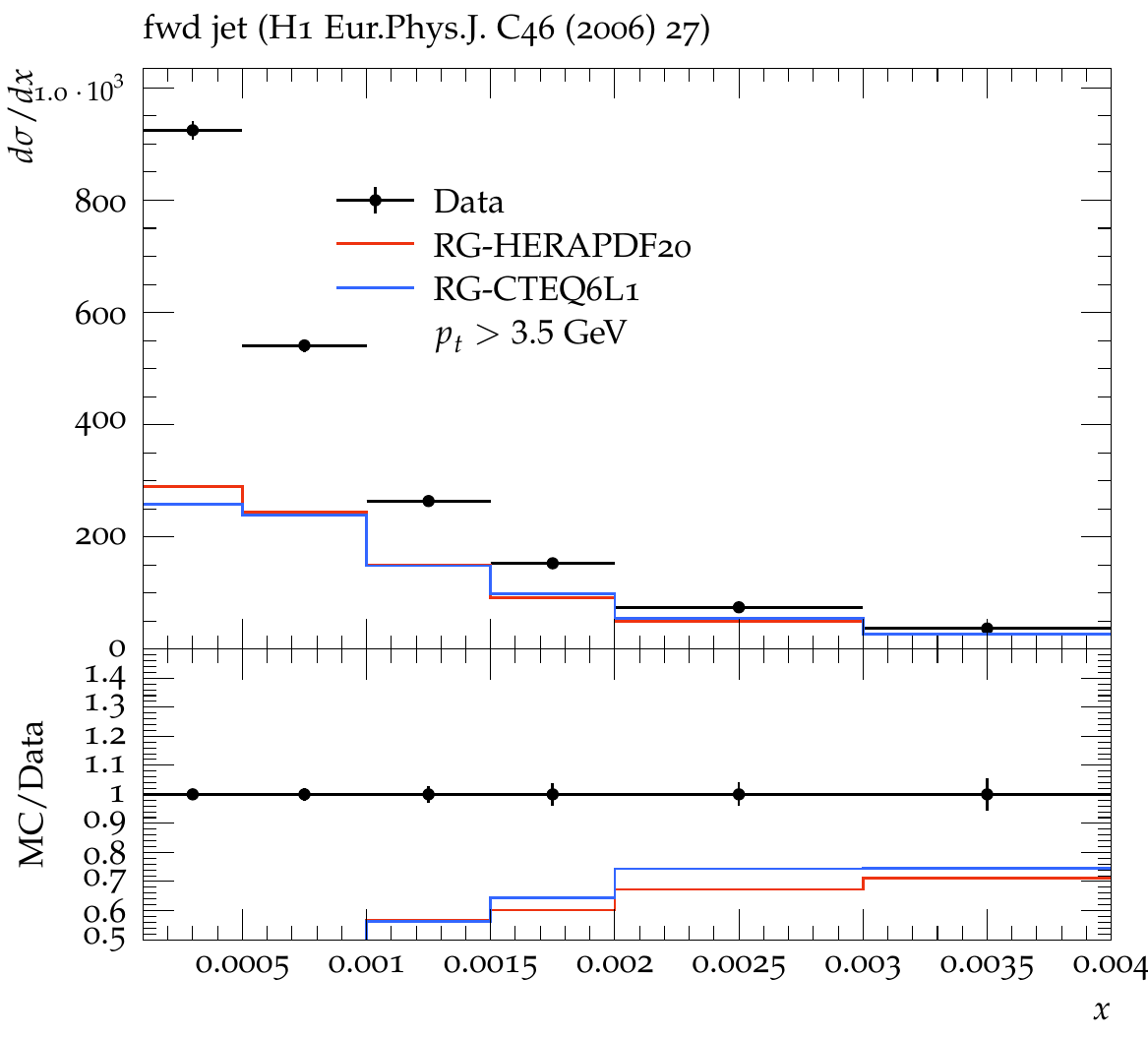}
\caption{Comparison of forward jet cross section as a function of $x$ as obtained from \rivet and the corresponding one from \rivethztool, showing only statistical uncertainties of the measurement.}
\label{fig:H1_2006_I690939_1}
\end{center}
\end{figure}

\begin{figure}[htbp]
\begin{center}
\includegraphics[width=0.5\linewidth]{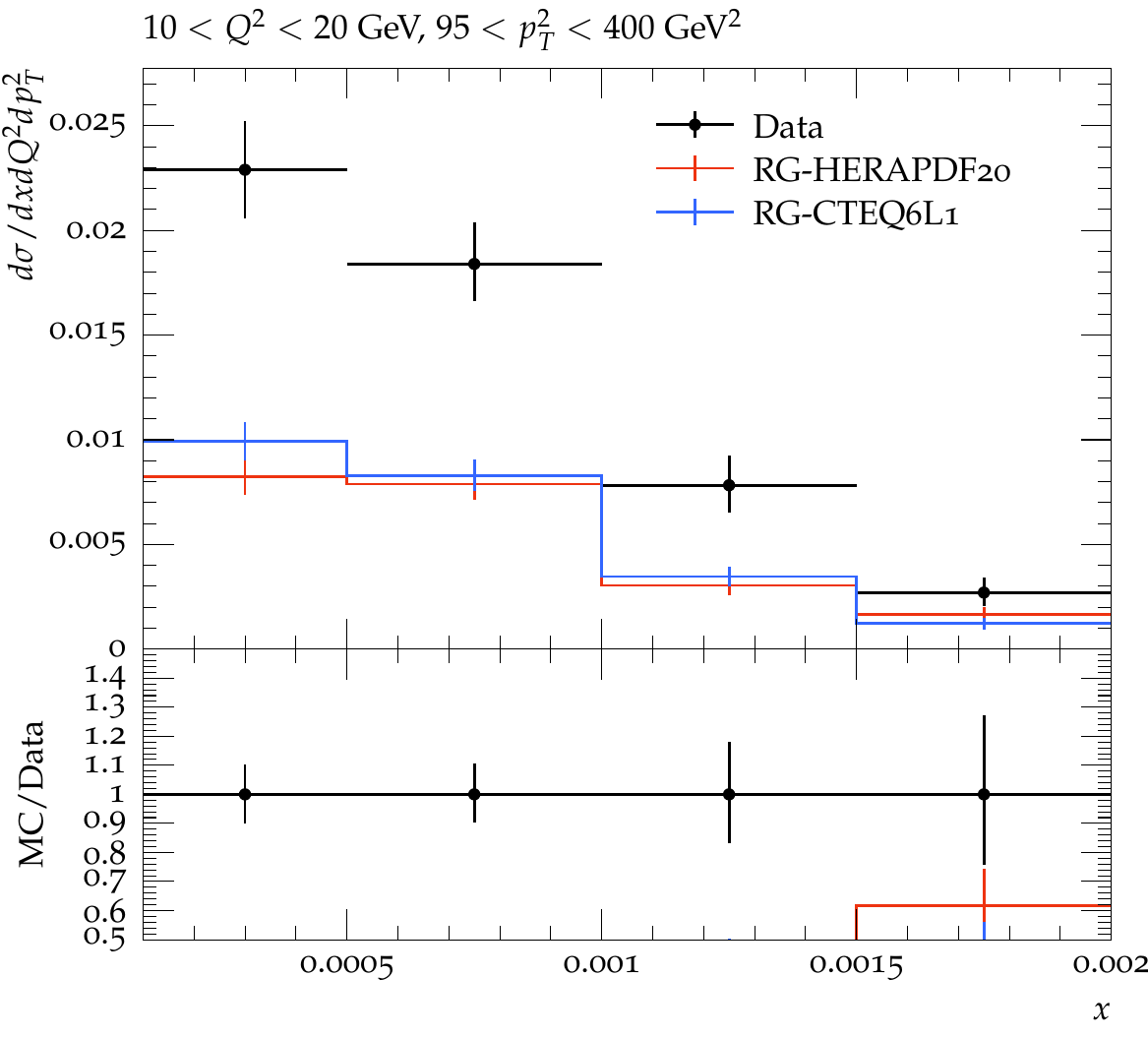}\includegraphics[width=0.5\linewidth]{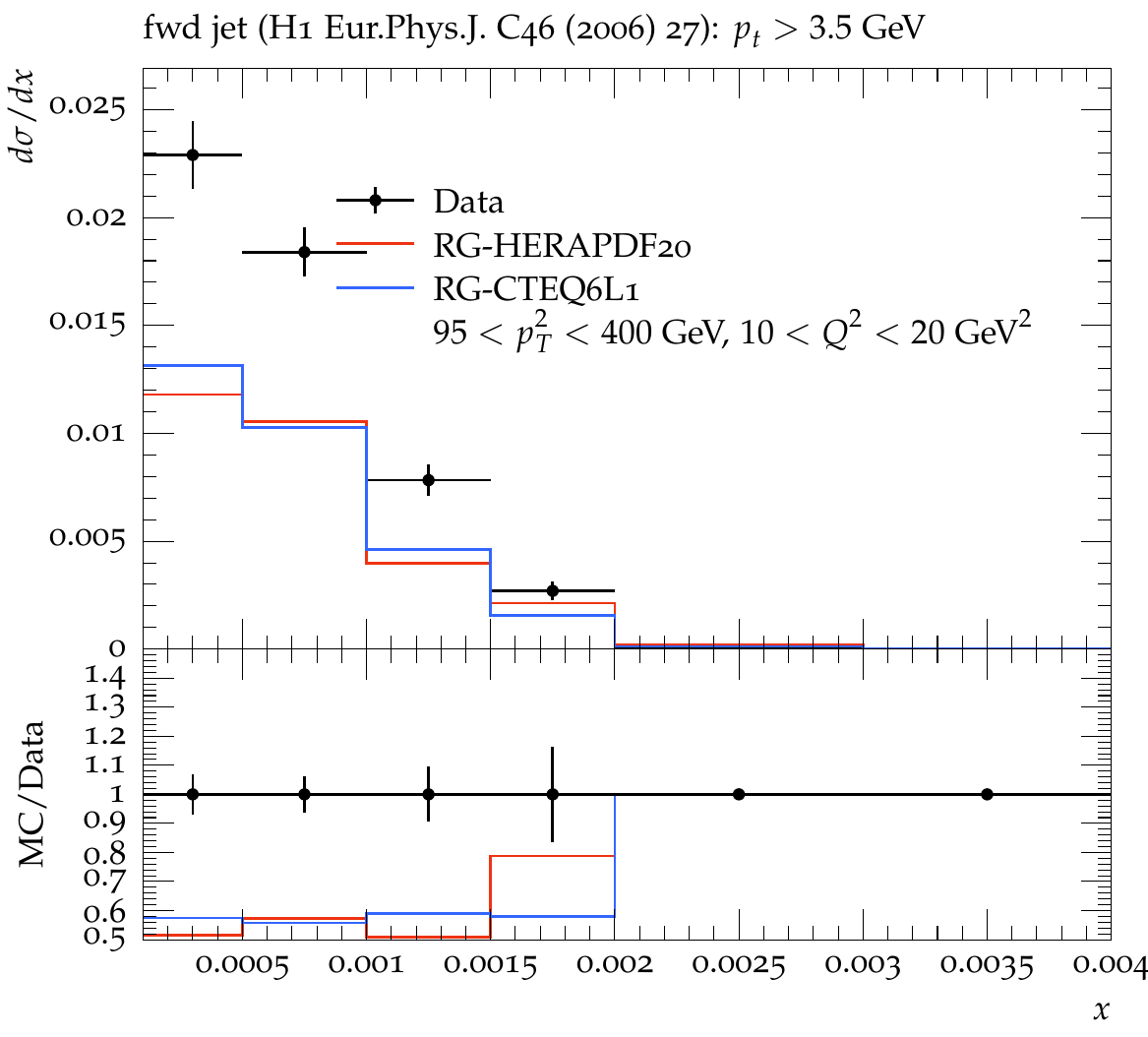}
\caption{Comparison of forward jet cross section as a function of $x$ in a specific bin in $Q^2$ and $p_T^2$ as obtained from \rivet and the corresponding one from \rivethztool. The differences between \rivet and \rivethztool  result from slightly different definitions of the used jet-algorithms}
\label{fig:H1_2006_I690939_2}
\end{center}
\end{figure}

\renewcommand{\thissection}{H1\_2006\_I690939, HZH0508055}
\index{HZH050805}
\index{H1\_2006\_I690939}
\markboth{\thischapter}{\thissection}
{\bf Abstract} (cited from  Ref.~\cite{Aktas:2005up}: "The production of forward jets has been measured in deep inelastic $ep$ collisions at HERA. The results are presented in terms of single differential cross sections as a function of the Bjorken scaling variable ($x_{Bj}$) and as triple differential cross sections $d^3 \sigma / dx_{Bj} dQ^2 dp_{t,jet}^2$, where $Q^2$ is the four momentum transfer squared and $p_{t,jet}^2$ is the squared transverse momentum of the forward jet. Also cross sections for events with a di-jet system in addition to the forward jet are measured as a function of the rapidity separation between the forward jet and the two additional jets."\\
The results of the \rivet plugin\footnote{Authors: Giorgia Bonomelli, Andrea Achilleos} are compared with those from \rivethztool  for the same kinematic range. 
Validation plots for the single differential cross section are shown in Fig.~\ref{fig:H1_2006_I690939_1} and for the triple differential cross section in Fig.~\ref{fig:H1_2006_I690939_2}.
The inclusive jet algorithm, used in this analysis, is not available in \fastjet, therefore small differences between the results obtained from  the \rivet plugin and from  \rivethztool  are visible in 
Figs.~\ref{fig:H1_2006_I690939_1} and \ref{fig:H1_2006_I690939_2}. 

\subsection{Measurement of Event Shape Variables in Deep-Inelastic Scattering at HERA (H1) (H1\_2006\_I699835, HZH0512014)}
\begin{figure}[htbp]
\begin{center}
\includegraphics[width=0.5\linewidth]{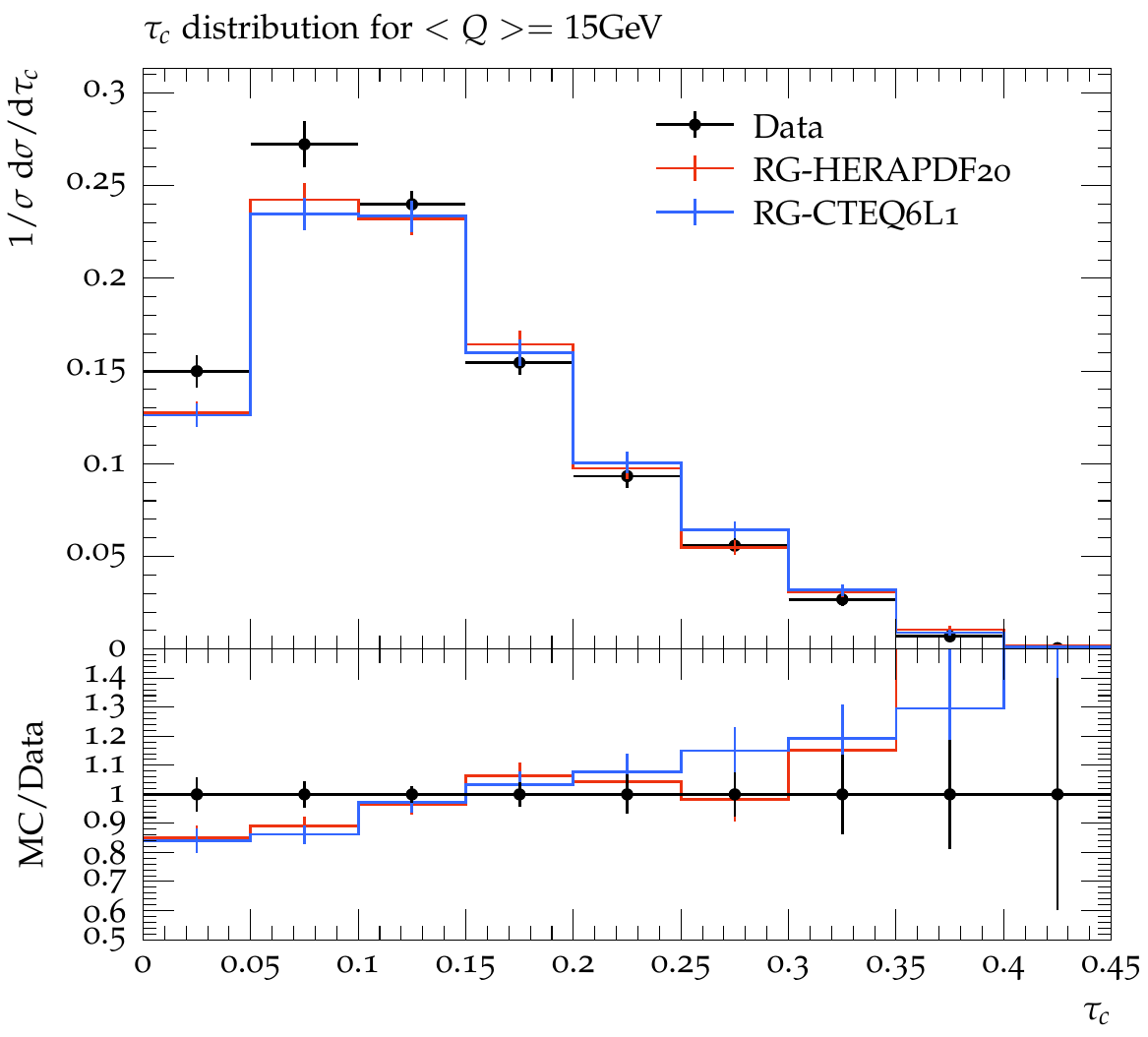}\includegraphics[width=0.5\linewidth]{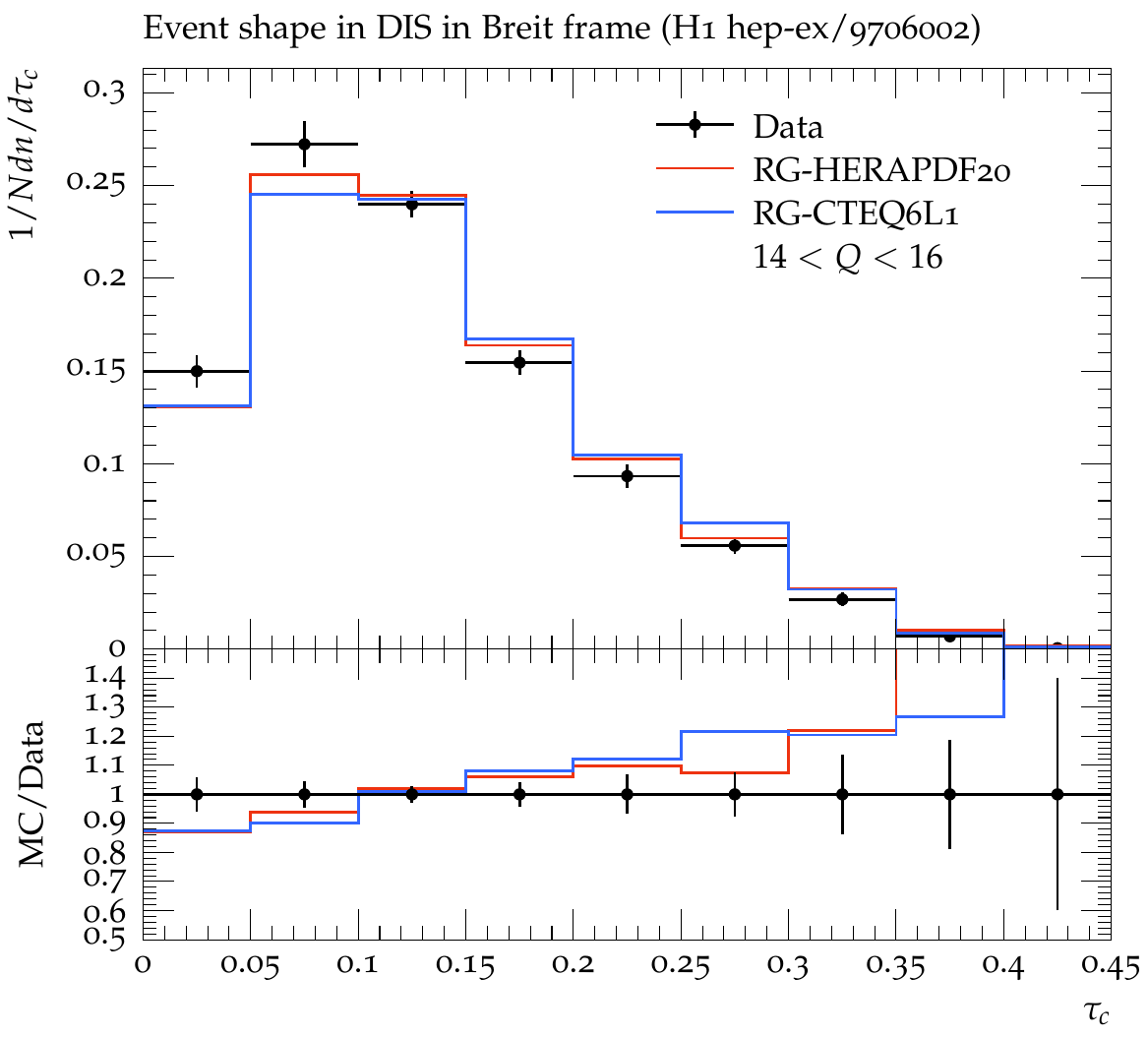}
\caption{Distribution of $\tau_c$  with respect to the thrust axis for $14 < Q < 16$~GeV. The left plot shows results obtained from \rivet, with two different pdfs (HERAPDF2.0 and CTEQ6L1). The right plot shows the corresponding results form \rivethztool, with only statistical uncertainties.}
\label{fig:H1_2006_I699835_1}
\end{center}
\end{figure}

\begin{figure}[htbp]
\begin{center}
\includegraphics[width=0.5\linewidth]{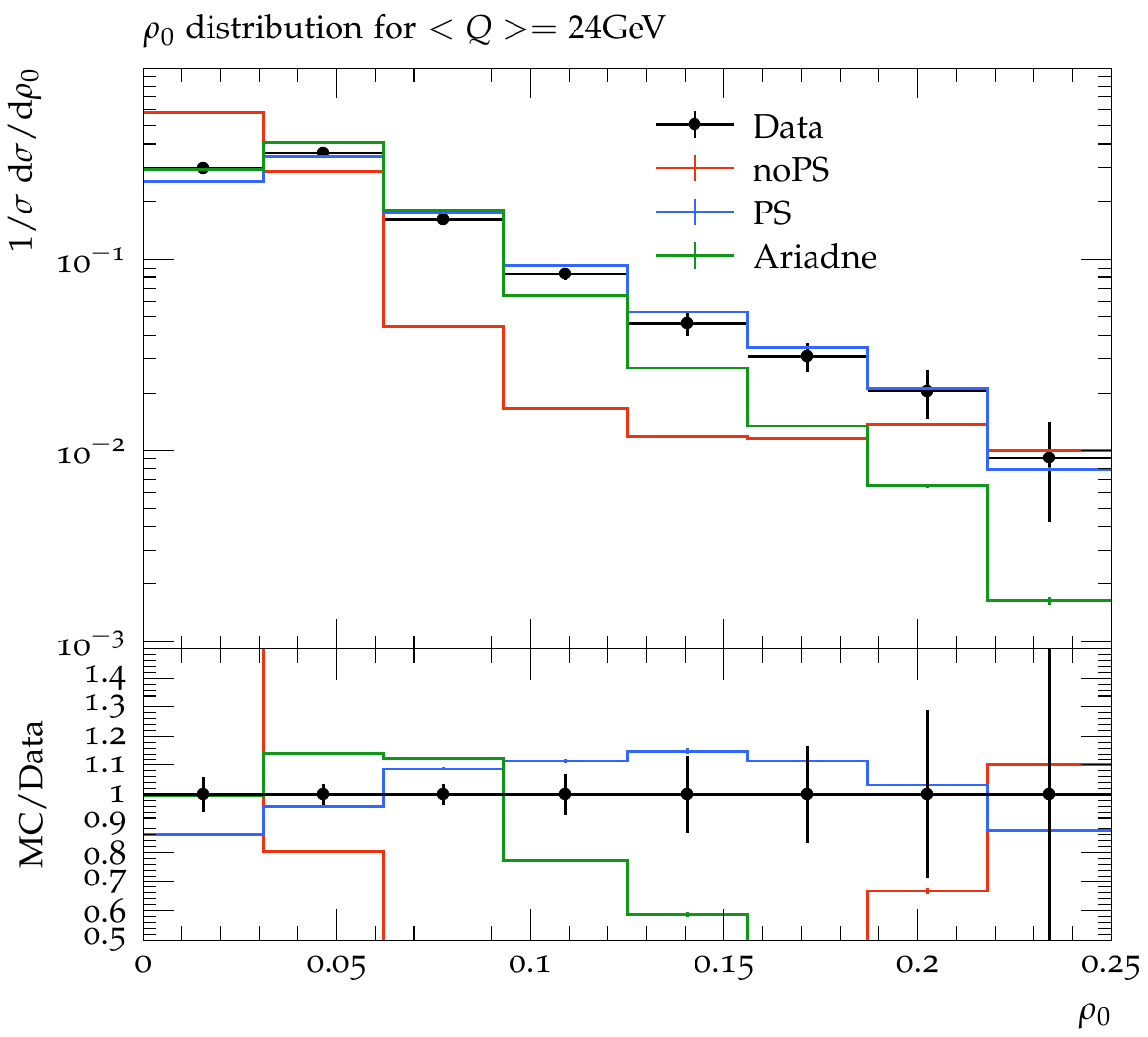}
\caption{Distribution of the normalised jet mass, $\rho_0$,  in the range  $20 < Q < 30$~GeV with no parton shower, with initial and final state parton showers and with the \ARIADNE\ parton shower generator, all obtained with \rivet.}
\label{fig:H1_2006_I699835_2}
\end{center}
\end{figure}

\renewcommand{\thissection}{H1\_2006\_I699835, HZH0512014 }
\index{HZh0512014 }
\index{H1\_2006\_I699835}
\markboth{\thischapter}{\thissection}
{\bf Abstract} (cited from Ref.~\cite{H1:2005zsk}): "Deep-inelastic $ep$ scattering data taken with the H1 detector at HERA and corresponding to an integrated luminosity of $106 \,{\rm pb}^{-1} $ are used to study the differential distributions of event shape variables. These include thrust, jet broadening, jet mass and the $C$-parameter. The four-momentum transfer $Q$ is taken to be the relevant energy scale and ranges between 14 GeV and 200 GeV. "\\
The results of the \rivet plugin\footnote{Author: Alejandro B.~Galv\'an} are compared with those from \rivethztool  for the same kinematic range. 
Validation plots are shown in Fig.~\ref{fig:H1_2006_I699835_1}. In  Fig.~\ref{fig:H1_2006_I699835_2} the measurement is compared with predictions using also the ARIADNE parton radiation.

\subsection{Measurement of inclusive production of \boldmath$D^{*\pm}$ mesons both with and without dijet production in DIS collisions at HERA  (H1) (H1\_2007\_I736052, HZH0701023)}
\renewcommand{\thissection}{H1\_2007\_I736052, HZH0701023}
\index{HZH0701023}
\index{H1\_2007\_I736052}
\markboth{\thischapter}{\thissection}
{\bf Abstract} (cited from  Ref.~\cite{Aktas:2006py}): "Inclusive $D^*$ production is measured in deep-inelastic $ep$ scattering at HERA with the H1 detector. In addition, the production of dijets in events with a $D^*$ meson is investigated. The analysis covers values of photon virtuality $2< Q^2 \leq 100$ GeV$^2$ and of inelasticity $0.05 \leq y \leq 0.7$. Differential cross sections are measured as a function of $Q^2$ and $x$ and of various $D^*$ meson and jet observables. " \\

\begin{figure}[htbp]
\begin{center}
\includegraphics[width=0.5\linewidth]{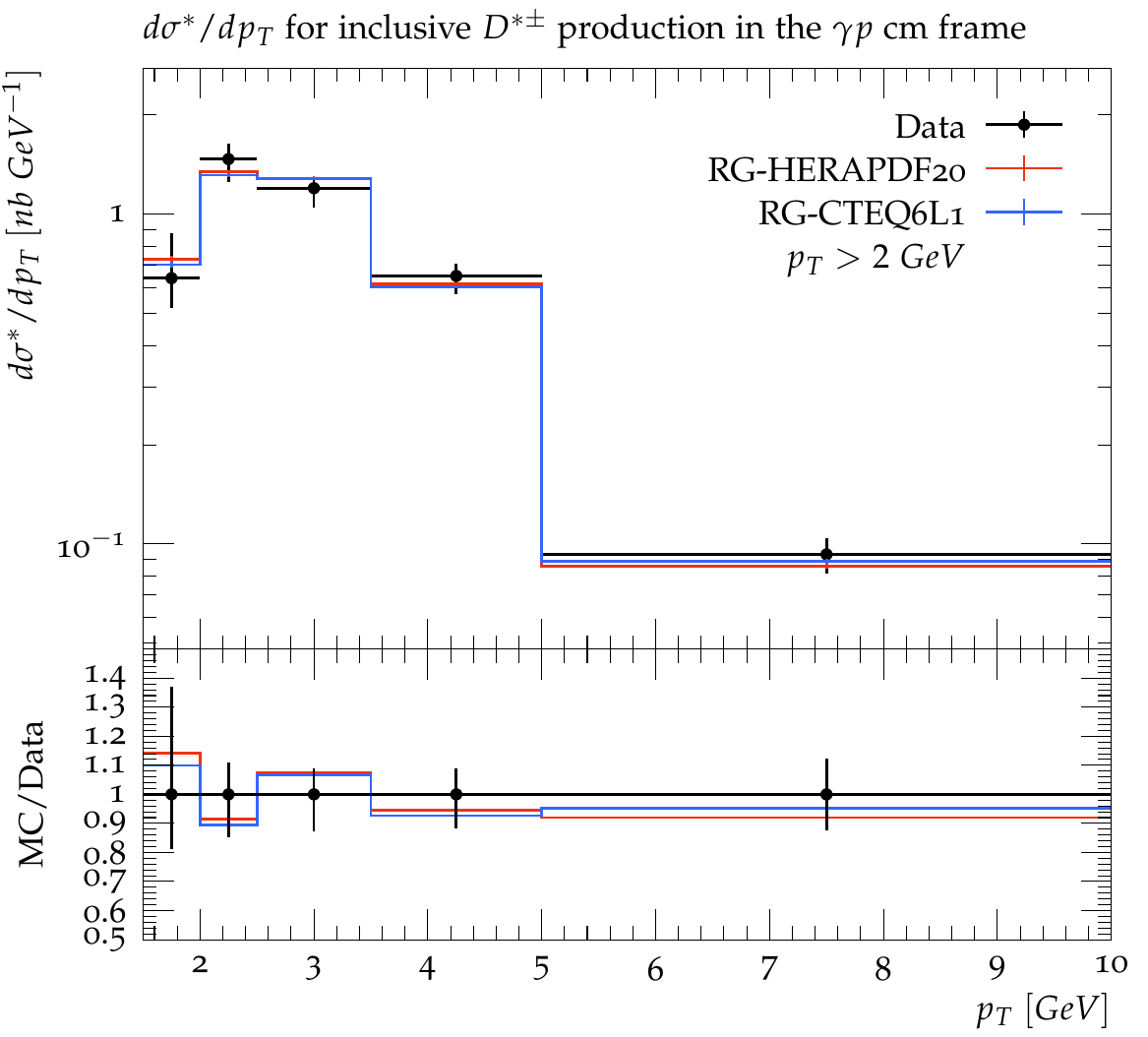}\includegraphics[width=0.5\linewidth]{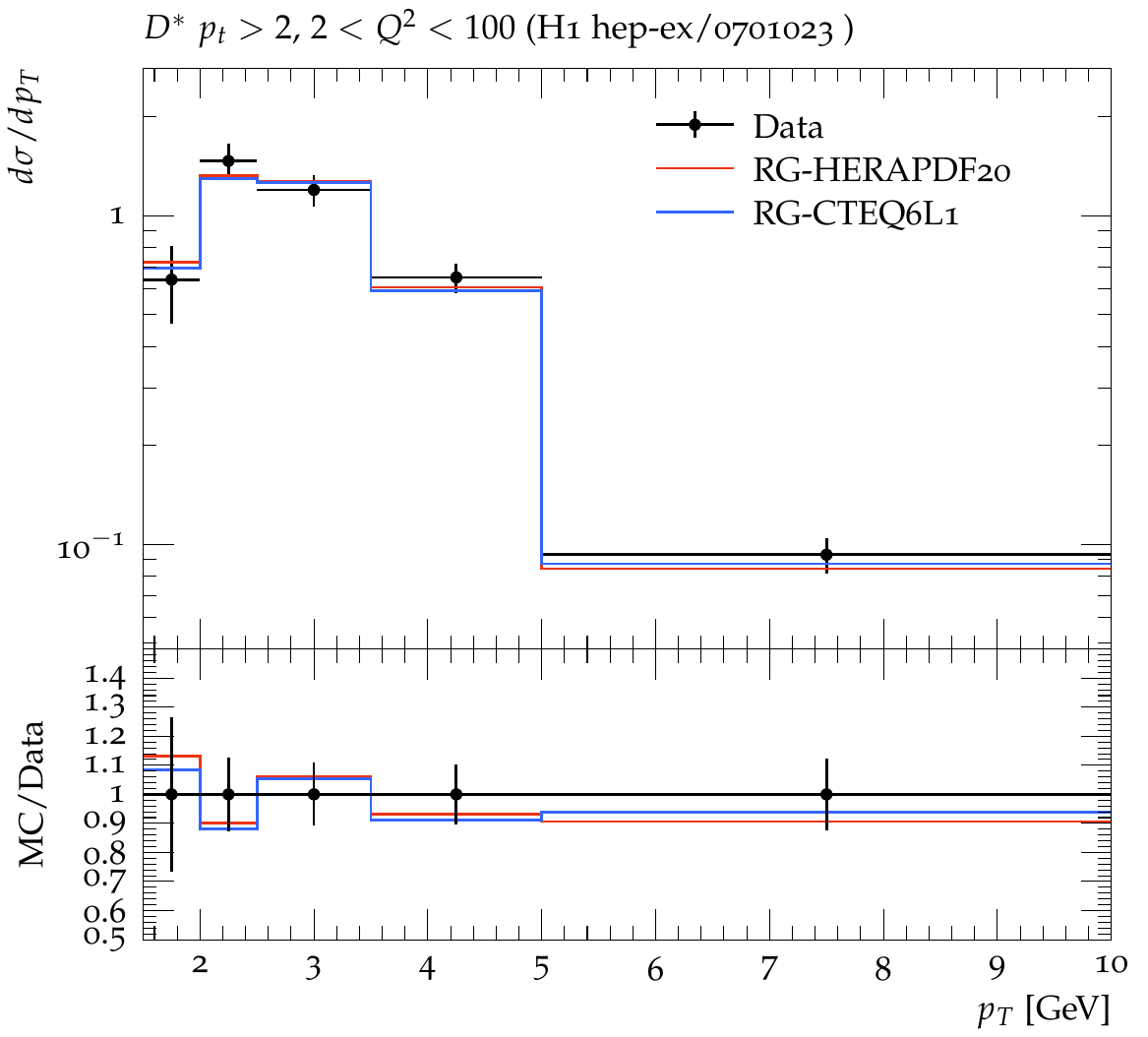}
\caption{Comparison of results obtained from \rivet (left) and from \rivethztool (right). Differential cross sections for inclusive $D^{*\pm}$ meson production with the additional requirement $p_T \textgreater 2.0$ \GeV\ for the $D^{*\pm}$ meson in the $\gamma p$ center-of-mass frame as a function of $p_T$.}
\label{fig:H1_2007_I736052_1}
\end{center}
\end{figure}

The results of the \rivet plugin\footnote{Author: Maksim Davydov} are compared with those from \rivethztool  for the same kinematic range. 
Validation plots are shown in Fig.~\ref{fig:H1_2007_I736052_1}. 

\subsection{Inclusive dijet cross sections in neutral current deep inelastic scattering at HERA (ZEUS) (ZEUS\_2010\_I875006)} 
\begin{figure}[htbp]
\begin{center}
\includegraphics[width=0.5\linewidth]{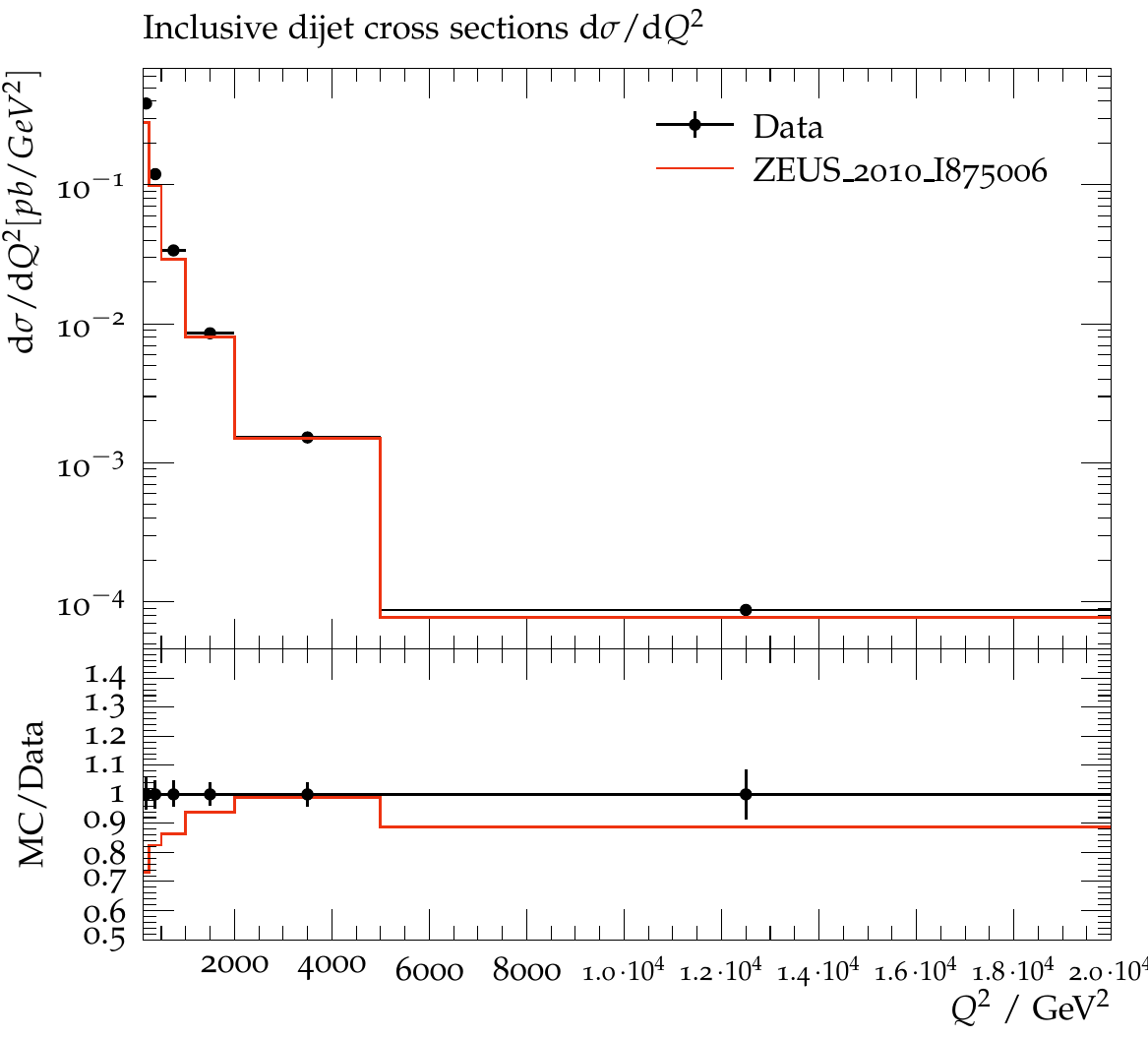}\includegraphics[width=0.5\linewidth]{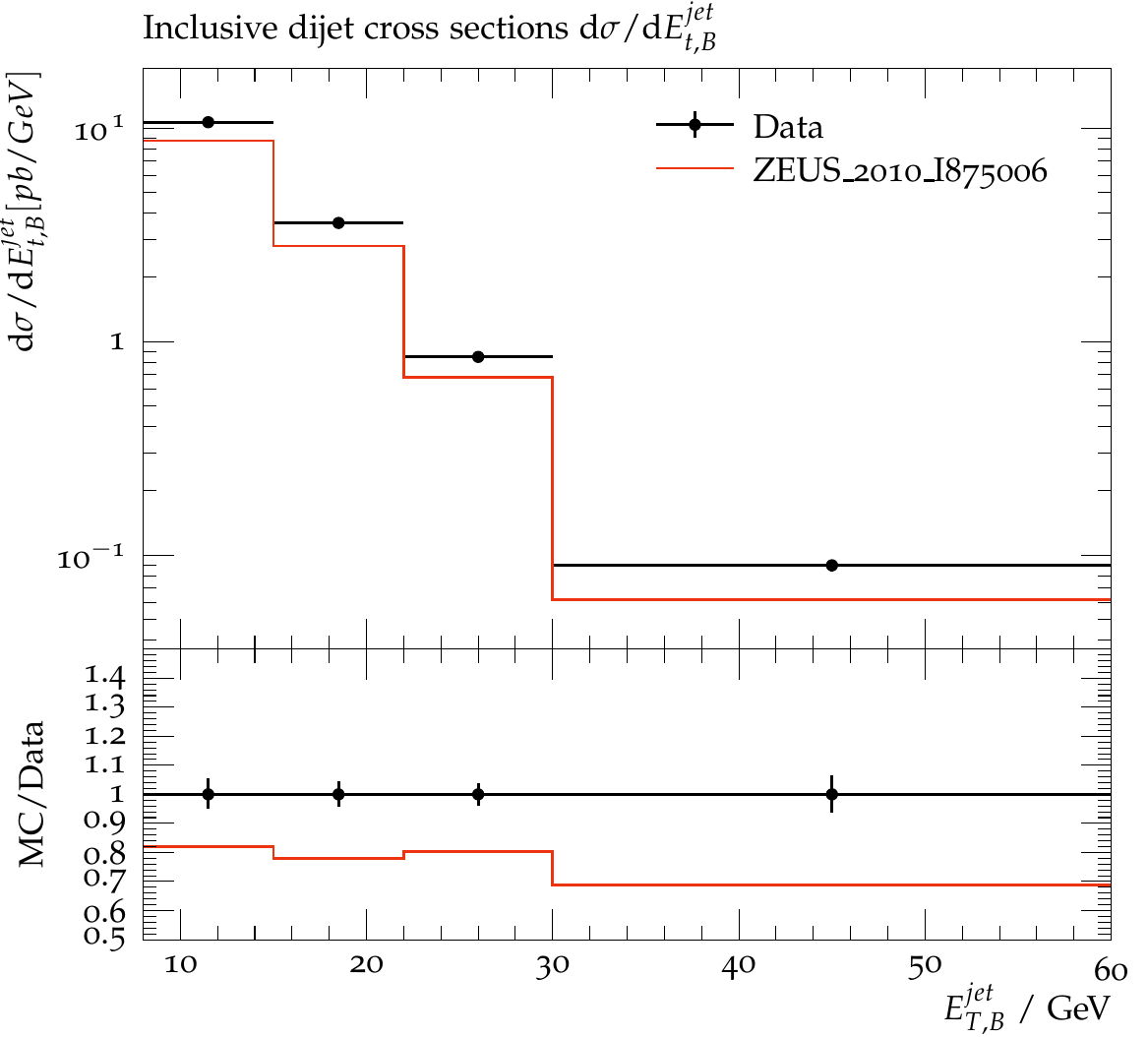}
\caption{Comparison of results obtained from \rivet using HERWIG MC. Differential cross sections for the production of dijets in neutral current DIS as a function of $Q^2$ (left) and jet transverse energy in the Breit frame, $E_{T, B}^{jet}$ (right).}
\label{fig:ZEUS_2010_I875006}
\end{center}
\end{figure}

\renewcommand{\thissection}{ZEUS\_2010\_I875006}
\index{ZEUS\_2010\_I875006}
\markboth{\thischapter}{\thissection}
{\bf Abstract} (cited from  Ref.~\cite{ZEUS:2010vyw}): "Single- and double-differential inclusive dijet cross sections in neutral current deep inelastic $ep$ scattering have been measured with the ZEUS detector using an integrated luminosity of 374\,pb$^{-1}$. The measurement was performed at large values of the photon virtuality, $Q^2$, between 125 and 20000 GeV$^2$. The jets were reconstructed with the $k_T$ cluster algorithm in the Breit reference frame and selected by requiring their transverse energies in the Breit frame, $E_{T,B}^{jet}$, to be larger than 8 GeV. In addition, the invariant mass of the dijet system, $M_{jj}$, was required to be greater than 20 GeV. "\\
The results of the \rivet plugin\footnote{Author: Jacob Shannon} are  shown in Fig.~\ref{fig:ZEUS_2010_I875006} using the \herwig~7.2 Monte Carlo generator~\cite{Bellm:2015jjp} with proton proton parton density function MMHT 2014~\cite{Harland-Lang:2014zoa}.

\section{Outlook and Conclusion}

The summer student program 2021 at DESY was run as a fully online program, with video sessions every day for several hours for 8 weeks.  
In a dedicated effort of summer students in 2021 at DESY several older HERA analyses have been coded into the \rivet package, making use of \rivethztool for validation, where available. In some cases, inconsistencies of the data stored on HEPData were found and corrected after approval by the responsible persons from the experiments.
\begin{tolerant}{2000}
Through these re-codings of analysis logic in the predominant modern analysis-preservation framework, the summer students in 2021 have made a very significant contribution to preservation and re-interpretation of measurements from the HERA experiments.
\end{tolerant}

\bibliographystyle{mybibstyle-new.bst}
\raggedright

\bibliography{/Users/jung/Bib/hannes-bib}

\end{document}